\documentclass[preprint,aps,pre,preprintnumbers,amsmath,amssymb,nofootinbib,floatfix]{revtex4-1}
\usepackage[utf8]{inputenc}
\usepackage[colorlinks=true, urlcolor=blue, anchorcolor=blue,citecolor=blue,filecolor=blue,linkcolor=blue,menucolor=blue]{hyperref}
\usepackage{graphicx,color}
\usepackage{caption}
\usepackage{subcaption}
\usepackage{amsmath,amssymb}
\usepackage{epstopdf}
\usepackage{soul}
\usepackage{tabularx}
\usepackage{enumitem}
\usepackage{graphicx}%
\usepackage{subcaption}%
\usepackage{multirow}%
\usepackage{amsmath,amssymb,amsfonts}%
\usepackage{physics}
\usepackage{bm}
\usepackage{mathrsfs}
\usepackage[title]{appendix}%
\usepackage{xcolor}%
\usepackage{textcomp}%
\usepackage{booktabs}%
\usepackage{import} %
\usepackage{float} %
\usepackage{hhline} %

\def\al#1\eal{\begin{align}#1\end{align}}
\newcommand{\bs}{\mathbf{s}}

\newcommand{\bR}{\mathbf{R}}

\begin{document}

\title[Article Title]{
Emergent memory in cell-like active systems}

\author{M. Besse}
\affiliation{Laboratoire Jean Perrin, CNRS/Sorbonne Université, Paris, France}

\author{R. Voituriez}
\affiliation{Laboratoire Jean Perrin, CNRS/Sorbonne Université, Paris, France}

\begin{abstract}
    Active  systems across scales, ranging from molecular machines to human crowds~\cite{Toner:2005,Jlicher:2007rx,Bricard:2013aa,Marchetti:2013wd,Prost:2015la,Bechinger:2016aa,TrepatSahai2018NatPhys,Alert:2020aa,Gu:2025aa}, are usually modeled as assemblies of self-propelled particles driven by internally generated forces~\cite{TailleurCates2008RTP,cates-arcmp-2015,Bechinger:2016aa}. However, these models often assume memoryless dynamics and no coupling of internal active forces to the environment. Here, guided by the example of living cells, which have recently been shown  to  display multi-timescale memory effects~\cite{dAlessandro:2021wx,Alraies:2024aa,Kalukula:2025aa,Arya2025.07.02.662855,Jacques:2023aa}, we introduce a general theoretical framework that goes beyond this paradigm by incorporating internal state dynamics and environmental sensing into active particle models. We show that when the self-propulsion of an agent  depends on internal variables with their own complex dynamics—modulated by local environmental cues— environmental memory spontaneously emerges and gives rise to new classes of behaviours. These include memory-induced responses, adaptable localization in complex landscapes, suppression of motility-induced phase separation~\cite{cates-arcmp-2015}, and enhanced jamming transitions~\cite{VanHecke2010JammingReview,BerthierBiroli2011RMP}. Our results demonstrate how minimal information processing capabilities, intrinsic to non-equilibrium agents with internal states like living cells, can profoundly influence both individual and collective behaviours. This framework bridges cell-scale activity and large-scale intelligent motion in cell assemblies, and opens the way to the quantitative analysis and design of systems ranging from synthetic colloids~\cite{Bricard:2013aa,Bechinger:2016aa,Chen:2025aa} to biological collectives~\cite{TrepatSahai2018NatPhys,Alert:2020aa,Gu:2025aa} and robotic swarms~\cite{Rubenstein2014KiloBots,swarmrobotics}.
    \end{abstract}

\maketitle

\section*{Introduction}

Living organisms must sense and respond to environmental signals to fulfill basic yet vital functions such as resource uptake, mating, or escape from  hazards. Such constraints apply down to the scale of single cells, which have been shown -- depending on cell types -- to display a broad range of specialized sensory mechanisms that can initiate dedicated motile responses to external cues of various nature,  prominent examples being chemotaxis~\cite{Parent:1999aa,Levine2006,Levine2013Chemotaxis} or rigidity~\cite{Discher:2005jt,Trichet:2012vn} and topography~\cite{Le-Berre:2013fk,Reversat:2020aa} sensing.  
From a physics perspective, the dynamics of motile cells -- or larger scale organisms --  can be generically modelled by active random walks~\cite{Selmeczi:2008vn,Maiuri:gf,Bechinger:2016aa}. In this context activity refers to the fact that motile cells irreversibly consume energy to self-propel, and such active self-propulsion has been shown to have spectacular consequences at the scale of cell collectives, such as collective motion, cell turbulence, or microphase separation, with applications far beyond cell biology~\cite{Toner:2005,Marchetti:2013wd,cates-arcmp-2015,Alert2022ActiveTurbulence}.

Activity or nonequilibriumness can fundamentally be rooted in the breaking of time reversal symmetry of  microscopic processes. In the context of active matter and in particular of living cells, the focus has been so far on the out-of-equilibrium molecular processes enabling force generation -- a prominent example being molecular motors and treadmilling protein filaments, which at the cellular scale can  lead to self-propulsion~\cite{Jlicher:2007rx,Prost:2015la,Maiuri:gf,Liu:fk}. At large scales, the dynamics of  self-propelled particles can be minimally captured by a class of active random walks of the form
\al
\gamma \dot{x} = p + F_{\rm ext} + \xi \,, \label{eq:force balance} 
\eal
where $x$ refers to the position of the particle (e.g. a cell). This equation expresses force balance on the cell, assumed to be overdamped with linear viscous friction of coefficient $\gamma$ and subjected to: $(i)$ a self-propulsion force $p$, here identified as  cell polarity for simplicity~; $(ii)$ a generic external force $F_{\rm ext}(x)$ that recapitulates the mechanical interactions with the environment other than friction (geometric confinement, external force fields, or other cells) ; $(iii)$ a noise term $\xi$ of both thermal and active origin that can thus be colored. Importantly, so far most of models of active particles~\cite{TailleurCates2008RTP,Bechinger:2016aa,Martin2021AOUPReview} have considered  descriptions of the cell polarity $p$ that are  memoryless -- with the exception of recent  data-driven phenomenological  models of cell migration~\cite{Mitterwallner2020NonMarkovianMotility,Klimek2024NonMarkovianClassification,Klimek2025IntrinsicVariance,Bruckner2021LearningPNAS,Bruckner2024LearningReview} -- and fully independent of the cell environment -- with the exception of specific aligning interactions either with other particles as in the Vicsek model~\cite{vicsek1995novel}, or with an external field as in chemotaxis models~\cite{Levine2006,Celani:2010,Levine2013Chemotaxis}.

However, activity is also known to impact various processes that regulate cell shape, polarity, adhesion properties or receptors activity, and more generally all biochemical networks that control the cell state~\cite{Geiger2009FAsensing,Chugh2018ActinCortex,Ridley2003MigrationReview,Dupont2011YAPTAZ}. Beyond self-propulsion per se, activity can thus control cell dynamics via information processing, i.e. by adapting the cell state in response to its complex interactions with the environment. In general, the cell state $\bs$ can be thought of as a high dimensional vector that recapitulates all of its internal degrees of freedom, either biochemical (protein levels, gene expression status) or physical (cell shape, intracellular organelle organisation), and is in practice intractable as a whole. Many works have focused on the dynamics of subcomponents of $\bs$, including the response to external stimuli of various nature (chemical or physical), notably at the level of protein regulation networks~\cite{Barkai1997Robustness,Purvis2013SignalingDynamics} or gene expression networks~\cite{Dupont2011YAPTAZ,Balazsi2011CellularDecision}. The dynamics of the cell state can be formally written without loss of generality
\al
\dot \bs=\bR(\bs,\bm{\mathcal{E}}_x),\label{eq:sdot}
\eal where $\bR$ encodes  the cell response to the set $\bm{\mathcal{E}}_x$ of external stimuli of the environment sensed by the cell at  position $x$. Observations and models have shown that complex temporal patterns emerge, at time scales ranging from seconds to days depending on the function $\bR$ that can take a broad range of non linear forms~\cite{Purvis2013SignalingDynamics}. 

In general, cell polarity $p$ as defined above is determined by the internal state variables $\bs$ of the cell, so that one can write $p(\bs)$ (see Fig.~\ref{fig:protocols}a). Its dynamics is thus expected to be complex, meaning notably that $(i)$ it displays strong memory effects involving a potentially broad range of time scales induced by the complex dynamics of the internal degrees of freedom $\bs$ and $(ii)$ it depends on the environment stimuli $\bm{\mathcal{E}}_x$ experienced by the cell along its path $\{x_{t'}\}_{t'\le t}$ until time $t$. We argue that properties $(i)$ and $(ii)$ are intrinsic features of cell dynamics, which more generally apply to any nonequilibrium agent endowed with internal degrees of freedom and sensors of its environment. Strikingly, increasing experimental evidence reports that migrating cells of various types can indeed display   memory effects at different time scales~\cite{dAlessandro:2021wx,Alraies:2024aa,Kalukula:2025aa,Arya2025.07.02.662855,Jacques:2023aa}. While the underlying cellular or molecular mechanisms can be of very different nature, such as footprint deposition in the environment~\cite{dAlessandro:2021wx,Barbier-Chebbah:2022aa,Jacques:2023aa}, nuclear deformations~\cite{Alraies:2024aa,Arya2025.07.02.662855} or cell cortex dynamics~\cite{Kalukula:2025aa}, all these observations support the generic scenario that we propose. Features (i) and (ii) have so far been left aside in theoretical models of cell migration; we present in this paper a systematic theoretical framework to bridge this gap, which leads us to introduce a generic class of non Markovian active particles. Our analysis of minimal models of this class provides proofs of principle  that  memory abilities and the non-equilibrium nature of the dynamics have deep,  underexplored consequences both at the single-cell and collective levels: emergence of environment-sensitive response that can lead to controllable localization in complex environments, suppression of motility induced phase separation or enhanced jamming transition. Although our analysis is based on linear response and primarily explores simple single-exponential memory kernels, we expect that more complex memory functions and possibly nonlinear responses — which in principle could be inferred from experimental data~\cite{ferretti2020building,Mitterwallner2020NonMarkovianMotility,Klimek2024NonMarkovianClassification} — will reveal an even broader spectrum of behaviours and lead to experimentally testable predictions ; these  extensions are left for future work. Beyond cell behaviour models, our work opens the way to the description of a broad class of  active agents -- living or artificial -- with minimal information processing abilities, such as active colloids~\cite{Bricard:2013aa,Bechinger:2016aa,Chen:2025aa}, animal or human crowds~\cite{Gu:2025aa} or robot swarms~\cite{Rubenstein2014KiloBots,swarmrobotics}.  

\section*{General framework: non Markovian active particles}

\begin{figure}
\centering
\begin{subfigure}[b]{\textwidth}
  \includegraphics[width=\textwidth]{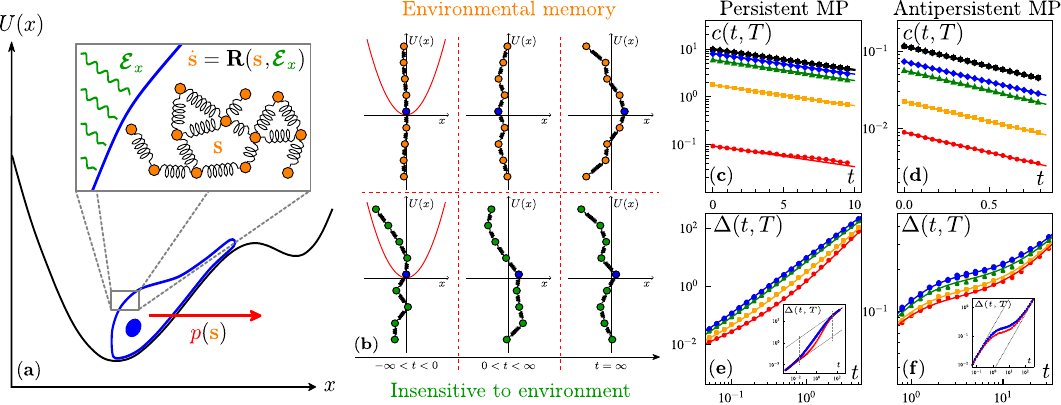}
\end{subfigure}
\caption{$\mathbf{(a)}$ Minimal model of crawling cell (blue) interacting with an external environment $U$. Its internal degrees of freedom $\mathbf{s}$ (orange beads) evolve in response to the local environment $\bm{\mathcal{E}}_x$ (green). To linear order, the response function $\mathbf{R}$ can be modelled by a network of (non reciprocal) harmonic springs. The  self-propulsion force $p(\mathbf{s})$ drives the motion of the cell through the landscape. $\mathbf{(b)}$ Trap-and-release protocol applied to  examples of linear spring networks: shaken Rouse chain (see SM IV for definition), which displays environmental memory (top) and equilibrium  Rouse chain, with environment-insensitive dynamics (bottom). For $t<0$, the particle (blue monomer of position $x$)  is confined by an harmonic trap, then released at $t=0$.  $\mathbf{(c)}$ Memory-induced transient dynamics of the polarity correlator $c(t,T)=\langle p(t+T)p(T) \rangle$ for persistent MPs. Colours from red to blue denote increasing observation times after release $T/\tau_m = 0, 0.1, 0.5, 1$. Blue also corresponds to stationary state. Symbols show numerics, lines analytics (SM VI). Black symbols: AOUPs, insensitive to the protocols. $\mathbf{(d)}$ Same analysis for antipersistent MPs. Black symbols: equilibrium dimers, also protocol-insensitive. $\mathbf{(e)}$ Mean squared increments $\Delta(t,T)=\langle \left( x(t+T)-x(T) \right)^2 \rangle$ for persistent MPs, highlighting again protocol-dependent transients (analytics in SM VI). Curves for the control protocol collapse on the stationary (blue) line. Inset: crossover from short-time diffusion $D$ to long-time  diffusion $D_\star$. $\mathbf{(f)}$ Same as in $\mathbf{(e)}$ for antipersistent MPs. Simulation details in SM VIII.}
\label{fig:protocols}
\end{figure}

Without loss of generality, we start from the formal dynamics of a cell of position $x$ and polarity $p$ defined by Eqs.~\ref{eq:force balance}-\ref{eq:sdot}, which are intractable in general because of the high dimensionality of $\bs$ and of the unspecified functions $\bR(\bs,\bm{\mathcal{E}}_x)$ and $p(\bs)$ (see Fig.~\ref{fig:protocols}a) ; we keep one-dimensional notations for simplicity but the theory naturally extends to any space dimension, as illustrated below. To make progress, our central hypothesis is twofold. First, for the sake of simplicity, we primarily focus on the cell  response to mechanical stimuli and thus postulate that $\bR(\bs,\bm{\mathcal{E}}_x)$ depends on the environment stimuli $\bm{\mathcal{E}}_x$ only via the external forces $F_{\rm ext}(x,t)$ and the (rescaled) friction force $\dot x$. Note that an explicit dependence on external fields, as in the case of chemotaxis can in principle be taken into account similarly by redefining an effective external force \cite{Jakuszeit:2025aa}. Second,  we focus on the linear response of the cell state to perturbations as a first step. Expanding Eqs.~\ref{eq:force balance}-\ref{eq:sdot} (see SM II) then leads to the following generalized Langevin equation (GLE), which is at the core of this paper:
\al
\int_{-\infty}^t \dot{x}(t')K(t-t') = F_{\rm ext}(x(t)) + \sqrt{2 D} \eta(t), \quad \langle \eta(t) \eta(t') \rangle = G(\abs{t-t'}). \label{eq:noneqGLE}
\eal

Several comments are in order. $(i)$ As expected~\cite{zwanzig1961memory,mori1965transport,Zhao:2024aa}, integrating out the dynamics of the internal degrees of freedom $\bs$ makes non Markovian the dynamics of the cell's position $x$, as it is apparent from the convolution term involving the memory kernel $K$ and from the Gaussian colored noise $\eta$. Note however that  these two terms, while leading both to non Markovian dynamics of $x(t)$,  differ strikingly. The convolution term integrates past events of the dynamics and is thus sensitive to the history of stimuli $\{\bm{\mathcal{E}}_{x_{t'}}\}_{t'\le t}$ experienced by the cell, thus endowing the cell with a memory of its environment. For clarity, a process that satisfies Eq.~\ref{eq:noneqGLE} with a non trivial kernel (ie $K(t)\not\propto\delta(t)$) will be called hereafter endowed with environmental memory. In contrast the  noise $\eta$, even if correlated,  is independent of the environment.

$(ii)$ Importantly, Eq.~\ref{eq:noneqGLE} conserves its plain physical meaning of (overdamped) force balance. In particular $F_{\rm ext}(x)$ has the meaning of a standard deterministic force, while $\eta$ models fluctuating forces of both thermal and active origin. It is thus well-suited to  describe generic cell/environment or cell/cell interactions as we show below (see SM II for details). In the following we assume for simplicity that $F_{\rm ext}(x)$ derives from a potential $U(x)$, which typically models the interaction of the cell with its environment or with other cells. 

$(iii)$ Because the dynamics of $x$ and $\bs$ as defined by Eqs.~\ref{eq:force balance}-\ref{eq:sdot} are in general out of equilibrium, there is no restriction on the functional form of $K$, which is in particular independent of $G$. To single out the effect of the memory kernel $K$, we assume in the following that $G(t)= \delta(t)$. This assumption is in practice not very restrictive as discussed in SM III. If the functional form of $K$ can be either inferred from experimental data~\cite{ferretti2020building,Mitterwallner2020NonMarkovianMotility,Klimek2024NonMarkovianClassification} or derived from examples of explicit dynamics for $\bs$ (see SM II-IV), we assume hereafter that $K$ is given. More precisely, we first consider simple exponential forms, which as we proceed to show feature a rich phenomenology and  ensure both environmental memory ($K(t)\not\propto\delta(t)$) and the breaking of detailed balance ($K \not\propto G$). The analysis of more complex kernels is left for further works and opens the way to an even  broader spectrum of behaviours.

$(iv)$ Classical active particle models, such as active Ornstein Uhlenbeck particles (AOUPs)~\cite{Martin2021AOUPReview}, can be cast in the form of Eq.~\ref{eq:noneqGLE} by taking  time $K(t) \propto \delta(t)$ and $G(\abs{t})\propto e^{-|t|/\tau}$. This clearly shows that AOUPs have no environmental memory as defined above, despite being out of equilibrium and having a correlated propulsion force. AOUPs thus stand in contrast to the class of generic models (with $K(t)\not\propto\delta(t)$) that is the focus of this paper. 

$(v)$ Formally comparable GLEs have been studied  in various contexts, in particular that of complex viscoelastic fluids~\cite{microgle}, where however fluctuation-dissipation relations impose strong constraints ($K(t)\propto G(t)$) that we relax in this paper. Since the pioneering works of Zwanzig~\cite{zwanzig1961memory} and Mori~\cite{mori1965transport}, various examples of  non-equilibrium GLEs have also been derived, e.g. in the context of active polymers~\cite{Vandebroek2015ActiveViscoelasticPolymer} or active glasses~\cite{berthier2013nonequilibrium,Szamel2015GlassyDynamicsSPP}, or introduced as phenomenological and data-driven models of cell migration~\cite{Mitterwallner2020NonMarkovianMotility,Klimek2024NonMarkovianClassification,Klimek2025IntrinsicVariance}. In contrast to these earlier works, we argue here that non-equilibrium GLEs generically provide minimal models of active particles with internal degrees of freedom, and we show that their coupling to environmental cues leads to new emerging behaviours.  

We now discuss more precisely the form of memory kernel $K$ that we study in this paper. A standard way to handle non-locality in GLEs  is to approximate the memory kernel $K$ via its Prony series~\cite{prony}, namely a weighted sum of exponentials
\al
K(t) \propto \gamma \delta(t) + \sum_{i=1}^{N} \kappa_i e^{- \alpha_i t }, \label{eq:prony_series}
\eal
where $\alpha_i>0$ and $\kappa_i > -\alpha_i$ for stability reason (see SM V), but the signs of $\kappa_i$'s are not prescribed. This allows to recast the GLE of Eq.~\ref{eq:noneqGLE} into a set of coupled Langevin equations up to the introduction of auxiliary variables $z_i$:
\al
\gamma \dot{x} = - \sum_{i=1}^{N} \kappa_i (x-z_i) - U'(x) + \sqrt{2D} \eta, \quad \gamma_i' \dot{z}_i = \abs{\kappa_i} (x-z_i), \quad \gamma_i'=\frac{\abs{\kappa_i}}{\alpha_i} \,, \label{eq:prony_embedding}
\eal
where $\eta$ is a white noise of unit variance. It can be interpreted as the dynamics of an overdamped particle of position $x$ at (rescaled) temperature $T=D$ in a conservative potential $U(x)$ that is coupled via linear springs of stiffness $\kappa_i$ to $N$ overdamped independent internal degrees of freedom $z_i$ of viscosity $\gamma_i'$.
Given that the internal states $z_i$'s are athermal and the springs potentially nonreciprocal when $\kappa_i<0$, the model is clearly out of equilibrium. This embedding in principle provides a Markovian description of the dynamics for any choice of kernel $K(t)$. In the following we set $\gamma$ to $1$ up to a rescaling of time and a redefinition of parameters. For the sake of simplicity, we mainly consider hereafter the minimal truncation of the memory kernel
\al
K(t) = \delta(t) + \kappa e^{- \alpha t}, \label{eq:minimal_kernel}
\eal
which is equivalent  to the following model with a single internal degree of freedom $z$:
\al
\begin{cases}
\dot{x} = - \kappa (x-z) - U'(x) + \sqrt{2D} \eta, \\
\dot{z} = \alpha (x-z),
\end{cases} \label{eq:MP}
\eal
defining what we call below a memory particle (MP). The polarity or self-propulsion force is then defined by $p=- \kappa (x-z)$ by identification with Eq.~\ref{eq:force balance}. Note that for $\kappa\not=0$ one has $K(t) \not\propto \delta(t)$ so that this model is both active and endowed with environmental memory, which as we proceed to show induces a rich phenomenology.

In absence of external force ($U'=0$), the analysis of this model is straightforward and shows that depending on the sign of $\kappa$, the MP exhibits either persistent ($\kappa<0$) or antipersistent ($\kappa>0$) behavior over times shorter than the typical memory time $\tau_m = (\alpha + \kappa)^{-1}$, which defines the corresponding Brownian length scale  $\ell_D = \sqrt{D \tau_m}$. The dynamics become diffusive on longer timescales with an effective diffusion coefficient $ D_\star/D = 1 - \kappa(2\alpha+\kappa)/(\alpha+\kappa)^2 = 1/(1+\kappa/\alpha)^2 \,$, which in turn defines the cross-over length scale    $\ell_\star = \sqrt{D_\star \tau_m}$. Note that $\ell_\star>\ell_D$ for persistent MPs and $\ell_\star<\ell_D$ for antipersistent MPs.

Importantly, in the absence of interactions with the environment ($U'=0$), the position process $x(t)$ of a persistent MP can be mapped exactly\footnote{Note however that this matching does not hold for the polarity process $p_{\rm \scriptscriptstyle MP}=-\kappa(x-z)$ and $p_{\rm \scriptscriptstyle AOUP}=v\xi$ due to noise cross-correlations.} to the position process $X(t)$ of a  classical thermal persistent active Ornstein–Uhlenbeck particles (AOUPs)  defined as 
\al
\begin{cases}
\dot{X} = v \xi - U'(X) + \sqrt{2 D} \eta, \label{eq:AOUP} \\
\tau \dot{\xi} = - \xi + \sqrt{2} \lambda,
\end{cases}
\eal
where $\xi$ is an Ornstein-Uhlenbeck process of relaxation time $\tau$ and $\lambda,\eta$ are  independent white noises with unit variance. The correlation functions of the two Gaussian processes $x(t)$ and $X(t)$  are indeed identical upon choosing (see SM V)
\al
\tau = \frac{1}{\alpha+\kappa}, \quad v^2 = -\frac{\kappa(2\alpha+\kappa)}{(\alpha+\kappa)^2} D. \label{eq:matching_pMP}
\eal

Regarding antipersistent MPs, in  absence of interaction ($U'=0$), the position process $x(t)$  can be mapped exactly to the position process $X(t)$ of either  an antipersistent version of AOUP (see SM V for  definition) or of an equilibrium dimer with asymmetric friction
\al
\begin{cases}
	\dot{X} = -\kappa'(X-Z) -U'(X) + \sqrt{2 D} \eta, \label{eq:antipersistent_eqdim} \\
	\dot{Z} = r \kappa'(X-Z) + \sqrt{2 D r} \eta_Z,
\end{cases}
\eal
where $Z$ is a particle with a fluid friction coefficient $1/r$. $Z$ is harmonically coupled to $X$ via an harmonic spring of stiffness $\kappa'$ but it is insensitive to the direct influence of the environment $U$.  The mapping conditions to ensure equivalence of correlation functions in absence of force are given by (see SM V)
\al
\kappa'=\frac{\kappa(2\alpha+\kappa)}{\alpha+\kappa}, \quad r =\frac{\alpha^2}{\kappa(2\alpha+\kappa)} \,. \label{eq:matching_apMP}
\eal
Hereafter these matching conditions are assumed to hold, so that the position process of MPs cannot be distinguished from their matched models of (anti)persistent AOUPs or equilibrium dimers in absence of interactions. This provides important benchmark models that are either active without environmental memory (AOUPs) or endowed with environmental memory but at equilibrium (equilibrium dimers). Importantly, even if the position processes in absence of interaction of MPs and their matched models are identical, the dynamics of their internal degree of freedom  intrinsically differ. We show in the next sections that MPs, because they are both out of equilibrium and endowed with environmental memory, show strikingly different behaviors in the presence of interactions compared to their benchmark models. Last, to capture longer memory effects, we also consider memory kernel with infinitely long memory, whose prototypical examples are powerlaws of the form $K(t) \propto \delta(t) + A/t^{1/2}$. We show in SM IV that such kernel can be built from the dynamics of a tagged monomer of an infinitely long shaken Rouse polymer as represented in Fig.~\ref{fig:protocols}b. It is thus a linear chain instance of the random network of harmonic couplings depicted in Fig.~\ref{fig:protocols}a.

\section*{Memory-induced response}

The impact of interactions with the environment on  the dynamics of MPs can be simply illustrated by the following prototypical trap-and-release  protocol pictured in Fig.~\ref{fig:protocols}b, which allows for an explicit quantification of memory effects in the dynamics. A particle is trapped in an harmonic potential $U(x)=kx^2/2$ for all $t<0$ ; at time $t=0$, it is suddenly released from the trap, which mimics a perturbation of the environment. Its subsequent free evolution ($U=0$ for $t>0$) is then compared to the control protocol,  namely  the stationary dynamics in absence of perturbation ($U=0$ for all $t$). We analytically show (see SM VI) that the history of environmental perturbations does impact the future dynamics of MPs (either persistent or antipersistent), by inducing a transient regime controlled by the memory timescale $\tau_m$. In stark contrast,  AOUPs, antipersistent AOUPs, or equilibrium dimers display  stationary dynamics for $t>0$ that are independent of the perturbation in the past (presence or absence of a trap for $t<0$), see Fig.~\ref{fig:protocols}. In the case of a power-law memory kernel of the form $K(t) \propto \delta(t) + A/t^{1/2}$, the relaxation to the stationary dynamics follows a power-law scaling $\propto T^{-5/2}$ (see SM VI), making the impact of environmental memory arbitrarily long.

This memory-induced response of MPs is quantitatively evidenced in Fig.~\ref{fig:protocols}, which shows the time-dependent mean-squared increments (MSI) of the position $\Delta(T,t) = \langle ( x(t+T) - x(T) )^2 \rangle$,
and the  two-time correlator of the polarity (defined below Eq.~\ref{eq:MP})
$ c(t,T) = \langle p(t+T)p(T) \rangle \,. $ Explicit analytical expressions are derived in SM VI. Importantly, our analysis shows that such memory-induced response requires both an environmental memory ($K(t)\not\propto \delta(t)$) and non-equilibrium dynamics, which makes it a key  feature of MPs. In contrast, AOUPs do not display such response, because they have no environmental memory, the dynamics of $p$ being fully independent of the environment. In turn, equilibrium dimers do have environmental memory as stated above~; however, their equilibrium nature imposes that at steady state the distribution of the internal degree of freedom $Z$ (conditioned on the position of the particle $X$) is given by the Boltzmann measure $P(Z|X)\propto e^{-\frac{\kappa'}{2D}(Z-X)^2}$, which is independent of the environment $U(X)$. More generally, the equilibrium condition imposes strong constraints on the dynamics of the internal degrees of freedom, so that, even if the particle is endowed with environmental memory, their distribution is  at  steady state independent of the environment. This highlights the need of non-equilibrium dynamics and environmental memory for  memory-induced responses. Indeed for MPs, the stationary distribution $P(z|x)$ is not an equilibrium distribution and now depends on the environment $U(x)$ (see SM VI,VII). The dynamics upon trap release at $t=0$ is thus controlled by the relaxation of the internal degree of freedom $z$ to a different steady state, which is a hallmark of environmental memory that we discuss in this paper.

\section*{Environment sensing}
As discussed above in the context of harmonic traps, a striking property of MPs, made possible by both their environmental memory and non-equilibrium nature, is that the distribution of their  internal degrees of freedom $z$ depends on $U(x)$ even at steady state and is thus sensitive to the environment, in contrast to AOUPs or equilibrium dimers. This provides a minimal mechanism of environment sensing for MPs, which  has strong consequences on their dynamics, as we illustrate in this section by analyzing the steady state dynamics in a given confining potential $U$. 

\begin{figure}
\centering
\begin{subfigure}[b]{0.6\textwidth}
  \includegraphics[width=\textwidth, clip]{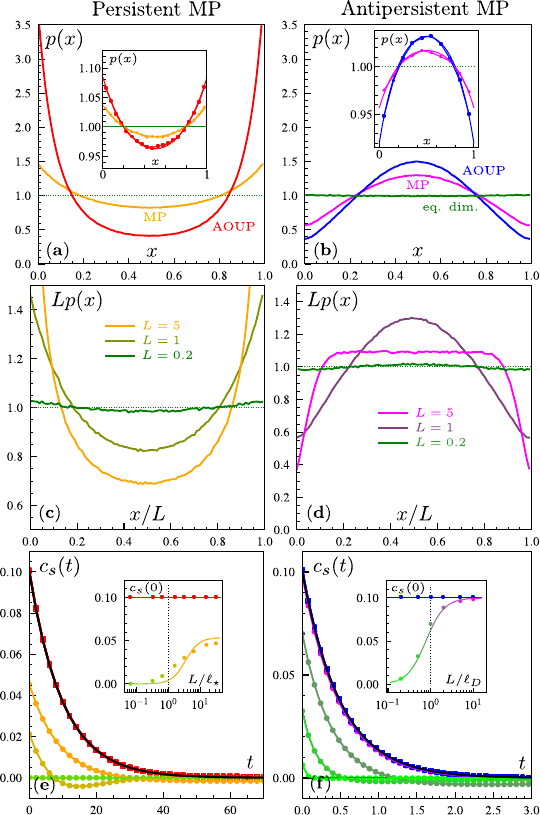}
\end{subfigure}
\caption{$\mathbf{(a)}$ Stationary distribution $p(x)$ in a hard box of size $L=1$ for persistent MPs (orange) and AOUPs (red). Green dashed line shows uniform Boltzmann measure. Inset: perturbative predictions from  SM VII (same colours) compared with simulations. $\mathbf{(b)}$ Same as $\mathbf{(a)}$ for antipersistent particles: antipersistent AOUPs (blue), MPs (magenta), equilibrium dimers (green). $\mathbf{(c)}$ Rescaled profiles $Lp(x/L)$ for persistent MPs for different box sizes $L$ (in units $\ell_\star$). $\mathbf{(d)}$ Same as $\mathbf{(c)}$ for antipersistent MPs. $\mathbf{(e)}$ Polarity correlator $c_s(t) = \langle p(t)p(0) \rangle$ for persistent AOUPs (red) and MPs (orange $\to$ green as $L$ decreases). The black curve shows predictions for AOUPs (SM VII). Inset: correlation amplitude $c_s(0)$ vs. $L/\ell_\star$. Orange-to-green and red curves respectively show the exact harmonic results for MPs (SM VII) and the AOUP prediction. $\mathbf{(f)}$ Same as $\mathbf{(e)}$ for antipersistent AOUPs (blue) and MPs (magenta $\to$ green). Simulation details in  SM VIII.}
\label{fig:harbox}
\end{figure}

The stationary distribution for an equilibrium particle, such as the equilibrium dimer of Eq.~\ref{eq:antipersistent_eqdim}, is straightforward to obtain via the Boltzmann measure for any type of interaction potential $U$. This is however much harder for nonequilibrium models such as AOUPs, for which exact results are known only for specific potentials $U$ and perturbative schemes must be introduced in the general case. This is also the case for MPs. We derive the exact stationary distribution in the case of an harmonic potential while we develop different perturbative methods in the  case of a generic potential in SM VII. For clarity we only discuss here the prototypical case of a confining hard box (i.e. an infinite-depth square well) of width $L$ as shown in Fig.~\ref{fig:harbox} and we highlight the strikingly new features of MPs. We first analyze perturbatively the regime of small active force (or polarity), by considering the small $\kappa$ limit in Eq.~\ref{eq:MP}. As detailed in SM VII, the stationary measure for such MP reads to first order in $\kappa$:
\al
p^{\rm \scriptscriptstyle MP}(x) = \frac{1}{L} - \frac{\kappa/\alpha}{L} \sech \left(\frac{L/2}{\ell_{D}^{0}}\right)  \left[\cosh \left(\frac{L/2-x}{\ell_{D}^{0}}\right) - \frac{\ell_{D}^{0} }{L/2} \sinh \left(\frac{L/2}{ \ell_{D}^{0}}\right)\right], \label{eq:hardbox}
\eal
where $\ell_{D}^{0}=(D/\alpha)^{1/2}$ is the limit $\kappa\to0$ of $\ell_D$. It shows that MPs have  nonuniform stationary states that are controlled and shaped by their memory kernel (and notably the ratio $L/\ell_D^0$), as shown in Fig.~\ref{fig:harbox}a,b. For example persistent MPs accumulate at the walls, which is qualitatively comparable to the behaviour of classical persistent self propelled particles such as AOUPs ; in contrast antipersistent MPs show a depletion zone at the confining walls and concentrate in the bulk of the domain. This behaviour is generic and not limited to square wells or to the  small active force limit, as shown in  SM VII for a generic potential $U$ and in different perturbative limits.

We now quantitatively compare the behavior of  MPs with (anti)persistent AOUPs and equilibrium dimers -- keeping the matching conditions of Eqs.~(\ref{eq:matching_pMP},\ref{eq:matching_apMP}) so that the dynamics are identical in absence of confining potential. In the case of the confining square well, our perturbative analysis shows  that the stationary distributions for MPs and their matched AOUPs are qualitatively similar  at first order, but differ quantitatively (see  SM VII for analytical expressions and Fig.~\ref{fig:harbox}). 
This confirms our earlier claim that MPs display distinctive dynamical features in the presence of interactions, even if they can be mapped to classical models in absence of interactions.  

This quantitative difference in steady state distributions is in fact due to a fundamental difference between MPs and (anti)persistent AOUPs and equilibrium dimers, rooted in their environmental memory and non-equilibrium dynamics. For MPs, the dynamics of the internal degree of freedom $z$ (and not only their steady state distribution as discussed above) depend on the environment $U(x)$. Because it is controlled by the internal degrees of freedom $z$, the dynamics of the polarity $p$ of MPs (defined below Eq.~\ref{eq:MP}), which is a measure of their activity, is thus sensitive to the environment and depends on $U(x)$. This is clearly illustrated in the case of a confining hard box of size $L$ by analyzing the two-time correlation function of the polarity at steady state  $
c_s(t) = \langle p(0)p(t) \rangle \ $ (see Fig.~\ref{fig:harbox}e,f~; an exact analytical form is derived in the case of a harmonic confinement in SM VII).
While the dynamics of $p$ (defined by $p=v\xi$ in Eq.~\ref{eq:AOUP}) are clearly independent of $L$ for AOUPs, we find for MPs that $c_s(t)$ is controlled by $L$. More precisely, in the limit of large $L$ ($L\gg \ell_\star,\ell_D$), the dynamics of $p$ is that of a free MP particle, and thus behaves analogously to the corresponding AOUP model.  However, its amplitude vanishes in the opposite limit of a strong confinement $L \ll \ell_\star,\ell_D$, so that the dynamics is effectively Brownian.\footnote{For persistent MPs, $\ell_D<L<\ell_\star$ defines a cross over regime of persistent dynamics with $L$-dependent effective self-propulsion $p$; the cross over regime is defined by $\ell_\star<L<\ell_D$ for antipersistent MPs.} This overall makes the activity of MPs environment-sensitive. Such environment-sensing mechanism results from the coupling of activity and environmental memory and is thus a unique feature of MPs: environmental memory allows for the integration of the successive interaction events with the confining walls via the memory kernel $K(t)$ and regulates the polarity accordingly. This mechanism has a direct consequence on the steady state distribution, as shown in Fig.~\ref{fig:harbox}c,d and in agreement with our analytical (perturbative) prediction of Eq.~\ref{eq:hardbox}: accumulation or depletion at the walls is significant for $L\gg \ell_D,\ell_\star$ as a result of activity, and vanishes for $L\ll \ell_D,\ell_\star$ because MPs become Brownian in this regime.

\section*{Localization  in complex environments}

\begin{figure}
\centering
\begin{subfigure}[b]{\textwidth}
  \includegraphics[width=\textwidth,clip]{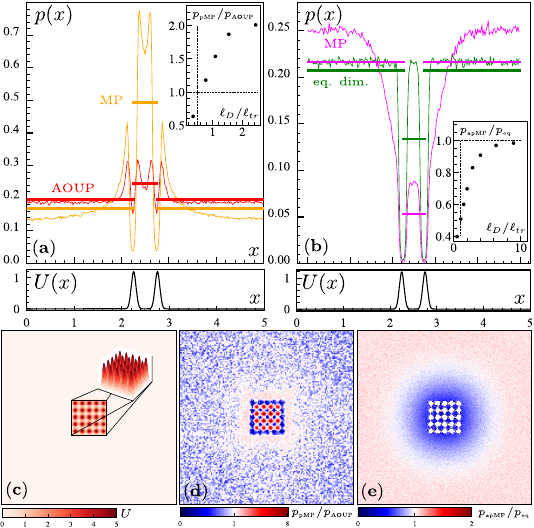}%
\end{subfigure}
\caption{$\mathbf{(a,b)}$ Localization of MPs in a $1$d isolated trap $U$ (black). $\mathbf{(a)}$ Persistent case: Averaged densities inside/outside the trap for persistent MPs (bold orange) and AOUPs (bold red). Thin lines: nonaveraged profiles. Inset: ratios of trapping probabilities for persistent MPs vs. AOUPs at different ratios $\ell_D/\ell_{tr}$. $\mathbf{(b)}$ Antipersistent case: Same as in $\mathbf{(a)}$ for antipersistent MPs (magenta) and equilibrium dimers (green, following Boltzmann measure). $\mathbf{(c,d,e)}$ Localization of MPs in a $2$d corrugated landscape of Gaussian pillars. $\mathbf{(c)}$ $2d$ heatmap of the landscape. Inset: 3D view. $\mathbf{(d)}$ Ratio of stationary distributions of persistent MPs vs. AOUPs in the landscape. $\mathbf{(e)}$ Same as $\mathbf{(d)}$ for antipersistent MPs vs. equilibrium dimers. Numerical details in  SM VIII.}
\label{fig:multibumps}
\end{figure}

The mechanism of environment sensing for MPs evidenced above, and in particular the fact that the activity of MPs depends on the local confinement, has important consequences in the case of more complex environments as we discuss in this paragraph. We start with a simple one-dimensional potential $U$ modeling an isolated trap of width $\ell_{tr}$ as pictured in Fig.~\ref{fig:multibumps}a,b. Our analysis above indicates that if $\ell_{tr}\ll \ell_D $,  MPs are effectively  Brownian inside the trap, while they behave as  (anti)persistent free MPs outside the trap. Such environment-sensitive dynamics is expected to have a strong impact on the  probability  of barrier crossing, which should be  larger  (resp. lower) for persistent (resp. antipersistent) MPs than for Brownian particles, as suggested by the accumulation (resp. depletion) phenomenon at the barrier described in the previous section. Our numerical simulations (see Fig.~\ref{fig:multibumps}a,b) indeed confirm this analysis and show that in this geometry  barrier crossing probabilities for MPs are asymmetric and can  thus lead to either an excess condensation in the trap for persistent MPs, or on the contrary to depletion in the trap for antipersistent MPs. 

Note that an effective accumulation in the trap can also be observed for persistent AOUPs, and merely stems from their classical (symmetric) accumulation on both sides of the barrier~; depletion in the trap for equilibrium dimers is effectively also observed, simply as the result of the finite width of the barrier which induces local exclusion. The mechanism of localization for MPs is genuinely different, as it arises from their environment-dependent regulation of activity, made possible by their environmental memory. It is clearly  illustrated by the asymmetric distribution  on each side of the barriers (in contrast to AOUPs and equilibrium dimers), and by the fact that  localization (either inside or outside the trap) can be controlled and significantly enhanced by tuning the control parameter $\ell_D $  of the memory kernel as shown in the insets of Figs.~\ref{fig:multibumps}a,b. More precisely, in the case of persistent MPs, localization in the trap can be dramatically enhanced if the typical outward crossing probability -- controlled by the Brownian dynamics and thus of the order $e^{-U_0/D}$ -- is small, whereas the inward crossing probability  is significantly larger because of self propulsion ($v\gg U_0/\sigma$, where $\sigma$ is the barrier width and $v$ the polarity of a free MP as defined in Eq.~\ref{eq:matching_pMP}). This analysis, even if in a simplistic geometry, suggests that the environmental memory of MPs (which is in practice parameterized by the memory kernel $K(t)$) can lead to a specific steady state localization of the particle in heterogeneous environments, either favoring or disfavoring rugged domains. 

To confirm our prediction in a more complex heterogeneous  energy landscape, we consider the two-dimensional example  depicted in Fig.~\ref{fig:multibumps}c. This corresponds to a distribution of Gaussian bumps with width $\sigma$, arranged on a rectangular lattice, resulting in a locally rugged topography. 
As suggested by our one-dimensional analysis, we  find that the persistent MPs can present a significant excess localization at steady state in the rugged region, as compared to their matched persistent AOUPs, as shown in Fig.~\ref{fig:multibumps}d. On the contrary antipersistent MPs accumulate outside the rugged region as seen in Fig.~\ref{fig:multibumps}e. We anticipate that these results apply to more complex energy landscapes, and to more complex choices of memory kernels $K(t)$ ; this could open the way to a refined control of steady state localization of MPs in complex environments.

\section*{Suppressed Motility Induced Phase Separation}

\label{sec:noMIPS}

Importantly, as stated above, Eq.~\ref{eq:noneqGLE} has the plain physical meaning of (overdamped) force balance, and thus straightforwardly applies not only to MPs interacting with a fixed environment, but also to interacting MPs. The mechanism of environment sensing -- which leads to an environment dependent activity -- can have in this context deep large-scale  consequences, which we illustrate here. We start with the emblematic example of motility induced phase separation (MIPS), which usually happens for persistent self-propelled particles (e.g. AOUPs) interacting via short-range repulsive interactions. MIPS has no classical equilibrium counterpart and has been proposed to explain a variety of phase-separating phenomena observed in living and artificial  active matter systems in absence of attractive interactions~\cite{cates-arcmp-2015}.
MIPS has been shown to be critically controlled by activity (here parametrized by the persistence time $\tau$) and particle density (or packing fraction) ; because, as we have shown, environmental memory allows MPs to adapt their activity to their local environment -- we anticipate that it can also control MIPS for persistent MPs. 

\begin{figure}
\centering
\begin{subfigure}[b]{\textwidth}
  \includegraphics[width=\textwidth,clip]{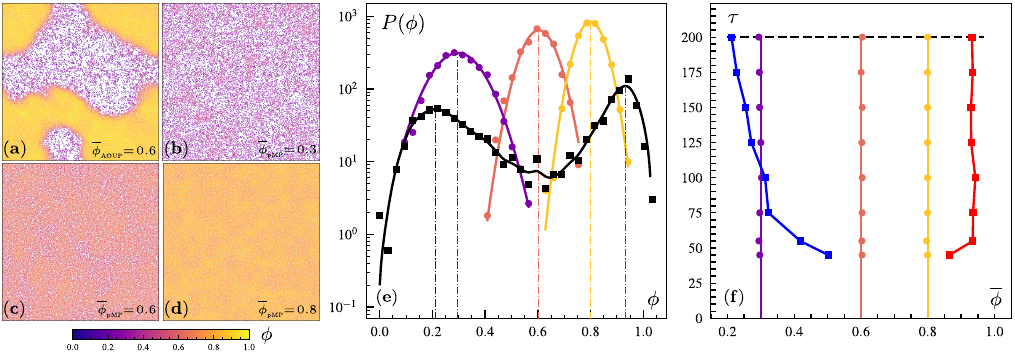}%
\end{subfigure}
\caption{$\mathbf{(a)}$ MIPS in assemblies of interacting persistent AOUPs ($\overline{\phi}=0.6, \, \tau=200$) colored by local packing fraction. Colorbar applies to all panels. $\mathbf{(b,c,d)}$ Persistent MPs under the same conditions at $\overline{\phi}=0.3,0.6,0.8$, showing no phase separation. $\mathbf{(e)}$ Probability distribution of local packing fraction $P(\phi)$ of configurations $\mathbf{(a-d)}$: black bimodal for AOUPs, unimodals (purple, salmon, gold) for persistent MPs ($\overline{\phi}=0.3,0.6,0.8$). $\mathbf{(f)}$ AOUP phase diagram: gas/liquid binodals (blue/red squares) delimit MIPS region. Overlaid: most probable local packing fractions of persistent MPs at different $\overline{\phi}=0.3,0.6,0.8$ (purple, salmon, gold); round dots show no phase separation. Dashed line $\tau=200$ marks conditions of $\mathbf{(a-e)}$. Numerical details in  SM VIII.}
\label{fig:disappearance_mips}
\end{figure}

To do so, we consider a large assembly of two-dimensional persistent MPs endowed with repulsive short-range interactions modeled via an harmonic soft-potential of the form $u(r) = 
k(2 r_0-r)^2/2$ for  $r < r_0$ and $u(r)=0$ otherwise,
where $r_0$ corresponds to the radius of the particles and $k$ to the stiffness of the potential. The microscopic parameters are chosen so that the corresponding persistent thermal AOUPs (which we recall have identical free dynamics) display MIPS at a given threshold packing fraction (Fig.~\ref{fig:disappearance_mips}) as it is classically reported~\cite{Fodor2016PRL,Martin2021AOUPReview}. Strikingly, in spite of the persistent behavior of  MPs in absence of interactions, we find no sign of phase separation  when increasing the packing fraction for a broad range of values of activity ($\tau$), as reported in Fig.~\ref{fig:disappearance_mips}f. Instead, persistent MPs consistently  behave as a Brownian liquid upon increasing the packing fraction, as confirmed by the single peak, Gaussian local density   distribution  displayed in Fig.~\ref{fig:disappearance_mips}e. Environmental memory thus suppresses MIPS by preventing cluster formation, which is the classical route to phase separation. Qualitatively, upon a local increase in density, the activity of MPs is reduced and their dynamics becomes Brownian, thus leading to a classical diffusive relaxation of density fluctuations, which suppresses the spinodal instability of MIPS.  This qualitative reasoning is supported by the following scaling argument. MIPS for \emph{athermal} AOUPs requires  $\ell_p\gg r_0$, where $r_0$ is the typical particle radius~\cite{Martin2021AOUPReview}. Using the matching conditions of free MPs and AOUPs of Eq.~\ref{eq:matching_pMP}, this condition is equivalent for MPs to $\ell_\star \gg r_0$. This means that for MIPS to happen, the cross over length $\ell_\star$ should be much larger than the typical interparticle distance in dense clusters. However, we have shown that in this regime of effective confinement MPs behave as Brownian particles~; cluster formation is thus prevented by diffusive spreading. This eventually shows that environmental memory, in the case of persistent MPs, endows the particles with a density dependent activity controlled by the memory kernel $K(t)$, and can thus allow to avoid MIPS.

\section*{Enhanced jamming transition}

As a last example of the impact of environmental memory on the dynamics of interacting MPs, we analyse the liquid-to-solid  (or jamming in a loose sense)  transition, which characterizes the dynamical arrest of particles with repulsive interactions upon increasing the packing fraction~\cite{VanHecke2010JammingReview,BerthierBiroli2011RMP}. As discussed above, antipersistent MPs confined in a square well show at steady state a depletion near hard walls, in contrast to the uniform equilibrium Boltzmann measure obeyed by equilibrium dimers. This indicates that antipersistent MPs have a larger effective interaction radius compared to their matched equilibrium dimers, and suggests that the jamming transition for antipersistent MPs with repulsive interactions should occur for smaller packing fractions and be controlled by the characteristic length $\ell_D$.
\begin{figure}
\centering
\begin{subfigure}[b]{\textwidth}
  \includegraphics[width=\textwidth,clip]{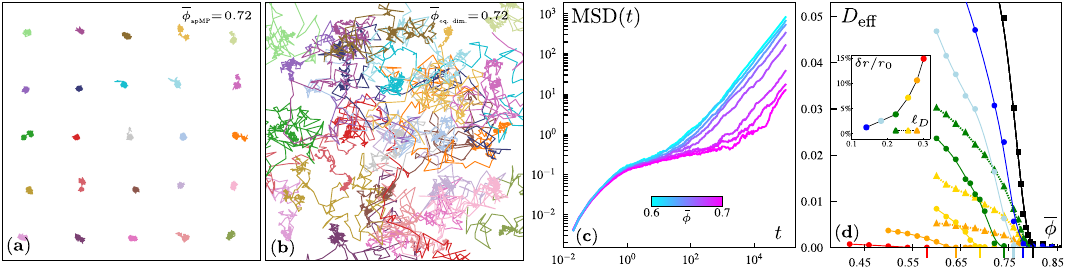}%
\end{subfigure}
\caption{$\mathbf{(a,b)}$ Selected trajectories of antipersistent MPs and matched equilibrium dimers at $\overline{\phi}=0.72$. MPs are jammed while equilibrium dimers remains mobile. $\mathbf{(c)}$ Mean-squared displacement for dense assemblies of antipersistent MPs. Colors to increasing $\overline{\phi}$ from $0.6$ to $0.7$ (cyan to magenta). All curves show short- and long-time diffusive scaling and a plateau develops at higher densities, indicating transient caging and the onset of jamming. $\mathbf{(d)}$ Effective diffusion coefficient $D_{\rm eff}$ vs. $\overline{\phi}$ for Brownian particles (bold black line/squares), equilibrium dimers (dotted lines/triangles) and antipersistent MPs (bold lines/dots). Colors indicates $\ell_D$. Colored ticks below the $x$-axis mark the jamming densities for antipersistent MPs (colored) and equilibrium brownian or dimers (black). Inset: effective particle radius $r_0+\delta r$ vs. $\ell_D$ for antipersistent MPs (dots) and equilibrium dimers (triangles); $\delta r$ is defined so that their jamming densities matches the one of Brownian particles. More details in  SM VIII.}
\label{fig:jamming_unjamming}
\end{figure}

We confirm this prediction in Fig.~\ref{fig:jamming_unjamming} using numerical simulations of two-dimensional dense systems of particles with the same soft-sphere repulsive potential as defined above. More precisely, we first determine the jamming transition for equilibrium dimers, namely the critical packing fraction beyond which a jammed state is observed. As expected, this threshold is the same as the one for Brownian particles in the low temperature regime that we consider in the simulations, no matter the kinetic parameters of the equilibrium dimers. This stems from the fact that the thermal energy in the system ($D$ in Eq.~\ref{eq:antipersistent_eqdim}) is sufficiently small compared to the energy required for exchanging  neighbors ($\propto k r_0^2$). In this regime, interactions are almost hard core and the jamming transition is known to occur for a critical packing fraction independent of temperature~\cite{VanHecke2010JammingReview,BerthierBiroli2011RMP}. Remarkably, as anticipated, we find that the jamming transition occurs at a significantly lower packing fraction for antipersistent MPs, which is controlled by the environmental memory via $\ell_D$ as shown in Fig.~\ref{fig:jamming_unjamming}d. This confirms our hypothesis that environmental memory, in the case of antipersistent MPs, endows the particles with an effective repulsive interaction controlled by $\ell_D$, and can thus be effectively tuned to control the jamming transition. We note that a shift of the jamming transition  has also been reported in a non Markovian spherical p-spin model~\cite{berthier2013nonequilibrium}, however originating from a different microscopic dynamics.

\section*{Discussion}

 Guided by the example of living cells and recent experimental observations that report memory effects of different natures in motile cells, we have developed a general framework for the dynamics of generic active agents endowed with internal degrees of freedom -- which characterize the cell state --  and minimal environmental sensing -- typically in  response to  chemical or mechanical stimuli. By combining exactly solvable limits, perturbative approaches, and numerical simulations, we established under minimal assumptions that such systems are generically out of equilibrium and equipped with environmental memory, which gives rise to  nontrivial dynamical responses to environmental cues and new classes of behaviour  absent in memoryless descriptions.

In particular, we demonstrated that environmental memory fundamentally reshapes both steady-state distributions and transient dynamics, producing tunable emergent interactions and collective phenomena. These effects enable controllable localization in complex landscapes, suppression of motility-induced phase separation, and enhanced jamming transitions. Although our analysis ii based on linear response and focused mostly on simple, single  exponential memory kernels, we anticipate that more complex memory functions and potentially non linear responses -- which in principle could be inferred from experimental data -- will unlock an even broader spectrum of behaviours and will  lead to experimentally testable predictions, whose exploration is left for future works. Our findings provide a proof of principle that even minimal information-processing capabilities -- intrinsic to nonequilibrium agents with internal states, including living cells and virtually any living or artificial self-propelled agent -- can profoundly shape both individual and collective dynamics. This framework thus links cell-scale activity to emergent, large-scale intelligent motion, and opens new perspectives of analysis and design of complex systems  ranging from synthetic colloids to biological collectives and robotic swarms.

{\bf Acknowledgements.}
Support from ERC synergy grant Shappincellfate is acknowledged. Simulations were performed on the  MeSU platform at Sorbonne Université.

\bibliography{biblio.bib}%

\end{document}


\title{Supplementary Material to\\ Emergent memory  in cell-like active systems}
\author{M. Besse}
\author{R. Voituriez}
\maketitle

\tableofcontents

\section{Notations}

\label{app:conv}

To alleviate integral notations, we adopt the following convention throughout the paper
\al
\int_{t=0}^{\infty} ... = \int_{0}^{\infty} ... \ dt \,.
\eal
The Laplace transform of a real-valued function $f(t)$ is then denoted by
\al
\cL[f](s)=\hat f(s)=\int_{t=0}^\infty f(t)e^{-st},
\eal
and its inverse $\cL^{-1}[\hat f](t) = f(t)$. The convolution between two functions $f$ and $g$ 
\al
(f * g)(t) = \int_{t'=0}^t f(t') g(t-t'),
\eal
is local in the Laplace space
\al
\cL[f*g] = \cL[f]\cL[g] = \hat f \hat g \,.
\eal
We furthermore define the two-point connected correlation function of $\phi$ a stochastic process or random field as
\al
\langle \phi(x,t) \phi(x',t') \rangle_c = \langle \phi(x,t) \phi(x',t') \rangle - \langle \phi(x,t) \rangle \langle \phi(x',t') \rangle \,.
\eal

\section{Generalized Langevin equations for memory particles}

We show in this section that the linearized dynamics of a generic overdamped,   nonequilibrium agent of position $x(t)$ with internal degrees of freedom $\bf s$ and self-propulsion force (polarity) $p$  satisfies the following Generalized Langevin Equation (GLE), given as Eq. 3 in the main text:
\al
\int_{t'=-\infty}^t K(t-t') \dot{x}(t') = F_{\rm ext}(t) +\xi(t), \quad \langle \xi(t)\xi(t') \rangle = G(|t-t'|), \label{eq:MoriZwanzig}
\eal
where $K$ is a memory kernel and $G$ the correlator of the noise.
For completeness we recall that force balance reads
\al
\gamma \dot{x} = p + F_{\rm ext} + \xi, \label{eq:force balance} 
\eal
and that we assume a generic first order dynamics for the state variables:
\al
\dot \bs=\bR(\bs,\bm{\mathcal{E}}_x),\label{eq:sdot}
\eal where $\bR$ encodes  the cell response to the set $\bm{\mathcal{E}}_x$ of external stimuli of the environment sensed by the particle at  position $x$. For simplicity, we restrict our analysis to mechanical stimuli so that $\bm{\mathcal{E}}_x=(\dot{x}, F_{\rm ext})$. In turn, the polarity is defined as a one-dimensional projection of $\bf s$ and denoted $p(\bf s)$.

\subsection{General  top-down  derivation}
\label{app:linearinversion}
\label{app:morizwanzig}

A general argument consists in linearizing the dynamics of $\bf s$ in Eq.~\ref{eq:sdot} and in defining $p(\bf s)$ as a projection of $\bf s$. The formal projection operator formalism defined by Zwanzig~\cite{zwanzig1961memory} and Mori~\cite{mori1965transport}, later extended to non equilibrium systems \cite{Vandebroek2015ActiveViscoelasticPolymer,berthier2013nonequilibrium,Szamel2015GlassyDynamicsSPP} can then be applied and leads to a GLE for the variable $x$ alone of the form of Eq.~\ref{eq:MoriZwanzig}.

Along that line, and without excessive formalism, we first present the main steps of the derivation.
As stated above, we linearize Eq.~\ref{eq:sdot} (around the steady state defined by $F_{\rm ext}=0$ that we assume exists) and make use of $\bm{\mathcal{E}}_x=(\dot{x}, F_{\rm ext})$. Linear response then yields on general grounds
\al
p(t) = \int_{t'=-\infty}^t dt'\left[ \dot{x}(t')R_1(t-t') + F_{\rm ext}(t')R_2(t-t')\right], \label{eq:Fint}
\eal
where $R_1$ and $R_2$ are given response functions (memory kernels) expressing the causality of the system and whose explicit functional form is undetermined at this stage. An explicit derivation is proposed in the next section. Note that here, for simplicity, we have assumed that all sources of noise are taken into account additively in the noise term $\xi$ in Eq.~\ref{eq:force balance}. We thus omit possible sources of noise in Eq.~\ref{eq:sdot}.

Making use of the linearity of Eq.~\ref{eq:force balance} and Eq.~\ref{eq:Fint}, one finds that  the dynamics of Eq.~\ref{eq:force balance} can be written, up to a linear inversion in  Laplace domain, 
\al
K * \dot{x} = F_{\rm ext} + \tilde{\xi}, \label{eq:linear_inversion}
\eal
where $K = \chi*(\delta-R_1)$, $\chi=\cL^{-1}[1/(1+\cL[R_2])]$,  and $\tilde{\xi}=\chi*\xi$ is a colored Gaussian noise. This is, upon redefinition of the memory kernel and noise, the expected form of Eq.~\ref{eq:MoriZwanzig} (Eq. 3 of the main text). Importantly, under this form, Eq.~\ref{eq:MoriZwanzig} preserves its plain physical meaning of overdamped force balance, and can thus be straightforwardly used to analyse the dynamics of either particles in  external fields, or of interacting particles. \\

\subsection{Constructive bottom-up derivation}

\label{app:explicitinternaldynamics}

In this section, we show how to derive Eq.~\ref{eq:MoriZwanzig} from Eq.~\ref{eq:force balance}, with an explicit construction of the memory kernel $K$. One of the advantages of this explicit derivation is that different types of memory kernels naturally emerge depending on the dynamics of the internal states. If the dynamics is reciprocal (as in App.~\ref{app:reciprocal}), the memory kernel is a sum of decaying real exponentials, whereas if the dynamics is nonreciprocal (as in App.~\ref{app:nonreciprocal}), the memory kernel also exhibits oscillatory terms arising from the presence of complex eigenvalues of the operator defining the linear dynamics of the system. \\

For convenience, we rewrite the dynamics of a cell (defined by Eq.~\ref{eq:force balance} and Eq.~\ref{eq:sdot}) using a vector $(x,\bs)$ composed of $x$, describing the position of the cell, and $\bs$, representing the high-dimensional vector of internal states. Upon setting $\gamma = 1$, this can be written as
\al
& \p_t \begin{pmatrix} x \\ \bs \end{pmatrix} = \bF_{\rm int}(x,\bs) + \bF_{\rm ext}(x,\bs) + \BL \,. \label{eq:markovianmodel}
\eal
where 
\al
& \bF_{\rm int}(x,\bs)= \begin{pmatrix} p(x,\bs) \\ \bR(x,\bs) \end{pmatrix}, \quad \bF_{\rm ext}(x,\bs)= \begin{pmatrix} -U'(x) \\ \mathbf{0} \end{pmatrix}, \quad \BL= \begin{pmatrix}  \sqrt{2D} \eta \\ \mathbf{0} \end{pmatrix} \,.
\eal
Here we assume for simplicity that $\eta$ is a unit Gaussian white noise, $p$ denotes the polarity of the cell, and $\bR$ is the response function of the internal state $\bs$ as defined above (Eq.~\ref{eq:force balance}-\ref{eq:sdot}). We also assume that the external forces are conservative and, more importantly, that they do not directly affect the internal state of the cell $\bs$. Note, however, that $\bs$ still senses the environment in this minimal version of the model because $\bR$ depends on $x$ ; the discussion below can in principle be extended  to the general case $\bR(\bs,\bm{\mathcal{E}}_x)$. The term $\BL$ models fluctuations arising from the environment; we have assumed for simplicity that it acts only on the dynamics of $x$ but additional stochastic components could also be included in the dynamics of $\bs$. 
The vector field $\bF_{\rm int}$ describes the high-dimensional coupling between $x$ and the internal state $\bs$. Although $\bF_{\rm int}$ cannot be determined explicitly, it must satisfy several properties summarized below. 

First, in the absence of external forces, the system must be invariant under space translations, namely
\al
\bF_{\rm int}\left(x+a,\mathbf{T}_a(\bs)\right)=\bF_{\rm int}(x,\bs), \quad \ \forall \ x,\bs,a,
\eal
where $\mathbf{T}_a$ is a suitable differentiable transformation such that $\mathbf{T}_0(\bs)=\bs$. Differentiating this condition with respect to $a$ yields
\al
\p_x \bF_{\rm int}(x,\bs)+\p_a \mathbf{T}_a(\bs) \cdot \p_{\bs} \bF_{\rm int}(x,\bs)=0, \label{eq:nonlinearconstraint}
\eal
which means that $(1,\p_a \mathbf{T}_a(\bs))$ lies in the kernel of the Jacobian of $\bF_{\rm int}$, implying that its determinant vanishes at any point $(x,\bs)$. 

Second, using the Helmholtz–Hodge decomposition and up to gauge-fixing terms, it is possible to rewrite $\bF_{\rm int}$ as
\al
\bF_{\rm int} = - \boldsymbol{\nabla} \Phi + \cR, \label{eq:HH}
\eal
where $\Phi$ is a scalar potential field (curl-free) and $\cR$ a solenoidal (divergence-free) field. The presence or absence of a solenoidal component in $\bF_{\rm int}$ has important consequences for the form of the memory kernel $K$, as we detail in the following sections. \\

To make further progress, we linearize $\bF_{\rm int}$ around a fixed point of the non fluctuating ($D \to 0$) dynamics, which we assume for simplicity is stable and satisfies $\bF_{\rm int}=\mathbf{0}$.  Linearizing the internal dynamics at small $D$ in Eq.~\ref{eq:markovianmodel}, we obtain
\al
\p_t \begin{pmatrix} x \\ \bs \end{pmatrix} = \CL \begin{pmatrix} x \\ \bs \end{pmatrix} + \bF_{\rm ext} + \BL,
\eal
where $\CL$ is the Jacobian of $\bF_{\rm int}$, whose (possibly complex) eigenvalues all have strictly negative real parts. Decomposing the Jacobian as
\al
\CL = \begin{pmatrix} \CL_{1,1} & \CL_{1,N} \\ \CL_{N,1} & \CL_{N,N} \end{pmatrix}, \label{eq:decomp_jac}
\eal
we can take advantage of the now linear dynamics of the internal state $\bs$ to integrate it out and obtain the reduced dynamics of $x$, up to an integration by parts, 
\al
\bs(t) = - \CL_{N,N}^{-1} \CL_{N,1} x(t) + \int_{t'=-\infty}^t \CL_{N,N}^{-1} e^{(t-t')\CL_{N,N}} \CL_{N,1} \dot{x}(t') \,.
\eal
This expression holds if $\CL_{N,N}$ is invertible, which is for instance the case if it is a sufficiently large random matrix without any specific symmetry or constraint. The reduced equation of motion for $x$ then reads
\al
\dot{x} = \left( \CL_{1,1} - \CL_{1,N} \CL_{N,N}^{-1} \CL_{N,1} \right) x + \int_{t'=-\infty}^t \CL_{1,N} \CL_{N,N}^{-1} e^{(t-t')\CL_{N,N}} & \CL_{N,1} \dot{x}(t') - U'(x) + \sqrt{2D} \eta \,. \nonumber \label{eq:reduced_dyn}
\eal
We see that translational invariance in $x$ requires
\al 
\CL_{1,1} - \CL_{1,N} \CL_{N,N}^{-1} \CL_{N,1} = 0 \iff \det{\CL}=0,
\eal
or equivalently, that $0$ is an eigenvalue of $\CL$, as already found in the more general nonlinear case of Eq.~\ref{eq:nonlinearconstraint}. Under this condition, the memory kernel $K$, as defined in Eq.~3 in the main text, reads
\al
K(t) = \CL_{1,N} \CL_{N,N}^{-1} e^{t \CL_{N,N}} \CL_{N,1} \,. \label{eq:explicit_K}
\eal
This is the main result of this section: it shows how $K$ can be explicitly constructed from the microscopic parameters of the internal dynamics. The behavior of $K$ is primarily shaped by the spectral properties of $\CL_{N,N}$, which we now characterize. \\

As we have already shown, $0$ must be an eigenvalue of $\CL$, which implies that $\CL_{1,1}$ is fixed once the other coefficients of $\CL$ are chosen. Apart from this constraint, there are no further restrictions on $\CL_{N,N}$. In the absence of precise information about the high-dimensional coupling between $x$ and $\bs$, we model $\CL_{N,N}$ as a large real random matrix without specific symmetry. It is thus   invertible and almost surely has $N$ distinct eigenvalues. Moreover, the spectral content of $\CL_{N,N}$ is exactly that of $\CL$ up to the eigenvalue $0$.

To obtain a more explicit form of $K$, additional information about the spectrum of $\CL$ is required. This chiefly depends on the type of dynamics assumed for $\bF_{\rm int}$, as discussed in the next section.

\subsubsection{Gradient dynamics}

\label{app:reciprocal}

We first study the simpler case where $\bF_{\rm int}$ obeys gradient dynamics ($\cR=0$ in Eq.~\ref{eq:HH}). This means that $\CL$ is a real symmetric matrix (as the Hessian of $\Phi$), and the internal dynamics of the cell are reciprocal. According to the spectral theorem, there exists an orthogonal transformation matrix $P$ such that
\al
\partial_t \begin{pmatrix} x \\ \bs \end{pmatrix} = - P^T \begin{pmatrix} 0 & 0 \\ 0 & \bL \end{pmatrix} P \begin{pmatrix} x \\ \bs \end{pmatrix} + \begin{pmatrix} -U'(x) \\ 0 \end{pmatrix} + \begin{pmatrix} \sqrt{2D} \eta \\ 0 \end{pmatrix},
\eal
where $\bL = {\rm diag}(\alpha_1,...,\alpha_N)$ is a diagonal matrix with all strictly positive coefficients, since we are working around a stable fixed point of the dynamics by definition. Because we consider $\CL_{N,N}$ as a sufficiently large random matrix without specific symmetries, its eigenvalues are all distinct. We also use the fact that $0$ must be an eigenvalue of $\CL$ to ensure translational invariance in space. Decomposing $P$ as
\al
P = \begin{pmatrix} P_{1,1} & P_{1,N} \\ P_{N,1} & P_{N,N} \end{pmatrix}, 
\eal
we obtain, using the decomposition in Eq.~\ref{eq:decomp_jac},
\al
& \CL_{1,1} = -P_{N,1}^T \bL P_{N,1}, \quad \CL_{1,N} = -P_{N,1}^T \bL P_{N,N}, \\  & \CL_{N,1} = -P_{N,N}^T \bL P_{N,1}, \quad \CL_{N,N} = -P_{N,N}^T \bL P_{N,N},
\eal
which implies that $\CL_{N,N}$ is necessarily a real symmetric matrix diagonalized by $\bL$ through the orthogonal matrix $P_{N,N}$. The memory kernel $K$, as given by Eq.~\ref{eq:explicit_K}, thus simplifies to
\al
K(t) = - P_{N,1}^T e^{- t \bL } \bL P_{N,1} = - \sum_{i=1}^N \kappa_i e^{-\alpha_i t}, \quad \kappa_i = \alpha_i [P_{N,1}]_i^2 >0 \,. \label{eq:kernel_gradient}
\eal
It is a sum of $N$ exponentially decaying terms with negative amplitudes and characteristic timescales $1/\alpha_i$, where the $\alpha_i$'s are the nonzero eigenvalues of the degenerate Jacobian of $\bF_{\rm int}$. As discussed next in App.~\ref{app:prony}, the reduced equation of motion for $x$ in the gradient case
\al
\dot{x} = - \int_{t=-\infty}^t P_{N,1}^T e^{-(t-t') \bL} \bL P_{N,1} - U'(x) + \sqrt{2D} \eta,
\eal
can be interpreted as the dynamics of a thermal bead $x$ in a potential $U$, harmonically coupled via springs of stiffness $\kappa_i$ to $N$ independent internal states $z_i$ with viscous friction coefficients $\gamma_i=\kappa_i/\alpha_i$ and vanishing temperature. The precise statistics of the $\kappa_i$ and $\alpha_i$ require further investigation. However, the motion of such a particle $x$ is generally expected to be antipersistent, as discussed later. Moreover, the smallest eigenvalue $\alpha_c$ dominates the long-time dynamics of the memory kernel $K$, and this work focuses specifically on this eigenvalue.

\subsubsection{Generic dynamics: nonreciprocity}

\label{app:nonreciprocal}

We now consider the generic case where $\CL$ is not necessarily symmetric ($\cR \neq 0$ in Eq.~\ref{eq:HH}), meaning that the internal forces are nonreciprocal. In particular, $\CL_{N,N}$, taken as a large real random matrix without symmetry, has now complex eigenvalues that come in conjugate pairs when $N$ is even. If $N$ is odd, there is one additional real eigenvalue.

From Eq.~\ref{eq:explicit_K}, it follows that $K$ is a weighted sum of exponentials $e^{-\alpha_i t}$, where the $\alpha_i$ are the complex eigenvalues of $\CL_{N,N}$. Since $K$ must be real -- as it results from the projection of an exponential of a real matrix -- it can be rewritten as
\al
K(t) & = \kappa_0' e^{-\alpha_0 t} \delta_{N\%2\equiv1} + \sum_{i=1}^{\lfloor N/2 \rfloor} \kappa_i' e^{-\alpha_i t} + \overline{ \kappa_i' e^{-\alpha_i t} }, \\
& \propto \kappa_0' e^{-\alpha_0 t} \delta_{N\%2\equiv1} + \sum_{i=1}^{\lfloor N/2 \rfloor} \kappa_i'' e^{-\Re{\alpha_i} t} \cos{\left( \Im{\alpha_i} t - \phi_i \right)}, \label{eq:kernel_nonreciprocal}
\eal
where, for any $i$, $\kappa_0$, $\kappa_i''$, and $\phi_i$ are real, while $\kappa_i'$ is complex. Unlike the memory kernel resulting from gradient dynamics in Eq.~\ref{eq:kernel_gradient}, $K$ is now oscillatory owing to the imaginary parts of the eigenvalues $\alpha_i$. Moreover, there is no constraint on the sign of the exponential amplitudes, which can lead to persistent motion of $x$.

In contrast to the gradient case, we have not derived an explicit expression for $K$ in terms of the microscopic parameters of the internal dynamics. Such an expression could be obtained using a Jordan block decomposition for real matrices, but this requires further investigation. Exploring the consequences of these oscillatory kernels is also left for future work. In this paper, we restrict ourselves to nonreciprocal dynamics with an internal state dimension $N=1$, that is, we consider only the $\kappa_0'$ term in Eq.~\ref{eq:kernel_nonreciprocal}. This corresponds to the model of a persistent memory particle (pMP) defined in Eq.~7 of the main text.

\section{Markovian embedding of memory kernels: Prony series}

\label{app:prony}

In this section, we show how decomposing a memory kernel into its Prony series,
\al
K(t) = \gamma \delta(t) + \sum_{i=1}^N \kappa_i e^{-\alpha_i t},
\eal
allows for a Markovian embedding in the form of Eq.~5 of the main text, which we recall here for completeness:
\al
\gamma \dot{x} = - \sum_{i=1}^{N} \kappa_i (x-z_i) - U'(x) + \sqrt{2D} \eta, \quad \gamma_i' \dot{z}_i = \abs{\kappa_i} (x-z_i), \quad \gamma_i'=\frac{\abs{\kappa_i}}{\alpha_i} \,.
\eal
Because the coupling between $z_i$ and $x$ is linear, $z_i$ can be expressed explicitly as a function of $x$:
\al
z_i(t) = e^{-\alpha_i (t-t_0)}  z_i(t_0) + \int_{t'=t_0}^t \alpha_i e^{-\alpha_i (t-t')} x(t') ,
\eal
where we assume that the dynamics start at $t_0$. This relation implies that
\al
x(t) - z_i(t) =  e^{-\alpha_i (t-t_0)} \big( x(t_0)-z_i(t_0) \big) + \int_{t'=t_0}^t e^{-\alpha_i (t-t')} \dot{x}(t') \,.
\eal
The dynamics of $x$ therefore obey the GLE:
\al
\int_{t'=t_0}^t \dot{x}(t') K(t-t') = -U'(x(t)) + \sqrt{2D} \eta(t) - \sum_{i=1}^N \tilde{K}(t-t_0) \big(x(t_0)-z_i(t_0)\big) , \label{eq:prony_results}
\eal
where $K(t) = \gamma \delta(t) + \tilde{K}(t)$ and $\tilde{K}(t)= \sum_{i=1}^N \kappa_i e^{-\alpha_i t}$. Depending on the context, one may take the limit $t_0 \to -\infty$, in which case the GLE becomes
\al
& \int_{t'=-\infty}^t \dot{x}(t') K(t-t') = -U'(x(t))  + \sqrt{2D} \eta(t) ,
\eal
and loses its dependence on the initial conditions as the system reaches a stationary state.

While this Markovian embedding via Prony series is standard, it is usually carried out simultaneously for both the memory kernel $K$ and the correlator of the colored noise $G$ in Eq.~3 of the main text, which naturally enforces the fluctuation–dissipation relation. Here, however, we emphasize that it is always possible to perform two separate embeddings for $K$ and $G$. This approach has the advantage of clearly disentangling the effects of the two terms -- particularly in nonequilibrium contexts -- and highlights the feedback loops that are present in the Markovian embedding for $K$ but absent in $G$.

\section{Powerlaw memory kernel: the Shaken Rouse Line (SRL) }

\label{app:embeddingSRL}

We show in this section how to build a Markovian embedding equivalent to a GLE of the form of Eq.~3
\al
& \int_{-\infty}^t \dot{x}(t')K(t-t') = -U'(x(t)) + \sqrt{2 D} \eta(t), \quad \langle \eta(t) \eta(t') \rangle = G(\abs{t-t'}), \label{eq:powerlawGLE}
\eal
with a power law memory kernel of the form $K(t) \propto \delta(t) + A/t^{1/2}$ and a white noise correlator $G(t) = \delta(t)$. The approach below can in principle be generalized to other types of power law kernels $K(t) \sim 1/t^{\alpha}$.\\

To do so, we consider a Shaken Rouse Line (SRL) defined as a one-dimensional interface $h$ whose equation of motion reads
\al
\p_t h(y,t) = \nu \, \partial_y^2 h(y,t) - \mu \, U'(h(y,t)) \delta(y) + \sqrt{2D} \eta(y,t) \,. \label{eq:SRL}
\eal
Here $\nu$ is similar to a diffusion coefficient, $\mu$ to a mobility coefficient, $U$ to an interaction potential and $\eta$ is a unitary and centered Gaussian white noise
\al
\langle \eta(y,t) \eta(y',t') \rangle = 2 D \, \delta(y) \delta(y') \delta(t-t') \,.
\eal
Several comments are in order.

\begin{figure}[ht]
	\includegraphics[width=0.5\textwidth,clip]{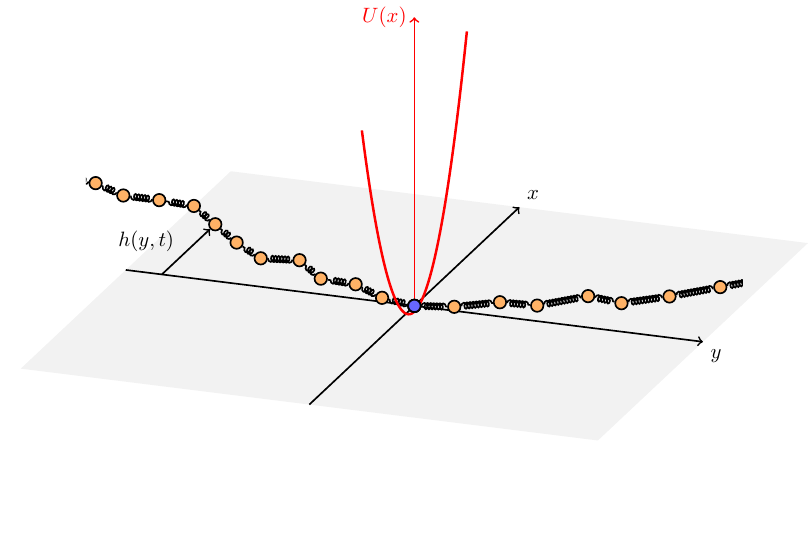}
	\caption{Sketch of a one-dimensional athermal shaken polymer, made of orange beads that are linked by the black harmonic springs. The orange beads are free to move only along the $x$-axis and their position is parameterized by $h(y,t)$. Note the $y \to -y$ symmetry. Only the central or tagged monomer in blue is subjected to thermal motion and to the force deriving from the red potential $U$. In the limit of an infinite number of beads, the polymer is analogous to the SRL introduced in Eq.~\ref{eq:SRL}. In this section, we are mainly interested in obtaining the reduced equation of motion for the position of the central monomer $x=h(0,t)$ that experiences the force $-U'$. The orange beads are analogous to the internal states $\bf s$.}
	\label{fig:SRL}
\end{figure}

$(a)$ This line or interface $h$ can be seen as the continuous limit of a one-dimensional infinitely long polymer made of beads that interact harmonically with their neighbors
\al
\zeta \dot{R_n} = k(R_{n+1}-2R_n+R_{n-1}) -U'(R_n) \delta_{n0} + \sqrt{2 k_B T \zeta} \, \eta_n, \label{eq:SRP}
\eal
where the $R_n$'s denote the positions of the beads, $\zeta$ a friction coefficient, $k$ a spring stiffness, $U$ an interaction potential, $k_B$ the Boltzmann constant, $T$ the temperature and $\eta_n$'s are Gaussian centered unitary white noises such that $\langle \eta_m(t)\eta_n(t') \rangle = \delta_{m0} \delta_{n0} \, \delta(t-t')$. Such a typical microscopic realization of the SRL is displayed in Fig.~\ref{fig:SRL}. Eq.~\ref{eq:SRL} can be recovered as the continuum limit of Eq.~\ref{eq:SRP} by $(i)$ setting $y=n a$ where $a$ is the distance between two neighboring monomers, $(ii)$ identifying the microscopic and macroscopic coefficients $\mu=a/\zeta, \nu = k a^2/\zeta, D=2 k_B T a^2 / \zeta$ and $(iii)$ taking the limit $a \to 0$ while keeping $\mu, k a$ and $T a$ constant.

$(b)$ The SRL is thus reminiscent of the Rouse polymer model~\cite{rouse1953theory} and its continuum limit which is the Edwards-Wilkinson equation describing the dynamics of free interfaces~\cite{edwards1982surface}. However, contrary to these models, only the central or ``tagged'' monomer is subjected to a noise $\eta$, the rest of the chain being athermal. The clearly drives the system out of equilibrium. We also stress out that the interaction potential $U$ only acts on the central monomer. 

$(c)$ This SRL  breaks time-reversal symmetry because of the energy injection in one point of the chain and its dissipation in the rest of the structure~; it is however the case if and only if the energy injected via the noise $\eta$ ($G$ in Eq.~\ref{eq:powerlawGLE}) is different to what can be dissipated by the structure of the chain ($K$ in Eq.~\ref{eq:powerlawGLE}). This is clearly realized with our choice of $\eta$ being a white noise.

$(d)$ In the case of an (infinitely) long Rouse polymer, the position of central monomer $x$ is known to be a microscopic and force-balance realization of a fractional Brownian motion (fBm) of Hurst exponent $H=1/4$.
As regards $x(t)=h(0,t)$ the position of the central monomer of a SRL, it is no longer a fBm as characterized in App.~\ref{app:testing_protocols}~; its mean-square displacement is for instance scaling as a log of time. \\

The rest of the section is dedicated to the derivation of the reduced dynamics of the central monomer of a SRL, $x(t)=h(0,t)$. To do so, we capitalize on the fact that the two half lines
\al
h^- = \{h(y,t)\}_{y\leq0}, \quad h^+=\{h(y,t)\}_{y\geq0}
\eal
evolve almost independently, except at their starting point where they are connected by the central monomer. We thus solve the dynamics of Eq.~\ref{eq:SRL} separately for $y>0$ and $y<0$, imposing for each half line to meet at the time-dependent boundary condition $x(t)=h^+(y=0^+,t)=h^-(y=0^-,t)$.

It is worth to note that the equation of motion for the SRL is invariant under the symmetry $y \to -y$ for any stochastic realization of $\eta$ (i.e. not in average). We can thus safely focus only on the evolution of one of the half lines, say $h^+$, which obeys a diffusion equation on the half space with a time-dependent boundary condition, namely
\al
\begin{cases}
	& \p_t h^+(y,t) = \nu \, \p^2_y h^+(y,t), \ \text{for} \ y>0, \\
	& h^+(y,0) = h^+_0(y), \ \text{for} \ y>0, \\
	& h^+(0,t) = x(t), \ \text{for} \ t \geq 0,
\end{cases} \label{eq:halfdiff}
\eal
where $x(t)$ is so far an arbitrary time-dependent forcing function, while $h^+_0$ corresponds to the initial profile of the interface. This linear problem can be solved by superposition principle: if $u$ solves
\al
& \begin{cases}
	& \p_t u(y,t) = \nu \, \p^2_y u(y,t), \ \text{for} \ y>0, \\
	& u(y,0) = 0, \ \text{for} \ y>0, \\
	& u(0,t) = x(t), \ \text{for} \ t>0,
\end{cases}
\eal
and $v$ solves
\al
& \begin{cases}
	& \p_t v(y,t) = \nu \, \p^2_y v(y,t), \ \text{for} \ y>0, \\
	& v(y,0) = h^+_0(y), \ \text{for} \ y>0, \\
	& v(0,t) = 0, \ \text{for} \ t>0,
\end{cases}
\eal
then $h^+=u+v$ solves Eq.~\ref{eq:halfdiff}. To solve for $u$ while accounting for the homogeneous Dirichlet boundary condition, we use a spatial Sine Fourier transform~\cite{debnath2016integral} defined as
\al
u_s(k,t) = \sqrt{\frac{2}{\pi}} \int_{x=0}^{\infty} u(x,t) \sin{kx} \,.
\eal
We then obtain
\al
\p_t u_s(q,t) & = - \nu \, q^2 u_s(q,t) + \sqrt{\frac{2}{\pi}} \nu q \, x(t), \\
u_s(q,0) & = 0,
\eal
whose solution reads in the Fourier space
\al
u_s(q,t) = \sqrt{\frac{2}{\pi}} \nu q \int_{t'=0}^t x(t') e^{- \nu q^2 (t-t') },
\eal
and back in the real space
\al
u(y,t) & = \frac{y}{2 \sqrt{\pi \nu }} \int_{t'=0}^t x(t') \frac{e^{-y^2 / 4 \nu (t-t')}}{(t-t')^{3/2}} \\
& = x(t) - x(0) \erf{\left(\frac{y}{2 \sqrt{\nu t}}\right)} - \int_{t'=0}^t \dot{x}(t') \erf{\left(\frac{y}{2 \sqrt{\nu (t-t')}}\right)} \,. \nonumber
\eal
As regards $v$, it is solved via the method of images as it corresponds to a reflecting wall problem with a Dirichlet boundary condition. This yields to
\al
v(y,t) = \int_{y'=0}^\infty \Big( G(y-y',t) - G(y+y',t) \Big) h_0^+(y'), 
\eal
where $G$ is the heat kernel, fundamental solution to the diffusion equation on $\mathbb{R}$, reading
\al
G(y,t)=\frac{1}{2 \sqrt{\pi \nu t}} \exp{ - \frac{y^2}{4 \nu t} } \,.
\eal
We call $G_m(y,y',t)=G(y-y',t)-G(y+y',t)$ the kernel made of the superposition of the two mirrored kernels. All in all, $h^+$ can be written under the form
\al
h^+(y,t) & = \ x(t) - x(0) \erf{\left(\frac{y}{2 \sqrt{\nu t}}\right)} \label{eq:h+} - \int_{t'=0}^t \dot{x}(t') \erf{\left(\frac{y}{2 \sqrt{\nu (t-t')}}\right)} + \int_{y'=0}^\infty G_m(y,y',t) h^+_0(y') \,.
\eal
Using the $y \to -y$ symmetry, $h^-$ can be simply deduced from $h^+$ as $h^-(y,t) = h^+(-y,t)$,  for $y\leq 0$. Eventually, to obtain the dynamics of the full chain, we impose that both half chain meet at $y=0$, namely that $h$ is continuous. \\

It is now possible to obtain the equation of motion for the central monomer of position $x(t)=h(0,t)$. To do so we would like to evaluate Eq.~\ref{eq:SRL} at $y=0$ which is not directly possible due to the Dirac delta distributions. Instead we integrate Eq.~\ref{eq:SRL} over the microscopic interval $[-a/2,a/2]$
\al
\frac{1}{a} \int_{y=-a}^a \p_t h(y,t) = \frac{1}{a} \left( 2 \nu \partial_y h(a/2,t) - \int_{y=-a}^a U'(h(y,t)) \delta(y) + \eta(y,t) \right) \label{eq:averaged_interface},
\eal
where we use that $h$ is an even interface and $\partial_y h$ is odd. Given that $h$ is by construction continuous in $y=0$ and that 
\al
\frac{2 \nu}{a} \partial_y h(0^+,t) = 2 \sqrt{ \frac{\nu}{ \pi a^2 } } \left( \frac{x(0^+)}{\sqrt{t}} - \int_{t'=0}^t \frac{\dot{x}(t')}{\sqrt{t-t'}} + \frac{2}{\sqrt{t}} \int_{y=0}^\infty h_0^+\left(2 y \sqrt{\nu t}\right) y e^{-y^2} \right), \nonumber
\eal
we observe that, if we take $a$ to be the typical distance between neighboring monomers of the polymer in Eq.~\ref{eq:SRP} and if we revert to the microscopic coefficients $\zeta, k$ and $k_B T$, it is possible to safely take the limit $a \to 0$ in Eq.~\ref{eq:averaged_interface} to obtain
\al
\int_{t'=0}^t \dot{x}(t') K(t-t') = -U'(x(t)) + \eta(t) - \tilde{K}(t) \left( x(0) - \mathcal{Z}(t) \right), \label{eq:reduced-dynamics}
\eal
where $\eta$ is a unitary centered Gaussian white noise and 
\al
& K(t) = \delta(t) + \tilde{K}(t), \ \tilde{K}(t) = 2 \sqrt{ \frac{k}{\pi \zeta t} }, \ \mathcal{Z}(t) = 2 \int_{y=0}^\infty h_0^+\left(2 y \sqrt{\nu t}\right) y e^{-y^2} = \int_{y=-\infty}^\infty h_0\left(2 y \sqrt{\nu t}\right) \abs{y} e^{-y^2} \,.
\eal
Eq.~\ref{eq:reduced-dynamics} is the main result of this section. It notably captures the relaxation of the chain's initial conformation at time $t=0$ relative to $x(0)$ through the term $\mathcal{Z}$.

This result generalizes to the continuum ideas that have been developed for the Rouse polymer problem, in~\cite{panja2010generalized} for the equilibrium problem and explicitly in~\cite{vandebroek2016nonequilibrium,vandebroek2017generalized} for the discrete active Rouse polymer model. Our results notably agree with the ones obtained in~\cite{vandebroek2016nonequilibrium} in the limit of an infinite number of beads and provided temperature of all the beads except the central one is switched off~\footnote{More specifically we refer the reader to Chap.~6 and Eq.~$6.85$ in~\cite{vandebroek2016nonequilibrium}. While we treat the specific case of a shaken Rouse polymer, we do not a priori see any particular difficulty in generalizing our derivation in the continuum to the case of the thermal or active polymer treated in~\cite{vandebroek2016nonequilibrium}.}.

The similarity in the structure of Eqs.~\ref{eq:prony_results}~and~\ref{eq:reduced-dynamics} suggests another possible route for the markovian embedding of power law memory kernels. It consists in taking the limit $N \to \infty$ in Eq.~\ref{eq:prony_results} as well as choosing $\alpha_i$ and $\kappa_i$ to scale respectively as $b^{-i}$ and $b^{-i/2}$ with $b$ some positive number, as described in~\cite{goychuk2009viscoelastic}. In our case however, such strategy only works for the memory kernel $K$, even if a similar embedding can be designed for the correlator of the fluctuations $G$. As already noted, splitting the embedding of $K$ and $G$ offer more leeway to represent nonequilibrium GLE, while still allowing to recover the equilibrium case by setting $K=G$.

The advantages of our continuum derivation compared to discrete ones are three-fold. First we obtain a simpler expression -- and thus a clearer interpretation -- of the initial conditions in terms of the relaxation of the initial conformation of the chain. Second the (markovian) dynamics of these models are more tractable at the continuum level, especially thanks to the rich literature on fluctuating interfaces. Both points are leveraged in App.~\ref{app:testing_protocols}. Third, this continuum derivation could extend to other chains exhibiting general power law behaviours of the kernel $K(t)\sim 1/t^\alpha$, or even nonlinear ones, such as the KPZ equation~\cite{kardar1986dynamic}, where working in the continuum may be an asset.

\section{Matching the free dynamics of MPs, AOUPs, and equilibrium dimers}

\label{app:matching}

In this section, we characterize the dynamics of a free MP as defined in Eq.~7 of the main text, which we remind for completeness:
\al
\begin{cases}
\dot{x} = - \kappa (x-z) - U'(x) + \sqrt{2D} \eta, \\
\dot{z} = \alpha (x-z).
\end{cases} \label{eq:MP}
\eal
We show that it can be mapped onto free (anti)persistent AOUPs or equilibrium dimers depending on the values of $\alpha$ and $\kappa$.

\subsection{MPs} 

\label{app:MP}

We first note that the ``size'' $\delta = x - z$ of the dimer $(x, z)$ that constitutes the MP follows an Ornstein–Uhlenbeck process:
\al
\dot{\delta} = -(\kappa + \alpha) \delta + \sqrt{2D} \eta,
\eal 
whose dynamics remain bounded provided that $\alpha + \kappa \geq 0$. The condition $\kappa \geq -\alpha$ is therefore sufficient and necessary to prevent dimer dissociation (and in fact also to make the corresponding kernel $K$ semi-definite positive). We assume this condition to hold throughout the article. The dimer then exhibits a characteristic (memory) length due to thermal fluctuations $\ell_D = \sqrt{D \tau_m}$ where $\tau_m=(\alpha + \kappa)^{-1}$. It also corresponds to the dimer's size in the absence of interaction. 

We next solve Eq.~7 to obtain the evolution of $\dot{x}$ in the stationary state. We find in the Laplace domain
\al
\widehat{\dot{x}}(s) = \sqrt{2D} \hat{\chi}(s) \hat{\eta}(s), \quad \hat{\chi}(s) = \left(1 + \frac{\kappa}{s + \alpha}\right)^{-1} ,
\eal
which in the time domain reads
\al
\dot{x}(t) = \sqrt{2D} \int_{t'=-\infty}^t \chi(t - t') \eta(t'),
\eal
where $\chi(t) = \delta(t) - \kappa e^{-(\alpha + \kappa)t}$. This expression provides an explicit example of the procedure described in App.~\ref{app:linearinversion} for a generic memory kernel. The two-point velocity correlation function then reads
\al
\frac{1}{2D} \langle \dot{x}(t_1) \dot{x}(t_2) \rangle = 
\begin{cases}
     - \dfrac{\kappa(\kappa + 2\alpha)}{2(\alpha + \kappa)} e^{-(\alpha + \kappa) \abs{t_1 - t_2}}, & \text{if } t_1 \neq t_2, \\
     \delta(0), & \text{otherwise,} \label{eq:2pointspeedMP} 
\end{cases}
\eal
while the mean-square increments (MSI) -- sometimes more convenient to study in numerical simulations -- read
\al
\Delta(t) = \lim_{T \to \infty} \langle \left( x(T + t) - x(T) \right)^2 \rangle = 2Dt - 2 D \, \frac{ \kappa (2\alpha+\kappa)}{(\alpha+\kappa)^2} \tau_m \left( \frac{t}{\tau_m} - 1 + e^{-t/\tau_m} \right) \,. \label{eq:MSI_MP}
\eal
These memory particles exhibit qualitatively different behaviors depending on the sign of $\kappa$: they are persistent when the increments of $\langle \dot{x}(t_1) \dot{x}(t_2) \rangle$ are positively correlated ($\kappa<0$) and antipersistent when these correlations are negative ($\kappa>0$).

In the persistent regime $-\alpha < \kappa < 0$, three distinct dynamical regimes emerge, as shown in Fig.~\ref{fig:matching}a. At very short times, $t \ll -2(\alpha+\kappa)/[\kappa(2\alpha+\kappa)]$, the particle displays a diffusive behavior governed by the bare microscopic diffusivity $D$. At long times, $t \gg \tau_m=(\alpha + \kappa)^{-1}$, diffusion is recovered with a larger effective diffusivity $D_{\star}/D = 1 - \kappa(2\alpha + \kappa)/(\alpha + \kappa)^2$. Between these two limits, the dynamics becomes ballistic. Observing a clear ballistic window requires the MP to be strongly persistent, namely $-\kappa(2\alpha+\kappa)/2(\alpha+\kappa)^2 \gg 1$.

In the antipersistent regime, $\kappa>0$, no ballistic crossover appears. The dynamics remains diffusive at both short and long times, with the same short-time microscopic diffusivity $D$ and a reduced long-time diffusivity $D_{\star}/D = 1 - \kappa(2\alpha + \kappa)/(\alpha + \kappa)^2$. The corresponding mean-square increments are shown in Fig.~\ref{fig:matching}b.

\subsection{Comparison with persistent AOUPs} 

\label{app:pAOUP}

\begin{figure}[ht]
	\includegraphics[width=\textwidth,clip]{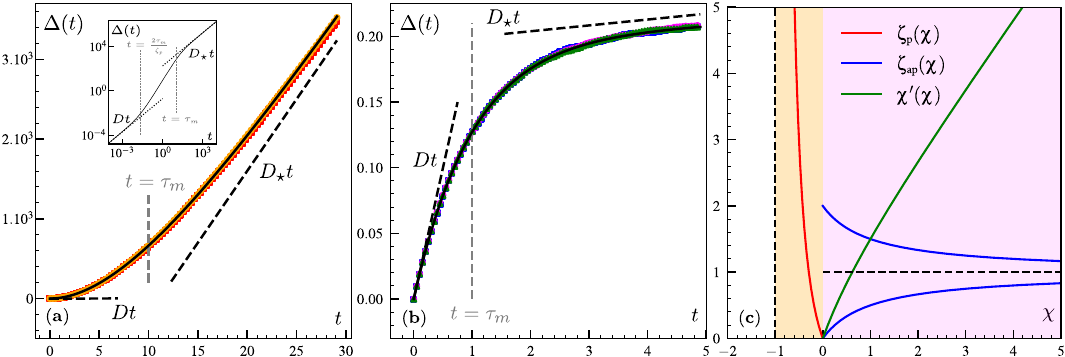}
	\caption{$\mathbf{(a)}$ Stationary increments of a free persistent MP (orange) and of its matched persistent AOUP (red) obtained from numerical simulations. Both models collapse onto the analytical prediction of Eq.~\ref{eq:MSI_pAOUP} (black). The dotted lines indicate the short- and long-time diffusive regimes, respectively characterized by diffusion coefficients $D < D_{\star} = D + v^2$. The grey vertical line marks the crossover time $\tau_m = (\alpha + \kappa)^{-1}$. Inset: analytical prediction of Eq.~\ref{eq:MSI_pAOUP} in log-log scale. The grey vertical lines indicates the two crossover times $t=2 \tau_m/\zeta_p$ and $t=\tau_m$. $\mathbf{(b)}$ Same comparison for antipersistent MPs (magenta), their matched antipersistent AOUPs (blue), and equilibrium dimers (green). In this regime, $D > D_{\star} = D - v_{\rm eff}^2$. The analytical prediction of Eq.~\ref{eq:MSI_apAOUP} is shown in black. $\mathbf{(c)}$ Graphical representation of the matching between the three free-particle models. The dimensionless parameters $\zeta_p$, $\zeta_{ap}$, and $\chi'$ are plotted as functions of the MP parameter $\chi$. Matching persistent (respectively antipersistent) AOUPs to persistent (respectively antipersistent) MPs corresponds to the red (respectively blue, with two branches) curve ; the green curve corresponds to equilibrium dimers. The shaded regions indicate persistent MPs ($\chi \in [-1,0]$, orange) and antipersistent MPs ($\chi \in [0,\infty[$, magenta).}
	\label{fig:matching}
\end{figure}

We now determine the free dynamics of persistent AOUPs, as defined in Eq.~8 of the main text, to establish their correspondence with free persistent MPs. We remind here for completeness the definition of a classical thermal persistent active Ornstein–Uhlenbeck particles (AOUPs)  :
\al
\begin{cases}
\dot{X} = v \xi - U'(X) + \sqrt{2 D} \eta, \label{eq:AOUP} \\
\tau \dot{\xi} = - \xi + \sqrt{2} \lambda,
\end{cases}
\eal
where $\xi$ is an Ornstein-Uhlenbeck process of relaxation time $\tau$ and $\lambda$ and $\eta$ are  independent white noises with unit variance.
We have
\al
\xi(t) = \frac{\sqrt{2}}{\tau} \int_{t'=-\infty}^t \lambda(t') e^{-(t - t')/\tau},
\eal
implying that $\xi$ is a Gaussian variable with
\al
\langle \xi(t_1) \xi(t_2) \rangle = \frac{1}{\tau} e^{-\abs{t_1 - t_2}/\tau} \,. \label{eq:amplixi}
\eal
The stationary two-point velocity correlation function of this Gaussian process thus reads
\al
\langle \dot{X}(t_1) \dot{X}(t_2) \rangle = 2D \delta(t_1 - t_2) + \frac{v^2}{\tau} e^{-\abs{t_1 - t_2}/\tau}, \label{eq:2pointspeedAOUP}
\eal
or, equivalently, in terms of the mean-square increments (MSI),
\al
\Delta(t) = \lim_{T \to \infty} \langle \left( X(T + t) - X(T) \right)^2 \rangle = 2Dt + 2v^2 \tau \left( \frac{t}{\tau} - 1 + e^{-t/\tau} \right) \,. \label{eq:MSI_pAOUP}
\eal
Comparing Eqs.~\ref{eq:2pointspeedMP},\ref{eq:2pointspeedAOUP} or Eqs.~\ref{eq:MSI_MP},\ref{eq:MSI_pAOUP}, a persistent MP can be mapped one-to-one onto a persistent AOUP by choosing
\al
\tau = \frac{1}{\alpha + \kappa}, \quad v^2 = -\frac{\kappa(2\alpha + \kappa)}{(\alpha + \kappa)^2} D \quad \iff \quad \alpha = \frac{1}{\tau} \sqrt{1 + \frac{v^2}{D}}, \quad \kappa = \frac{1}{\tau} \left( 1 - \sqrt{1 + \frac{v^2}{D}} \right), \label{eq:parameters_match}
\eal
as illustrated in the MSI of Fig.~\ref{fig:matching}a. Introducing the dimensionless parameters $\zeta_p = v^2 / D$ and $\chi = \kappa / \alpha$, we find that $\zeta_p$ depends only on $\chi$ and vice versa:
\al
\tau = \frac{\alpha^{-1}}{1 + \chi}, \quad \zeta_p = - \frac{\chi(2 + \chi)}{(1 + \chi)^2} \quad \iff \quad \alpha = \frac{\sqrt{1 + \zeta_p}}{\tau}, \quad \chi = -1 + \frac{1}{\sqrt{1 + \zeta_p}} \,. \label{eq:matchingfreepersistent}
\eal
In this work, we focus on the regime where the self-propulsion of persistent AOUPs and MPs dominates over thermal fluctuations, namely $\zeta_p \gg 1$. In this limit, their dynamics exhibits the three characteristic regimes discussed above, separated by two well-distinct crossover times, $2 \tau_m/\zeta_p$ and $\tau_m$, as illustrated in Fig.~\ref{fig:matching}a.

\subsection{Comparison with antipersistent AOUPs}

\label{app:apAOUP}

As discussed in Sec.~\ref{app:linearinversion}, a free dynamics driven by a memory kernel can be mapped onto one driven purely by correlated noise. For the antipersistent MP, this leads us to define an antipersistent Active Ornstein–Uhlenbeck Particle (AOUP) as
\al
\begin{cases}
	\dot{X} = - v \xi - U'(X) + \sqrt{2D} \eta, \label{eq:antipersistentOUP} \\
	\tau \dot{\xi} = - \xi + \sqrt{2} \eta,
\end{cases}
\eal
which turns out to be indistinguishable from an antipersistent MP, in the absence of interaction ($U'=0$). This antipersistent AOUP differs from the persistent AOUP only in that the noise term $\eta$ enters both equations -- with a minus sign in front of $v \xi$. Importantly and contrary to the persistent AOUP, this implies that this equation of motion is no longer invariant under the symmetry $(v,\xi) \to (v,-\xi)$~; this has notable consequences at linear level as discussed in this section but also in the presence of interactions as discussed in App.~\ref{app:interactions}. 

In the absence of interaction, it is possible to determine the MSI of an antipersistent AOUPs, which reads
\al
\lim_{T \to \infty} \langle \left( X(T + t) - X(T) \right)^2 \rangle = 2Dt + 2v(v-2\sqrt{D}) \tau \left( \frac{t}{\tau} - 1 + e^{-t/\tau} \right) \,.
\eal
In order for the particle to be antipersistent (at large times), we thus see that we need to impose $v<2 \sqrt{D}$, the anticorrelated drive ($\xi$) being otherwise too strong with respect to activating thermal noise ($\eta$). We thus introduce the dimensionless parameter $\zeta_{ap}=v/\sqrt{D}$ measuring the relative strength of the drive with respect to the thermal noise. Upon assuming $\zeta_{ap}<2$ (which we do in the rest of the paper), the MSI can be rewritten
\al
\lim_{T \to \infty} \langle \left( X(T + t) - X(T) \right)^2 \rangle = 2Dt - 2v_{\rm eff}^2 \tau \left( \frac{t}{\tau} - 1 + e^{-t/\tau} \right), \label{eq:MSI_apAOUP}
\eal
where $v_{\rm eff} = v \sqrt{2 / \zeta_{ap} - 1}$. The MSI of a free antipersistent AOUP is thus similar to that of a persistent AOUP, except for the redefinition of the effective speed $v_{\rm eff}$, which carries a minus sign. The macroscopic diffusion coefficient is therefore smaller than the microscopic one, reflecting the expected antipersistence of the particles. A numerical confirmation of this behavior is shown in Fig.~\ref{fig:matching}b.

Following the same reasoning as in the previous sections, an antipersistent MP can be mapped one-to-one onto an antipersistent AOUP in the regime $\alpha, \kappa \geq 0$, by choosing
\al
\tau = \frac{1}{\alpha + \kappa}, \ v = \frac{\kappa \sqrt{D}}{\alpha + \kappa} \ \text{or} \ \frac{(2\alpha + \kappa)\sqrt{D}}{\alpha + \kappa} \label{eq:matching_pAOUP_pMP} \Leftrightarrow \alpha = \frac{v / \sqrt{D} - 1}{\tau}, \ \kappa = \frac{2 - v / \sqrt{D}}{\tau} \ \text{or} \ \alpha = - \frac{v / \sqrt{D} - 1}{\tau}, \ \kappa = \frac{v / \sqrt{D}}{\tau},
\eal
depending whether $v / \sqrt{D}$ is greater or smaller than $1$. In terms of the dimensionless parameters $\zeta_{ap} = v / \sqrt{D} $ and $\chi = \kappa / \alpha$, we obtain
\al
\tau= \frac{\alpha^{-1}}{1 + \chi}, \quad \zeta_{ap} = \frac{\chi}{1 + \chi} \ \text{or} \  \frac{2 + \chi}{1 + \chi} \quad  \iff \quad \alpha = \frac{\tau}{1 + \chi}, \quad \chi = \frac{\zeta_{ap}}{1 - \zeta_{ap}} \ \text{or} \ \frac{2 - \zeta_{ap}}{\zeta_{ap}-1} \,.
\eal
A numerical confirmation of this matching is shown in Fig.~\ref{fig:matching}b. Unlike the persistent case, the mapping here is possible only if $\zeta_{ap} \leq 2$ and two different mappings are possible in order to match an antipersistent MP onto an antipersistent AOUP, as it is reflected in Fig.~\ref{fig:matching} (two blue branches). Conversely, given an antipersistent AOUP (with $\zeta_{ap}<2$), there is a unique way to match it onto an antipersistent MP, depending whether $\zeta_{ap}$ is greater or smaller than $1$. In this article, we choose to work with $\zeta_{ap}<1$.  

\subsection{Comparison with equilibrium dimers}

Free antipersistent MPs can also be mapped onto another class of particle models, namely equilibrium dimers whose two beads $X$ and $Z$ are subjected to distinct friction coefficients $\gamma_X$ and $\gamma_Z$
\al
\begin{cases}
	\gamma_X \dot{X} = -\kappa(X - Z) - U'(X) + \sqrt{2 \gamma_X k_B T} \eta,  \\
	\gamma_Z \dot{Z} = \kappa(X - Z) + \sqrt{2 \gamma_Z k_B T} \eta_Z \,.
\end{cases}
\eal
Here $k_B$ is the Boltzmann constant, $T$ is the temperature of the heat bath, and $\eta$ and $\eta_Z$ are independent unit Gaussian white noises. The dynamics can be equivalently written as
\al
\begin{cases}
	\dot{X} = -\kappa'(X - Z) - U'(X) + \sqrt{2 D} \eta, \\
	\dot{Z} = r \kappa'(X - Z) + \sqrt{2 D r} \eta_Z,
\end{cases}
\eal
where $\kappa' = \kappa / \gamma_X$, $r = \gamma_X / \gamma_Z$, and $D = k_B T / \gamma_X$. The MSI for the equilibrium dimers thus read
\al
\lim_{T \to \infty} \langle \left( X(T + t) - X(T) \right)^2 \rangle = 2Dt - 2v_{\rm eff}^2 \tau_{\rm eff} \left( \frac{t}{\tau_{\rm eff}} - 1 + e^{-t/\tau_{\rm eff}} \right), \label{eq:MSI_eqdim}
\eal
where $\tau_{\rm eff}^{-1}=(1+r)\kappa'$ and $v_{\rm eff} = \sqrt{D/(1+r)}$. Following the same reasoning as before, we find that equilibrium dimers and antipersistent MPs ($\alpha, \kappa > 0$) match in the absence of interactions when
\al
\kappa' = \frac{\kappa(2\alpha + \kappa)}{\alpha + \kappa}, \quad r = \frac{\alpha^2}{\kappa(2\alpha + \kappa)} \label{eq:matching} \quad \iff \quad \alpha = \kappa' \sqrt{r(1 + r)}, \quad \kappa = \kappa' \left( 1+r - \sqrt{r(1 + r)} \right) \,.
\eal
In terms of the dimensionless parameters $\chi=\kappa/\alpha, \chi' = \kappa' / \alpha$ and $r$, this reads
\al
\chi' = \frac{\chi(2 + \chi)}{1 + \chi}, \quad r = \frac{1}{\chi(2 + \chi)} \iff \chi' = \frac{1}{\sqrt{r(1 + r)}} \,.
\eal
Unlike the antipersistent AOUP, the equilibrium dimer can be matched onto an antipersistent MP without restriction, as illustrated in Fig.~\ref{fig:matching}c. This mapping is numerically confirmed in Fig.~\ref{fig:matching}b. \\

In this section, we have studied different models of particles and shown that the (anti)persistent ones were indistinguishable in the absence of interaction. This degeneracy is in fact lifted as soon as interactions are turned  on and qualitatively distinct behaviors emerge, as we discuss in App.~\ref{app:interactions}.

\section{Trap-and-release protocols}

\label{app:testing_protocols}

We show in this section how to determine the dynamics of particles described by a GLE in the trap-and-release protocols displayed in Fig.~$1$ of the main text. As a corollary, it allows to quantify the memory that a MP keeps from past environments and to show that both AOUPs (persistent or antipersistent) and equilibrium dimers are insensitive to these testing protocols, and thus have no environmental memory. As we will see, these protocols allow to probe how the stationary state of internal variables of GLEs depend on the  environment. \\

We recall that the trap-and-release protocols consist: $(i)$ in confining a particle in an harmonic trap during a sufficiently long time, so that the particle reaches a stationary state~; $(ii)$ in imposing a sudden change of environment by releasing the particle, once such a stationary state is reached. It is quantitatively defined by the following dynamics
\al
\int_{t'=-\infty}^t \dot{x}(t') K(t-t') = -k(t) x(t) + \sqrt{2D} \eta(t), \quad \langle \eta(t)\eta(t) \rangle = G(\abs{t-t'}), \label{eq:protocols}
\eal
where $k(t)=0$ for the control protocol while $k(t) = k \left(1-\Theta(t)\right)$ for the testing protocol, $\Theta$ standing for the Heaviside step function. 

It is useful to rewrite the memory kernel as $K(t)=\delta(t)+\tilde{K}(t)$. We  recall that $(i)$ for MPs $K(t)=\delta(t)+\kappa e^{-\alpha t}, \ G(t)=\delta(t)$~; $(ii)$ for persistent AOUPs $K(t)=\delta(t), \ G(t)=\delta(t)+e^{-\abs{t}/\tau}/\tau$~; $(iii)$ for antipersistent AOUPs, $K(t)=\delta(t), \ G(t)=\delta(t)-e^{-\abs{t}/\tau}/\tau$~; $(iv)$ for equilibrium dimers $K(t)=G(t)=\delta(t)+\kappa' e^{-r \kappa' t}$~; $(v)$ for the SRL $K(t)=\delta(t) + (\tau/t)^{1/2}, \ G(t)=\delta(t)$.

A few comments are in order. $(i)$ In this article, we consider for simplicity an harmonic trapping potential in the trap-and-release protocols, but the phenomenology that we find  applies to any sudden pertubation of the environment. This choice of an harmonic potential allows for explicit analytical results, but the numerical results as displayed in Fig.~$1$ of the main text are straightforward to obtain for any other choice of environment. $(ii)$ The quantification of memory effects in trap-and-release protocols can be analyzed via different observables such as two-time correlation functions of polarity $p$
\al
c(T,t) = \langle p(T+t)p(T) \rangle, \ \text{for} \ T,t>0,
\eal
or  the mean-square increments (MSI) of the position $x$ 
\al
\Delta(t,T) =  \langle \left( x(t+T)-x(T) \right)^2 \rangle, \ \text{for} \ T,t>0 \,.
\eal
In fact, in the case of an harmonic trap, the linear GLEs at stakes are Gaussian stochastic processes which are fully characterized by $c(T,t)$ or $\Delta(T,t)$. As regards $\Delta(T,t)$, it is important to compare particles conditioned so that $x(t=0)=0$~; this allows to avoid a bias due to a non vanishing initial force  in their free dynamics. This conditioning is however not necessary for the polarity as we discuss next and polarity thus appears as a more convenient observable when investigating environmental changes with more general interacting environments. Given that the dynamics we study can be cast under the form
\al
\dot{x} = p - U'(x) + \sqrt{2D} \eta, \quad \langle \eta(t)\eta(t') \rangle = \delta(t-t'),
\eal
characterizing the polarity $p$ is enough to completely determine the position $x(t)$.

In this section we investigate  trap-and-release protocols following two different routes: either via a Markovian embedding (App.~\ref{app:detourmarkovian}) which is helpful to develop intuition, or more directly within the non-Markovian GLE formalism (App.~\ref{app:GLEroute}).

\subsection{Markovian embedding}

\label{app:detourmarkovian}

We first assume we have access to the markovian representation of the different types of particles analyzed in this article. As suggested earlier, we mainly focus on the dynamics of the polarity $p$ and we will see that the effect of the testing protocols can be mainly read on the stationary distribution of the internal variables at time $t=0$.

\subsubsection{MPs}

\label{app:markov_trap_MP}

We first consider MPs in the first part of the trap-and-release protocols, where they are trapped in an harmonic potential of stiffness $k$. The control case (absence of any trap in the past) corresponds to $k=0$. Given that the markovian representation of MPs in the protocols
\al
\begin{cases}
\dot{x} = - \kappa (x-z) - k x + \sqrt{2D} \eta, \\
\dot{z} = \alpha (x-z) \,.
\end{cases}
\eal
is linear, the polarity $p=-\kappa(x-z)$ is explicit and it reads
\al
p(t)  = (x(0)-z(0)) \chi(t) + \sqrt{2D} \int_{t'=0}^t \chi(t-t') \eta(t'), \label{eq:evol_MP}
\eal
where $\chi(t)=-\kappa e^{-(\alpha+\kappa)t}$. The stationary state in the first part of the trap-and-release protocols is thus characterized by the following variance 
\al
\langle (x-z)^2 \rangle_- = \frac{D}{k+\alpha+\kappa}, \label{eq:variance_trapped_MP}
\eal
where $\langle \ \cdot \ \rangle_-$ denotes the average over past realizations of the noise $\{\eta(t)\}_{t \leq 0}$. Importantly, the variance is potential-dependent (via $k$) as expected out of equilibrium. 

We now want to determine the evolution of the MPs in the second part of the trap-and-release potential, once the trap has been released. To do so, we can reuse Eq.~\ref{eq:evol_MP} in the second part of the protocols (this time $k=0$), with the caveat that the exact relative positions between $x$ and $z$ at time $t=0$ is not known, only its statistics via Eq.~\ref{eq:variance_trapped_MP}. Upon averaging over $(i)$ the future realizations of the noise $\{\eta(t)\}_{t \leq 0}$ denoted by $\langle \ \cdot \ \rangle_+$ and $(ii)$ the initial conditions via $\langle \ \cdot \ \rangle_-$, the two-time correlator of polarity reads for $T,t>0$
\al
c(T,t) & = \langle \langle p(t+T)p(T) \rangle_+ \rangle_- \\
& = D \kappa^2 \left( \frac{ e^{-(\alpha +\kappa ) (t+2 T)}}{\alpha +\kappa +k}+\frac{2 \sinh (T (\alpha +\kappa )) e^{-((\alpha +\kappa ) (t+T))} }{\alpha +\kappa } \right) \,. \label{eq:corr_MP}
\eal
We find that, due to the presence of a trap, the correlator $c(T,t)$ is a function of $T$ and is thus non stationary, as it is displayed in Fig.~$1$ of the main text. This transient behavior is observed until $p$ (or $z$ in fact) reaches a new stationary state adapted to the absence of potential. It is a direct signature of  environmental memory in the system~; in this example memory effects are controlled by the time scale  $\tau_m=(\alpha+\kappa)^{-1}$.

By definition for the control protocol ($k=0$ in the past), the correlator $c(T,t)$ in Eq.~\ref{eq:corr_MP} no longer depends on $T$ and is thus stationary.

\subsubsection{AOUPs}

As regards the behavior of the polarity $p= \pm v \xi$ of AOUPs (be they persistent or antipersistent), it is completely independent of the environment $U$ in Eqs.~8~and~\ref{eq:antipersistentOUP} and thus insensitive to the trap-and-release protocols. A few $\tau$'s after the dynamics has started (at $t=-\infty$), the correlator of the polarity reads
\al
c(T,t) = \langle p(T)p(t+T) \rangle = \frac{v^2}{\tau} e^{-\abs{t}/\tau} \,. \label{eq:correlator_polarity_AOUP}
\eal
It is stationary and it stays unaffected for the rest of the dynamics. This is a direct consequence of the fact that, while AOUPs do feature an internal state $\xi$, $\xi$ is however not in a feedback loop with the environment $U$.

\subsubsection{Equilibrium dimers}

It turns out that the equilibrium dimer, while featuring an internal state $Z$ that is in feedback loop with the environment, is also insensitive to the trap-and-release protocols. This stems from the equilibrium nature of $Z$ as it is discussed in this section.

We proceed similarly as for MPs in App.~\ref{app:markov_trap_MP}. In the first part of the protocols, the markovian representation of equilibrium dimers reads
\al
\begin{cases}
	\dot{X} = -\kappa'(X-Z) - k X + \sqrt{2 D} \eta,\\
	\dot{Z} = r \kappa'(X-Z) + \sqrt{2 D r} \eta_Z \,.
\end{cases}
\eal
By linearity again, the polarity $p=-\kappa'(X-Z)$ reads
\al
p(t) = (X(0)-Z(0)) \chi(t) + \sqrt{2D} \int_{t'=0}^t \chi(t-t') \left( \eta(t') - \sqrt{r} \eta_Z(t') \right), \label{eq:evol_eqdim}
\eal
where $\chi(t) =  -\kappa' e^{-\kappa' (r+1) t}$. This implies notably that the equal-time correlator for the polarity in the stationary confined state reads 
\al
\langle (X-Z)^2 \rangle_- = \frac{D}{\kappa'},
\eal
where $\langle \ \cdot \ \rangle_-$ denotes the average over past realizations of the noises $\{\eta(t)\}_{t \leq 0}$ and $\{\eta_Z(t)\}_{t \leq 0}$. Importantly, this equal-time correlator is independent of the environment (i.e. of the spring stiffness $k$) as expected in equilibrium (see below) ; this stands in contrast to the case of  MPs, which are not at equilibrium.  

This very fact implies that equilibrium dimers are insensitive to the protocols, given that both the testing protocol and the control yield the same variance $\langle (X-Z)^2 \rangle_-$. To see it, we resort again to Eq.~\ref{eq:evol_eqdim} to find out the evolution for the polarity of the equilibrium dimer in the second part of the protocol. The two-time correlations, after averaging over the initial positions resulting from the stationary  state for $T<0$, read for $T,t>0$
\al
c(T,t) & = \langle \langle p(t+T)p(T) \rangle_+ \rangle_- \\
& =  D \kappa' e^{-\kappa' (r+1) t},
\eal
where $\langle \ \cdot \ \rangle_+$ denotes the average over future realizations of the noises $\{\eta(t)\}_{t \geq 0}$ and $\{\eta_Z(t)\}_{t \geq 0}$. Importantly, $c(T,t)$ is independent of $T$ and is stationary, which means that it is indeed insensitive to the trap-and-release protocols. More generally, it reflects the fact that $Z$ is in the same  stationary at time $t=0$ irrespective of the confining potential. This is a purely equilibrium effect that turns out to hold for any confining potential $U$. This stems from the fact that the joint equilibrium Boltzmann measure for the equilibrium dimer reads
\al
p(X,Z) \propto \exp{-\frac{\kappa'(X-Z)^2/2 + U(X)}{D}},
\eal
which implies that the marginal distribution for $\delta=X-Z$ satisfies to
\al
p(\delta=X-Z) \propto \exp{-\frac{\kappa'(X-Z)^2}{2D}}, \quad \langle (X-Z)^2 \rangle = \frac{D}{\kappa'},
\eal
which is independent of $U$.

\subsubsection{SRL}

We previously observed that the polarity or the MSI of a MP are displaying a transient behavior in the trap-and-release protocols, that is decaying exponentially fast and which is a direct consequence of the exponential decay of the memory kernel. We show in this section that this transient behavior (and therefore the time a MP keeps information about past environments) can be made arbitrary long upon considering power law memory kernels.

To gain some intuition about the lasting memory effects in presence of power law memory kernels, we first work with an explicit markovian model, whose reduced dynamics obeys a GLE with power law memory kernels. More precisely, we have seen in App.~\ref{app:embeddingSRL} that the shaken Rouse line $h$ defined as
\al
& \p_t h(y,t) = \nu \p_x^2 h(y,t) - k(t) h(y,t) \delta(y) + \eta(y,t), \\
& \langle \eta(y,t) \eta(y',t') \rangle = 2 D \delta(y) \delta(y') \delta(t-t'),
\eal
defines  a markovian embedding of a GLE with memory kernel $K(t) \sim \delta(t)+t^{-1/2}$ and noise correlator $G(t) \sim \delta(t)$. Here we take the interacting potential acting on the central monomer $x(t)=h(0,t)$ to be the trap-and-release potential, i.e. an harmonic trap of stiffness $k(t)=0$ for the control protocol and $k(t)=k(1-\Theta(t))$ for the trap-and-release protocol, $\Theta$ standing for the Heaviside step function.

We are interested in the behavior of the central monomer $x(t)=h(0,t)$ after its release from the trap, that we characterize by the MSI $\Delta(T,t)$ reading 
\al
\Delta(T,t) = \langle \left( x(T+t)-x(T) \right)^2 \rangle = \langle \left( h(0,T+t)-h(0,T) \right)^2 \rangle,
\eal
where $x(t)=h(0,t)$ is the position of the central monomer. It is useful to introduce $w_T(t)=h(0,T+t)-h(0,T)$ to rewrite the MSI as
\al
\Delta(T,t) = \langle w_T(t)^2 \rangle = \langle w_T(t) \rangle^2 + \langle w_T(t)^2 \rangle_c
\eal
where $\langle \ \cdot \ \rangle_c$ stands for the two-point connected correlation function. To go further, we need to determine the dynamics of $h$, which reads in the Fourier space for a constant stiffness $k$
\al
& \p_t h(q,t) = - \left( \nu q^2 + k \right) h(q,t) + \eta(q,t), \label{eq:interface_qspace} \\
& \langle \eta(q,t) \eta(q',t') \rangle = 2 D \delta(t-t') \,.
\eal
The linearity of this equation allows to solve $h(q,t)$ as
\al
h(q,t) = h(q,0) e^{-(\nu q^2 +k) t} + \int_{t'=0}^t \eta(q,t-t') e^{-(\nu q^2+k) t'}\,. \label{eq:hq}
\eal
It is a gaussian random variable determined by its first moment
\al
& \langle h(q,t) \rangle = h(q,0) e^{-(\nu q^2+k) t},
\eal
and its two-point connected correlation function
\al
\langle h(q_1,t_1) h(q_2,t_2) \rangle_c & = \langle h(q_1,t_1) h(q_2,t_2) \rangle - \langle h(q_1,t_1) \rangle \langle h(q_2,t_2) \rangle \\
& = \frac{2D}{2k + \nu (q_1^2 + q_2^2) } \left( e^{-\left[ \nu Q(t_1-t_2) + k \right] \abs{t_1-t_2}} - e^{-\nu \left( q_1^2 t_1 + q_2^2 t_2 \right) -k(t_1+t_2)} \right) , \nonumber \label{eq:hqhq}
\eal
where
\al
Q(t_1-t_2) =
\begin{cases}
	q_1^2, \ \text{if} \ t_1-t_2>0, \\
	q_2^2, \ \text{otherwise} \,.
\end{cases}
\eal
An expression for $h(y,t)$ can then be obtained by an inverse Fourier transform
\al
h(y,t) = \frac{1}{2 \pi} \int_{q=-\pi/a}^{\pi/a} h(q,t) e^{i q y} \,.
\eal
Here $a$ is a microscopic length or UV cutoff (see for instance~\cite{nattermann1992kinetic}) that we already encountered in App.~\ref{app:embeddingSRL} and below which the continuous model of the Rouse line does not properly describe the discrete Rouse polymer model.

\begin{figure}[H]
\centering
\begin{subfigure}[b]{0.5\textwidth}
  \includegraphics[width=\textwidth]{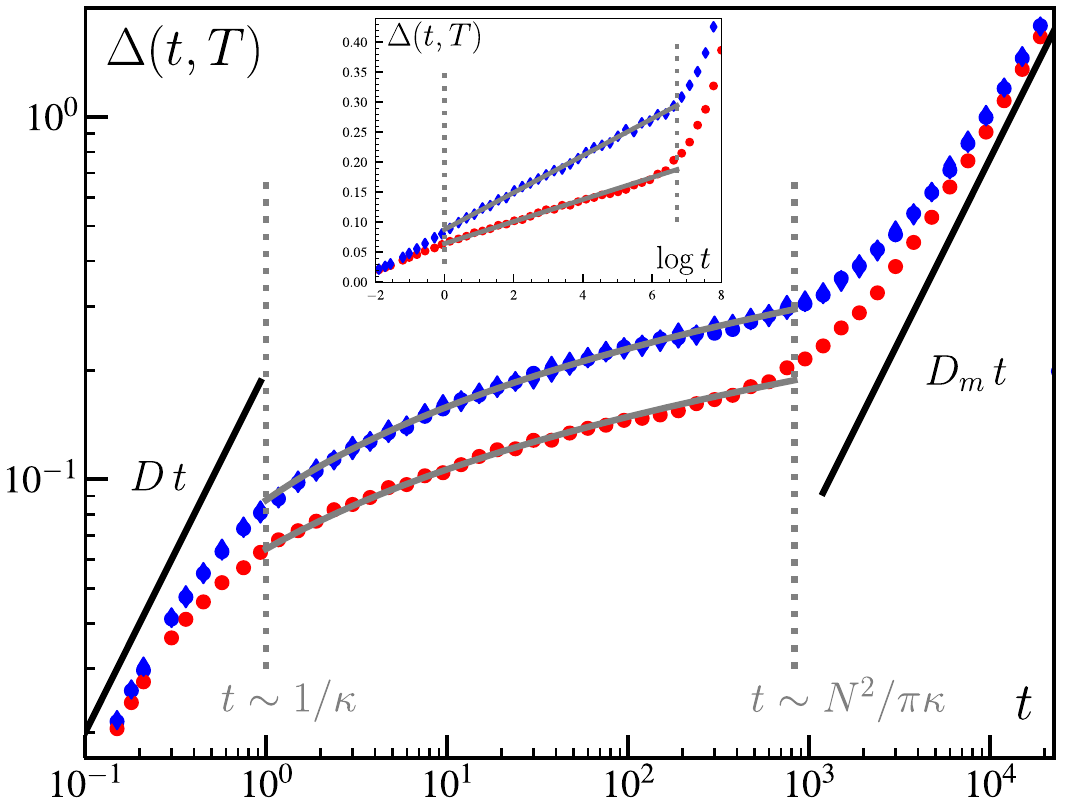}
\end{subfigure}
\caption{Memory-induced transient in the trap-and-release protocol for the Shaken Rouse Line (SRL). Mean-squared increments (MSI) $\Delta(t,T)=\langle \left( x(t+T)-x(T) \right)^2 \rangle$ are shown for the stationary state and control protocol (blue symbols) and after release from a trap imposed in the past (red dots). The difference between the two curves demonstrates a clear memory effect. Because the SRL is simulated as a discrete Rouse polymer with $N$ monomers, the MSI exhibits three regimes: $(i)$ an initial microscopic diffusive regime with diffusion coefficient $D$ (black line), $(ii)$ an intermediate logarithmic growth $\Delta(t)\!\sim\!\log t$ (grey line: fit) and $(iii)$ a final macroscopic diffusive regime with diffusion coefficient $D_m = D/N^2$ once all collective modes have relaxed. The corresponding crossover times, indicated by grey dashed lines, occur at the local relaxation time $t\!\sim\!1/\kappa$ and at the collective time $t\!\sim\!N^2/\pi \kappa$, where $\kappa$ is the spring stiffness of the discrete Rouse polymer. The inset shows $\Delta(t,T)$ plotted against $\log t$, confirming the quality of the logarithmic scaling. Simulation details in App.~\ref{app:simu}.}
\label{fig:SRL_protocols}
\end{figure}

With these results, it is now possible to determine $\Delta(t,T)$ in the trap-and-release protocols. The connected part of $\Delta(t,T)$ reads
\al
\langle w_T(t)^2 \rangle_{+,c} = \langle h(0,T+t)^2 \rangle_{+,c} - 2 \langle h(0,T+t)h(0,T) \rangle_{+,c} + \langle h(0,T)^2 \rangle_{+,c} ,
\eal 
where $\langle \ \cdot \ \rangle_+$ stands for the average over the future realizations of the noise $\{\eta(t)\}_{t \geq 0}$. For $t,T>0$, $k=0$ and we know via Eq.~\ref{eq:hqhq} that
\al
\langle h(0,T)h(0,T) \rangle_{+,c} & = \frac{1}{(2 \pi)^2} \int_{q_1,q_2=-\pi/a}^{\pi/a} \langle h(q_1,T)h(q_2,T) \rangle_{+,c} \\
& = \frac{1}{(2 \pi)^2} \int_{q_1,q_2=-\pi/a}^{\pi/a} \frac{2D}{\nu(q_1^2+q_2^2)} \left( 1 - e^{-\nu (q_1^2+q_2^2) T} \right) \\
& = \frac{D}{\pi \nu} \int_{\rho=0}^{\pi \sqrt{\nu T} /a} \frac{1-e^{-\rho^2}}{\rho} \\
& \underset{ \nu T \gg a^2}{\propto} \ln{\frac{\nu T}{a^2}},
\eal
and that 
\al
\langle h(0,T+t)h(0,T) \rangle_{+,c} & = \frac{1}{(2 \pi)^2} \int_{q_1,q_2=-\pi/a}^{\pi/a} \langle h(q_1,T+t)h(q_2,T) \rangle_{+,c}, \\
& = \frac{1}{(2 \pi)^2} \int_{q_1,q_2=-\pi/a}^{\pi/a} \frac{2D e^{-\nu q_1^2 t}}{\nu(q_1^2+q_2^2)} \left( 1-e^{-\nu (q_1^2+q_2^2) T} \right), \\
& = \frac{2D}{\nu (2\pi)^2} \int_{\theta=0}^{2 \pi} \int_{\rho=0}^{\pi/a} \frac{1-e^{-\nu \rho^2 T}}{\rho} e^{-\nu \rho^2 \cos^2{\theta} t}, \\
& = \frac{2D}{\nu (2\pi)^2} \int_{\theta=0}^{2 \pi} \int_{\rho=0}^{\pi \sqrt{\nu T} /a} \frac{1-e^{-\rho^2}}{\rho} e^{-\rho^2 \cos^2{\theta} \, t/T}, \\
& \hspace{-1.5em} \underset{ \nu T \gg a^2, \, \nu t}{\propto} \ln{\frac{T}{t}},
\eal
To understand these asymptotics, we follow  \cite{nattermann1992kinetic} and  split the $\rho$-integral over $[0,1]$ and $[1,\pi \sqrt{\nu T}/a]$, express the integral over the second interval in terms of the exponential integral and perform its Taylor series into small arguments. Collecting the different asymptotics we obtained, we find in the limit $T \to \infty$
\al
\langle w_T(t)^2 \rangle_{+,c} \underset{ \nu T \gg a^2, \, \nu t}{\propto} \ln{\frac{\nu (t+T)}{a^2}} - 2 \ln{\frac{T}{t}} + \ln{\frac{\nu T}{a^2}} \underset{ \nu T \gg a^2}{\propto} \ln{\frac{\nu t}{a^2}},
\eal
as  is numerically confirmed in Fig.~\ref{fig:SRL_protocols}. 

As regards the disconnected contribution in $\Delta(t,T)$, it can be determined using Eq.~\ref{eq:hq} and it reads
\al
\langle w_T(t) \rangle_+^2 & = \frac{1}{(2 \pi)^2} \int_{q_1,q_2=-\pi/a}^{\pi/a} h(q_1,0)h(q_2,0) \left( e^{- \nu q_1^2 (t+T)}- e^{-\nu q_1^2 T} \right) \left( e^{- \nu q_2^2 (t+T)} - e^{-\nu q_2^2 T} \right) \,.
\eal
In our protocols however, the initial conformation $h(x,0)$ of the line at $t=0$ is not known, only its statistical properties that are the ones of the stationary state reached by the line either in presence of an harmonic trapping (trap-and-release protocol) or in absence of potential (control protocol). We therefore average the disconnected contribution over the past realizations of the noise $\{\eta(t)\}_{t<0}$
\al
\langle \langle w_T(t) \rangle_+^2 \rangle_- & = \frac{1}{(2 \pi)^2} \int_{q_1,q_2=-\pi/a}^{\pi/a} \langle h(q_1,0)h(q_2,0) \rangle^- \left( e^{- \nu q_1^2 (t+T)}- e^{-\nu q_1^2 T} \right) \left( e^{- \nu q_2^2 (t+T)} - e^{-\nu q_2^2 T} \right), \label{eq:disconnected}
\eal
where $\langle \ \cdot \ \rangle_-$ stands for the average over past realizations of the noise $\{\eta(t)\}_{t \leq 0}$. To be able to characterize $\langle h(q_1,0)h(q_2,0) \rangle_-$, we first need to determine the stationary conformation of the line in the past. It is encoded by the equal-time correlator of $u(y,t)=h(y,t)-h(0,t)$ which reads
\al
\langle u(y_1,t)u(y_2,t) \rangle_- & = \frac{1}{(2 \pi)^2} \int_{q_1,q_2=-\pi/a}^{\pi/a} \frac{2D}{2k+\nu(q_1^2+q_2^2)} \left( 1 - e^{-\nu t (q_1^2+q_2^2)-2kt} \right) \left( e^{i (q_1 y_1+q_2 y_2)} - e^{i q_1 y_1} - e^{i q_2 y_2} +1 \right) \nonumber \\
& = \frac{2D}{2 \pi} \int_{\rho=0}^{\pi/a} \frac{1-e^{-(2k+\nu \rho^2) t}}{2k+\nu \rho^2} \left( J_0\left(\rho \sqrt{y_1^2+y_2^2}\right) - J_0\left(\rho \abs{y_1}\right) - J_0\left(\rho \abs{y_2}\right) + 1 \right) \,.
\eal
where $J_0$ is the $0$-th Bessel function of the first kind. We are now interested in the asymptotic conformation of the line, namely of $\langle u(y_1,t)u(y_2,t) \rangle_-$ in the limit of large $y_1,y_2$ assuming that $y_1/y_2$ is a nonvanishing constant equal to $\alpha$. If the SRL is confined in the harmonic trap ($k\neq0$), this former integral is convergent both in the limit of $a \to 0$ and of $y_1 \to \infty$. However, in the control case ($k=0$ in the past), the integral is now divergent in the limit $a \to 0$ and we have
\al
\langle u(y_1,t)u(y_2,t) \rangle_- \underset{\cC}{\propto} \frac{D}{2\pi \nu} \ln{\frac{y_1^2+y_2^2}{a^2}},
\eal
where the condition $\cC$ reads $\nu t \gg y_1^2+y_2^2 \gg a^2, y_1/y_2=\alpha$. This means that the equal-time correlator in the Fourier space in the stationary state $\langle h(q_1)h(q_2) \rangle^-$ decays either as $1/(q_1^2+q_2^2)$ in the control case ($k=0)$ or to $\delta(q_1+q_2)$ in the trap-and-release case ($k\neq 0$). 

Going back to Eq.~\ref{eq:disconnected}, this means that, for the control case, $\langle \langle w_T(t) \rangle_+^2 \rangle_-$ scales as a $\ln{\nu t/a^2}$ in the limit of $T\to\infty$. This was expected given that the control protocol must be stationary at any time, namely
\al
\Delta(t,T)=\langle w_T(t) \rangle^2 + \langle w_T(t)^2 \rangle_c=\lim_{T \to \infty} \langle w_T(t)^2 \rangle_c \propto \ln{\frac{\nu t}{a^2}} \,.
\eal
As regards the trap-and-release protocols, there is now a difference in scaling between $\langle w_T(t) \rangle^2$ and $\langle w_T(t)^2 \rangle_c$. More precisely, we have in the limit $T \to \infty$
\al
\langle w_T(t) \rangle^2 & \propto \frac{1}{(2 \pi)^2} \int_{q=-\pi/a}^{\pi/a} e^{-2 \nu q^2 T }\left( 1 - e^{- \nu q^2 t} \right)^2 \\
& \propto \frac{(\nu T)^{-1/2}}{(2 \pi)^2} \int_{q=-\pi \sqrt{\nu T} /a}^{\pi \sqrt{\nu T} /a} e^{-2 q^2}\left( 1 - e^{- q^2 t/T} \right)^2 \\
& \hspace{-1.5em} \underset{ \nu T \gg a^2, \, \nu t}{\propto} (\nu T)^{-1/2} \left(\frac{t}{T}\right)^2 \,.
\eal
As  expected, the  contribution $\langle w_T(t) \rangle^2$ quantifies the transient dynamics of the system and decays to $0$ for large $T$ in the trap-and-release case. This decay is  slow (scaling as $T^{-5/2}$) and yields significant, long lasting corrections to the asymptotic ($T\to\infty$) stationary scaling $\propto \ln{\frac{\nu t}{a^2}}$ in the control case. This shows that the SRL is not in a stationary state at $t=0$ in the trap-and-release protocol and it quantifies  the transient dynamics  that it observed in the MSI $\Delta(t,T)$ in Fig.~\ref{fig:SRL_protocols}. This finally illustrates the long lasting environmental memory of the SRL.

\subsection{GLE formalism}

\label{app:GLEroute}

We now show how to analyze the dynamics  the trap-and-release protocols directly at the GLE level. This approach shows that an explicit Markovian embedding (as discussed in the previous section) is not required to analyze the dynamics.

To do so, we rewrite, for $t>0$, Eq.~\ref{eq:protocols} as
\al
\int_{t'=0}^t \dot{x}(t') K(t-t') = \sqrt{2D} \eta(t) - \cH(t), \label{eq:history_decomposition} 
\eal
where $\cH(t)=\int_{t'=-\infty}^0 \dot{x}(t')K(t-t')$ is the (weighted) history of the GLE, which typically decays over time. It  quantifies the memory that a GLE keeps from its environment and is of stochastic nature as it depends on the past noise realizations $\{\eta(t)\}_{t < 0}$ (and not on the future ones $\{\eta(t)\}_{t > 0}$). 

We show in the following sections how to determine $\cH(t)$ for any choice of memory kernel $K$. We then discuss  explicit examples of dynamics that are   considered in this paper.

\subsubsection{General expressions of correlators}

We discuss in this section the procedure to determine $c(T,t)$ or $\Delta(T,t)$ in the trap-and-release protocols using the GLE formalism. Here we assume for the sake of simplicity that $G(t)=\delta(t)$. For positive times $T>0,t>0$ (namely after the release of the trap) we have
\al
& \langle p(T+t)p(T) \rangle = \int_{t_1'=0}^{T+t} \int_{t_2'=0}^{T} \tilde{K}(T+t-t_1') \tilde{K}(T-t_2') \langle \dot{x}(t_1') \dot{x}(t_2') \rangle,  \\
& \langle \left( x(T+t)-x(T) \right)^2 \rangle = \int_{t_1'=T}^{T+t} \int_{t_2'=T}^{T+t} \langle \dot{x}(t_1') \dot{x}(t_2') \rangle,
\eal
where we recall that $\tilde{K}(t)=K(t)+\delta(t)$. It is thus useful to determine $\langle \dot{x}(t_1') \dot{x}(t_2') \rangle$, which we do by solving first the following linear GLE, starting at $t=t_0$
\al
\int_{t'=t_0}^t \dot{x}(t') K(t-t') = -k x(t) + \phi(t),
\eal
where $\phi$ is an unspecified forcing function (for instance some random noise). This equation can be linearly inverted by going back and forth in the Laplace domain, either via an adapted Laplace transform (starting at $t=t_0$)~\cite{diterlizzi2020explicit} or by first using a time-translation $t \to t-t_0$ in order to shift time and then to use a standard Laplace transform (starting $t=0$). In both cases we obtain
\al
\dot{x}(t) = \int_{t'=t_0}^t \phi(t') \chi(t-t') - k x(t_0) \int_{0}^{t-t_0} \chi, \label{eq:GLEsol}
\eal
where $\chi(t)$ is the response function defined as the inverse Laplace transform of 
\al
\hat{\chi}(s) = \frac{s}{k+s \hat{K}(s)} \,. \label{eq:chi_s}
\eal
The imposed conditioning $x(0)=0$ that we mentioned earlier allows to remove the second term and to avoid  biased  dynamics for $t>0$. Using Eq.~\ref{eq:GLEsol} together with the history decomposition of Eq.~\ref{eq:history_decomposition} and the independence of $\{\eta(t)\}_{t>0}$ and $\{\cH(t)\}_{t>0}$, we obtain
\al
\langle \dot{x}(t_1) \dot{x}(t_2) \rangle^{+} & = \int_{t_1',t_2'=0}^{t_1,t_2} \chi_{+}(t_1-t_1') \chi_{+}(t_2-t_2') \cH(t_1')\cH(t_2') + 2D \int_{t'=0}^{ \min(t_1,t_2) } \chi_{+}(t_1-t') \chi_{+}(t_2-t'),
\eal
where $\langle \cdot \rangle_{+}$ denotes the average over the future realizations of $\{\eta(t)\}_{t>0}$ and $\chi_{+}$ is the response function associated to the future dynamics (take $k=0$ in Eq.~\ref{eq:chi_s}). At this point, $\langle \dot{x}(t_1) \dot{x}(t_2) \rangle_+$ is still a stochastic quantity due to $\cH$ which depends on the past realizations $\{\dot{x}(t)\}_{t<0}$. We thus further average $\langle \dot{x}(t_1) \dot{x}(t_2) \rangle_+$ over the past realizations of $\{\eta(t)\}_{t<0}$ that we denote by $\langle \cdot \rangle^{-}$. We obtain
\al
\langle \dot{x}(t_1) \dot{x}(t_2) \rangle = \langle \langle \dot{x}(t_1) \dot{x}(t_2) \rangle_{+} \rangle^{-} = P(t_1,t_2) + F(t_1,t_2) \,.
\eal
Here $F$ corresponds to the contributions from the future noise realizations $\{\eta(t)\}_{t>0}$ and reads
\al
F(t_1,t_2) & = 2D \int_{t'=0}^{ \min(t_1,t_2) } \chi_{+}(t_1-t') \chi_{+}(t_2-t'), \\
& = \psi(\abs{t_1-t_2}) - 2D \int_{t' = \min(t_1,t_2) }^{ \infty } \chi_{+}(t') \chi_{+}(\abs{t_1-t_2}+t'),
\eal
where $\psi_+(t)=2D \int_{t'=0}^\infty \chi_+(t') \chi_+(t+t')$ is the stationary correlation function to which $F$ converges asymptotically. As regards $P$, which denotes the contribution from the past of the protocols ($t<0$), it reads
\al
P(t_1,t_2) & = \int_{t_1',t_2'=0}^{t_1,t_2} \chi_{+}(t_1-t_1') \chi_{+}(t_2-t_2') \langle \cH(t_1')\cH(t_2') \rangle^{-},
\eal
where the nonstationary correlator $\langle \cH(t_1)\cH(t_2) \rangle^-$ reads
\al
\langle \cH(t_1)\cH(t_2) \rangle^- & = \int_{t_1',t_2'=-\infty}^{0} K(t_1-t_1') K(t_2-t_2') \psi_-(\abs{t_1'-t_2'}) \\
& = \int_{t_1',t_2'=t_1,t_2}^{\infty,\infty} K(t_1')K(t_2') \psi_-(\abs{\abs{t_1-t_2}-t_1'+t_2'}),
\eal
and the stationary correlation function $\psi_-$ is defined as
\al
\psi_-(\abs{t_1-t_2}) & = \langle \dot{x}(t_1)\dot{x}(t_2) \rangle^- \\
& = 2D \int_{t'=-\infty}^{\min(t_1,t_2)} \chi_{-}(t_1-t') \chi_{-}(t_2-t') \\
& = 2D \int_{t'=0}^\infty \chi_{-}(\abs{t_1-t_2}+t')\chi_-(t'),
\eal
$\chi_{-}$ being the response function associated to the past dynamics (this time, $k \neq 0$ in Eq.~\ref{eq:chi_s}). Eventually the observables of interest read 
\al
& \langle p(T+t)p(T) \rangle = \int_{t_1'=0}^{T+t} \int_{t_2'=0}^{T} \tilde{K}(T+t-t_1') \tilde{K}(T-t_2') \left(  P(t_1',t_2') + F(t_1',t_2') \right), \nonumber \\
& \langle \left( x(T+t)-x(T) \right)^2 \rangle = \int_{t_1'=T}^{T+t} \int_{t_2'=T}^{T+t} F(t_1',t_2') + P(t_1',t_2') \,.  \label{eq:MSI_GLE}
\eal
These equations are the main results of this section ; they are general in the sense that they hold for any memory kernel $K$. 

Importantly, in these equations, the ``future'' contribution $F$ is completely blind to what happened in the past and thus insensitive to the protocols. It is only the ``past'' contribution $P$ that differs in  the trap-and-release and control protocols. It is worth to note that for the control protocol we have by definition
\al
P(t_1,t_2) = 2D \int_{t' = \min(t_1,t_2) }^{ \infty } \chi_{+}(t') \chi_{+}(\abs{t_1-t_2}+t') \,.
\eal
We show in the next sections how to compute explicitly $P, F$ and $\Delta$ for the different models considered in this article.

\subsubsection{MPs}

As an application of the previous formalism, we determine $P, F$ and $\Delta$ for MPs, which amounts to take $K(t)=\delta(t)+\kappa e^{-\alpha t}$ and $G(t)=\delta(t)$. We assume $T,t>0$. The two response functions $\chi_+$ ($k=0)$ and $\chi_-$ ($k\neq0$) as defined in Eq.~\ref{eq:chi_s} read 
\al
\chi_{+}(t) & = \delta(t)-\kappa e^{-(\alpha+\kappa)t}, \\
\chi_{-}(t) & = \delta(t)+e^{- K t /2} \left[ \frac{A}{\Delta} \sinh{\left(\frac{\Delta t}{2}\right)}-B \cosh{\left(\frac{\Delta t}{2}\right)} \right]
\eal
where $K=\alpha+\kappa+k$ is a generalized stiffness, $\Delta = \sqrt{\Delta - 4 \alpha k}$ a discriminant, $A=k^2-k(\alpha-2\kappa)+\kappa(\alpha+\kappa)$ and $B=k+k$. This form holds true as long as $\Delta>0$. For $\Delta<0$ (but $K>0$), this expression holds upon changing the functions $\cosh$ and $\sinh$ into $\cos$ and $\sin$. With the help of a computer algebra system, we find for the future contribution $F$
\al
& F(T+t,T) = \frac{D  e^{-2 (\alpha +\kappa ) (2 t+3 T)}}{(\alpha +\kappa )^3} \left(2 \left(\kappa  (2 \alpha +\kappa )+\alpha ^2 t (\alpha +\kappa )\right) e^{2 (\alpha +\kappa ) (2 t+3 T)} \right. \\
& \left. - \kappa ^2 e^{2 (\alpha +\kappa ) (t+2 T)} + 2 \kappa ^2 e^{(\alpha +\kappa ) (3 t+4 T)}-\kappa ^2 e^{4 (\alpha +\kappa ) (t+T)}-2 \kappa  (2 \alpha +\kappa ) e^{3 (\alpha +\kappa ) (t+2 T)}\right) , \nonumber
\eal
while for the past contributions we successively have
\al
\langle \cH(t_1)\cH(t_2) \rangle^{-} = \frac{D \kappa^2 }{\alpha + \kappa + k} e^{-\alpha (t_1+t_2)}, \quad P(t_1,t_2) = \frac{D \kappa^2 }{\alpha + \kappa + k} e^{-\alpha (t_1+t_2)},
\eal
where both expressions are found to explicitly depends on the confining potential via $k$, as it is expected out of equilibrium. These two last expressions also hold for the control protocol, upon imposing $k=0$.

Using Eq.~\ref{eq:MSI_GLE}, it is then possible to determine the MSI. For the control protocol, it reads
\al
\Delta(T,t) & = \frac{2 D} {(\alpha + \kappa)^3} \left(\kappa(2\alpha + \kappa) + \alpha^2  t (\alpha + \kappa) - \kappa (2\alpha + \kappa) e^{-t(\alpha + \kappa)} \right) \\
& = 2Dt + 2 v^2 \tau_m \left( \frac{t}{\tau_m} -1 + e^{-t/\tau_m} \right),
\eal
where $v^2=- D \kappa (2\alpha+\kappa)/(\alpha+\kappa)^2$ and $\tau_m=(\alpha+\kappa)^{-1}$ as in Eq.~\ref{eq:MSI_MP}. This expression does not depend on $T$ as expected from the stationarity of the process. In the testing protocols however, the MSI read
\al
& \Delta(T,t) = D \tau_m^{3} \Bigg( \frac{\kappa^2 \tau_m^{-1} \left(e^{t/\tau_m} - 1 \right)^2 e^{-2(t+T)/\tau_m}}{\tau_m^{-1}+k} + 2\alpha^2 t/\tau_m \nonumber \\
& \quad \quad \quad \quad \quad \quad \quad \quad \quad  + \kappa \Bigg[ 4\alpha\left(1 - e^{-t/\tau_m} \right) + \kappa \left(-2 e^{-t/\tau_m} - e^{-2(t + T)/\tau_m} + 2e^{-(t + 2 T)/\tau_m} -e^{-2 T/\tau_m} + 2 \right) \Bigg] \Bigg),
\eal
where we recall $\tau_m=(\alpha+\kappa)^{-1}$. We now see that $\Delta(T,t)$ depends on $T$, at least up to a typical time scale $\tau_m = (\alpha+\kappa)^{-1}$, and quickly collapses on the MSI of the control protocol for  $t>\tau_m$, as it can be observed in Fig.~$1$ of the main text. This non stationary behaviour of the MSI in the trap-and-release protocol is a direct illustration of environmental memory  in the dynamics.

\subsubsection{AOUPs}

We already mentioned that AOUPs (whether persistent or antipersistent) can be cast under the GLE formalism upon choosing $K(t)=\delta(t)$ and $G$ a suitable exponential memory kernel. This means that these particles do not feature any environmental memory, have a trivial history $\cH \equiv 0$ and thus are insensitive to the testing protocols (see Eq.~\ref{eq:protocols}).

\section{Static interaction potential}

\label{app:interactions}

This Appendix is devoted to the analytical study of MPs (and their equivalent models in free space) in the presence of a static interaction potential. We mainly make use  of Fokker–Planck equations, and discuss how to integrate over   the internal degrees of freedom to derive the stationary distribution of the position $x$ of the particle.  We first discuss this procedure in the simple case of an external harmonic potential, where all results can be obtained exactly, in particular the stationary distribution and the   two-time correlation functions, which turn out to capture the main difference between MPs (environmental memory) and AOUPs (correlated noise). We then address the case of nonharmonic interaction potentials using perturbative methods~; we find that while MPs and AOUPs have   qualitatively comparable stationary distributions of the position  at first order in perturbation, their analytical forms quantitatively differ.

\subsection{Marginal Fokker–Planck equation}

\label{app:conditionalproba}

As discussed in the main text, the stationary distribution of position $x$ is a key observable for characterizing the behavior of non-Markovian particles—such as MPs, AOUPs, or equilibrium dimers—in a given environment. To obtain such distributions, we make use of  multivariate Fokker–Planck equations and marginalize over the hidden or internal degrees of freedom (i.e., $z$, $\xi$, or $Z$), as explained below.

The Langevin equation describing the motion of an MP in a potential $U$, Eq.~7, admits the following Fokker–Planck representation for the joint probability distribution $p(x,z)$:
\al
\p_t p + \bG \cdot \bJ = 0, \quad \bJ = (J_x, J_z),
\eal
where
\al
\begin{cases}
	J_x = - \kappa (x-z) p - U'(x) p - D \, \p_x p,\\
	J_z = \alpha (x-z) p \,.
\end{cases}
\eal
Obtaining the stationary distribution $p(x)$ amounts to: $(i)$ solving $\p_x J_x + \p_z J_z = 0$, and $(ii)$ integrating out the hidden variable $z$ over $\mathbb{R}$. To do so, we condition the joint probability $p(x,z)$ on the realization of $x$,
\al
p(x,z) = p(z|x) p(x),
\eal
and marginalize the Fokker–Planck equation over $z \in \mathbb{R}$ to obtain
\al
0 = - \p_x \left( \Big[ -U'(x) + \mu^{\rm \scriptscriptstyle MP}(x) \Big] p(x) \right) + D \p_x^2 p(x), \label{eq:marginalizedFP_MP}
\eal
where 
\al
\mu^{\rm \scriptscriptstyle MP}(x) = - \kappa \int_{z=-\infty}^\infty (x-z) p(z|x),
\eal
is the mean force felt by $x$ due to its coupling with $z$. Here we implicitly assume that $p(x,z)$ vanishes at the boundaries $z \to \pm \infty$. Equation~\ref{eq:marginalizedFP_MP} admits the stationary solution
\al
p(x) \propto \exp\left\{ -\frac{1}{D} \left( U(x) - \int^x \mu^{\rm \scriptscriptstyle MP} \right) \right\}.
\eal
As we will see, $\mu^{\rm \scriptscriptstyle MP}$ does not necessarily derive from a potential, so the resulting stationary measure departs from the equilibrium Boltzmann distribution—an expected outcome given the lack of detailed balance. 

The above reasoning applies \emph{mutatis mutandis} to AOUPs (persistent or antipersistent) and to equilibrium dimers. The stationary distribution again reads
\al
p(x) \propto \exp\left\{ -\frac{1}{D} \left( U(x) - \int^x \mu \right) \right\}, \label{eq:reducedpstat}
\eal
where the mean self-propulsion force $\mu$ depends on the model:
\al
\mu^{\rm \scriptscriptstyle pAOUP}(X) = v \int_{\xi=-\infty}^\infty \xi \, p(\xi|X),
\eal
for persistent AOUPs,
\al
\mu^{\rm \scriptscriptstyle apAOUP}(X) = - v \int_{\xi=-\infty}^\infty \xi \, p(\xi|X),
\eal
for antipersistent AOUPs, and
\al
\mu^{\rm \scriptscriptstyle eq. dim.}(X) = - \kappa' \int_{Z=-\infty}^\infty (X-Z) \, p(Z|X),
\eal
for equilibrium dimers. The latter case is illustrative, since the joint stationary measure is explicitly known because it is an equilibrium measure:
\al
p(X,Z) \propto \exp\left\{ - \frac{1}{D} \left[ \frac{\kappa'}{2} (X-Z)^2 + U(X) \right] \right\} \,.
\eal
It implies that
\al
p(Z|X) \propto \exp\left\{ - \frac{\kappa' (X-Z)^2}{2D} \right\},
\eal
which means $\mu^{\rm \scriptscriptstyle eq. dim.} = 0$, since the integrand is an even function of $(X-Z)$. We thus find back a previous result for this equilibrium model: $p(Z|X)$ is independent of the potential $U$. This is not the case for the other nonequilibrium models, which we analyze in the following sections. This constitutes an important difference between equilibrium versus nonequilibrium memory.

\subsection{Harmonic potential}

\label{app:OU}

Before addressing interactions in a generic potential, it is useful to consider the harmonic case: although not always representative of generic potentials, it is analytically tractable and provides valuable intuition. Here it turns out that it captures the main qualitative difference between MPs and AOUPs, which lies in the two-time correlation function rather than in the stationary probability distribution. We first focus on the stationary distribution of MPs (comparing it to their control models) before analyzing their two-time correlation functions.

\subsubsection{Stationary probability distribution}

The MPs defined in Eq.~7 with a harmonic potential $U(x) = k x^2 / 2$ can be rewritten as a two-dimensional Ornstein–Uhlenbeck stochastic process:
\al
\dot{\bY} = - \boldsymbol{\beta} \bY + \boldsymbol{\sigma}, \quad \bY = \begin{pmatrix} x \\ z \end{pmatrix}, \quad \boldsymbol{\beta} = \begin{pmatrix} \kappa + k & -\kappa \\ -\alpha & \alpha \end{pmatrix}, \quad \boldsymbol{\sigma} = \begin{pmatrix} \sqrt{2D} & 0 \\ 0 & 0 \end{pmatrix} \label{eq:2dOU}.
\eal
To ensure the stability of the dimer, we assume $\kappa>-(\alpha+k)$.\footnote{This condition arises from imposing the positivity of the real parts of the eigenvalues via the positivity of the trace and determinant of $\boldsymbol{\beta}$.} The stationary measure of the Ornstein-Uhlenbeck process is known to read~\cite{gardiner2004handbook} 
\al
p(\bY) = \frac{1}{2\pi \sqrt{\det \boldsymbol{\omega}}} \exp\left( -\frac{1}{2} \bY^{\rm T} \boldsymbol{\omega}^{-1} \bY \right),
\eal
where $\boldsymbol{\omega}$ satisfies the Lyapunov equation $\boldsymbol{\beta} \boldsymbol{\omega} + \boldsymbol{\omega} \boldsymbol{\beta}^{\rm T} = 2 \mathbf{D}$, with $\mathbf{D} = \boldsymbol{\sigma} \boldsymbol{\sigma}^{\rm T}/2$. We find
\al
p(x,z) \propto \exp\left\{ -\frac{(\alpha + \kappa + k)\left[ k z^2 + \alpha (x - z)^2 \right]}{2 \alpha D} \right\}, \label{eq:jointprobaOU}
\eal
which, after marginalization over $z$, yields
\al
p(x) \propto \exp\left\{ -\frac{U(x)}{D_{\rm \scriptscriptstyle MP}} \right\}, \quad 
\frac{D_{\rm \scriptscriptstyle MP}}{D} = \frac{k + \alpha}{k + \alpha + \kappa}. \label{eq:pstat_MP_harmonic}
\eal
The stationary distribution is, by definition, Gaussian just as for harmonically trapped persistent (resp. antipersistent) AOUPs defined in Eq.~\ref{eq:AOUP} (resp. Eq.~\ref{eq:antipersistentOUP}). However, although these models were indistinguishable in the absence of interactions, the effective diffusion coefficients now differ between the models (see Table~\ref{tab:effectivediffusion})~\footnote{It is worth to note that due to the close functional form of the effective diffusion coefficients, the different models behaves similarly in the different parameter-limiting regimes.}. This shows that, within an harmonic trap, there is a quantitative difference between dynamics driven by memory with respect to ones driven by correlated noise. For the equilibrium dimer, there is  a qualitative difference with respect to antipersistent AOUPs as it is invariant under time-reversal symmetry~; its stationary distribution is thus the expected Boltzmann distribution of a thermal Brownian particle in a trap ($D_{\rm eq. dim.}=D$).

\renewcommand{\arraystretch}{4.}

\begin{table}[h!]
    \centering
    \caption{ Effective diffusion coefficients for (anti)persistent MPs and AOUPs in terms of the ($v,\tau$)-variables of the AOUPs. The last row corresponds to the effective diffusion coefficients for antipersistent MPs expressed in terms of the parameters of the matched equilibrium dimer.}
    \begin{tabular}{|c|c|c|}
        \hhline{~|-|-|}  
        \multicolumn{1}{c|}{} & MP & AOUP \\ \hline
        \centering persistent & \(\displaystyle \frac{D_{\rm pMP}}{D} = \frac{k \tau + \sqrt{1 + v^2/D}}{ k \tau + 1} \) & \(\displaystyle \frac{D_{\rm pAOUP}}{D} = \frac{ k \tau + 1 + v^2/D }{ k \tau + 1 } \) \\ \hline
        antipersistent & \(\displaystyle \frac{D_{\rm apMP}}{D} = \frac{ \pm \left( k \tau - 1\right) + v/\sqrt{D}}{ k \tau + 1} \) & \(\displaystyle \frac{D_{\rm apAOUP}}{D} = \frac{ k \tau + (- 1 + v/\sqrt{D})^2 }{ k \tau + 1 } \) \\ \hline
        equilibrium dimer parametrization & \(\displaystyle \frac{D_{\rm apMP}}{D} = \frac{k + \kappa' \sqrt{r(1+r)}}{ k + \kappa' (1+r) } \) & \multicolumn{1}{c}{} \\
        \hhline{|-|-|~}  
    \end{tabular}
    \label{tab:effectivediffusion}
\end{table}

\renewcommand{\arraystretch}{1.}

If we come back to $D_{\rm \scriptscriptstyle MP}$, we see that when $\kappa$ vanishes, $x$ effectively behaves as an equilibrium particle in the potential $U(x)$ as expected. In turn, for $\kappa$ large, the stationary distribution can significantly differ from the equilibrium case, and be either broader (for pMPs) or narrower (for apMPs). The same behavior happens in the limit of a slow dynamics of the internal variable $z$ ($\alpha \to 0$), while the equilibrium case is recovered in the limit of a fast relaxation of the internal state ($\alpha \to \infty$).

\subsubsection{Two-time correlation function}

\begin{figure}[ht]
	\includegraphics[width=\textwidth,clip]{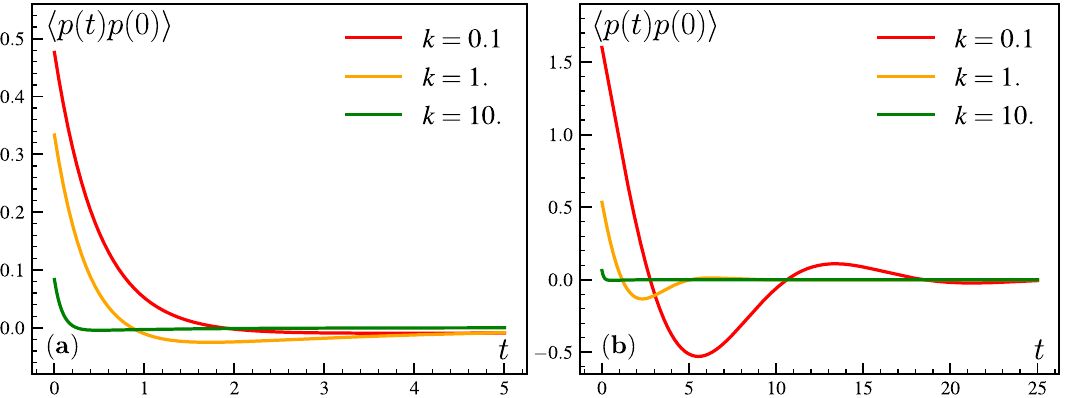}
	\caption{ $\mathbf{(a)}$ Stationary polarity correlation function $c(0,t)=\langle p(t)p(0) \rangle$ for an harmonically confined MP in the weakly self-propelled regime ($\Delta>0$). Curves corresponds to different trap stiffnesses $k$ (i.e. trap ``sizes''). For finite $k$, $c(0,t)$ exhibits an asymptotic exponential decay but is not strictly monotonic: initial positive correlations are followed by anticorrelations. Note the semi-quantitative agreement with the numerical hard box results of Fig.~$2$ in the main text. The anticorrelations likely originate from successive ``trap-climbing'' events (or wall collisions in the hard-box case), illustrating that MPs retain memory of past interactions with the confining potential. In the large-trap limit ($k\to 0$), $c(0,t)$ approaches a pure exponential decay, as in AOUPs. In the strong-confinement limit ($k\gg \alpha+\kappa$), $c(0,t)$ vanishes uniformly, showing that confinement suppresses the polarity of self-propulsion, consistent with our hard-box simulations. $\mathbf{(b)}$ Same as in $\mathbf{(a)}$, but for the strongly self-propelled regime ($\Delta<0$). The overall phenomenology remains similar, with the addition of oscillations arising from the complex eigenvalues of the nonreciprocal matrix $\boldsymbol{\beta}$. These oscillations reflect the fact that fast MPs can retain memory of multiple successive trap-climbing events, which restores positive correlations after two such events.}
	\label{fig:shape_pp}
\end{figure}

We now turn to the determination of the correlation functions of MPs (in fact of the polarity) since it is where the main difference between MPs and AOUPs lies. As we already mentioned, the polarity of AOUPs is independent of the environment and its Gaussian correlations in the stationary state read
\al
\langle p(t)p(0) \rangle = \frac{v^2}{\tau} e^{-\abs{t}/\tau} \,.
\eal
For MPs however, $p$ now explicitly depends on $U(x)=k x^2/2$. For a vectorial Ornstein-Uhlenbeck process as the MPs written into the form of Eq.~\ref{eq:2dOU}, its gaussian correlation in the stationary state are known to read~\cite{gardiner2004handbook} 
\al
\langle \bY(t)\bY(0) \rangle = \boldsymbol{\omega} e^{- \boldsymbol{\beta}^T t}, \ \text{for} \ t >0,
\eal
where $\bY, \boldsymbol{\omega}$ and $\boldsymbol{\beta}$ were introduced in the previous section. The correlations for the polarity $p$ are then obtained as
\al
\langle p(t)p(0) \rangle & = \begin{pmatrix} -\kappa \\ \kappa \end{pmatrix}^T \cdot \langle \bY(t)\bY(0) \rangle \cdot \begin{pmatrix} -\kappa \\ \kappa \end{pmatrix} \\
& = \frac{D \kappa^2}{k+\alpha+\kappa} \left( \left( e^{- \boldsymbol{\beta}^T t} \right)_{11} - \left( e^{- \boldsymbol{\beta}^T t} \right)_{12} \right) \,.
\eal
Using the trace-determinant form of the exponential of two-dimensional matrices, we find
\al
e^{- \boldsymbol{\beta}^T t} = e^{-Kt/2} \left( \cosh{\left(\frac{\Delta t}{2}\right)} \mathbf{I}_2 - \frac{2}{\Delta} \sinh{\left(\frac{\Delta t}{2}\right)} \left[ \boldsymbol{\beta}^T - \frac{K}{2} \mathbf{I}_2 \right]  \right), \label{eq:exp_matr}
\eal
where $K=\alpha+\kappa+k$ is a generalized stiffness and $\Delta=\sqrt{ K^2 - 4 \alpha k}$ a discriminant. This form holds true as long as $\Delta>0$. In the strongly propelled regime 
\al
\Delta<0 \iff - (k^{1/2} - \alpha^{1/2}) > \kappa > -(k + \alpha)
\eal
where the dynamics is still asymptotically stable but where $\boldsymbol{\beta}$ display complex eigenvalues (remember the $(x,z)$-dimer is nonreciprocal), Eq.~\ref{eq:exp_matr} still holds upon changing the functions $\cosh$ and $\sinh$ into $\cos$ and $\sin$~\footnote{We do not discuss the specific case $\Delta=0$ since it corresponds to a fine-tuning of the spring stiffness $k$ with respect to the dynamical parameters $\alpha$ and $\kappa$.}. The correlator of polarity eventually reads
\al
\langle p(t)p(0) \rangle & = \frac{D \kappa^2}{k+\alpha+\kappa} e^{-K t/2} \left( \cosh{\left(\frac{\Delta t}{2}\right)} + \frac{K}{\Delta} \sinh{\left(\frac{\Delta t}{2}\right)} \,. \right) \label{eq:correl_pp}
\eal
The correlator now depends on the stiffness of the potential $k$. While expected, this environmental dependency of the polarity for MPs is important because it qualitatively new compared to models of particles driven by correlated noise (AOUPs etc.). Its shape is plotted for different parameters values in Fig.~\ref{fig:shape_pp} but it is exponentially decaying to $0$ at large times and its maximal value is 
\al
\langle p(0)^2 \rangle = \frac{D \kappa^2}{k+\alpha+\kappa}, \label{eq:correl_pp_0}
\eal
as found in the previous section. Importantly, this implies that the polarity uniformly vanishes if the confinement is strong enough, namely if $k \gg \alpha+\kappa = \tau_m^{-1}$.

This harmonic case was important to treat as it allowed us to access to the correlation functions of the dynamics of MPs in presence of interaction, which is a notoriously difficult task for generic interactions, even at equilibrium~\footnote{Except of course if the correlation functions are independent from $U$ as it is the case for the polarity of AOUPs.}. It is in these correlation functions that the main qualitative difference between AOUPs and MPs is hidden. We anticipate that this conclusion  holds qualitatively, thereby illustrating the fundamental difference between MPs and  AOUPs.  

\subsection{General perturbation theory and hierarchy of moments}

\label{app:hierarchies}

We now consider a  generic interaction potential $U$. Exact results are not accessible in the general case, and we introduce  perturbative methods. To do so, we follow \cite{maitra2020enhanced} and derive the multivariate  Fokker-Planck equation describing the dynamics of the joint distribution of $x,z$. Integration over the internal variable $z$  yields an infinite hierarchy of equations, which we close perturbatively in the next section.

\subsubsection{MPs}

We derive two different hierarchies of moments for MPs (as defined in Eq.~\ref{eq:MP}), each suited to a specific perturbative limit: small memory time $\tau_m=(\alpha+\kappa)^{-1}$ or small polarity (self-propulsion)  amplitude $\kappa$, which we analyze below. \\

\paragraph{Hierarchy in the $\{x,\delta\}$-space.}

The first hierarchy is obtained from the Fokker–Planck representation in terms of the variables $\{x, \delta = x - z\}$ of Eq.~\ref{eq:MP}, which reads
\al
\p_t W(x,\delta,t) & = \CL_{\rm MP}^1 W(x,\delta,t), \label{eq:FPMP1}
\eal 
where
\al 
\CL_{\rm MP}^1 & = \p_x \left( U'(x) \bullet \right) + D \p^2_x \bullet + \kappa \, \delta \, \p_x \bullet + (\kappa+\alpha) \p_\delta (\delta \bullet) + U'(x) \p_\delta \bullet + 2 D \p_x \p_\delta \bullet + D \p_\delta^2 \bullet \,. \nonumber
\eal
We seek the stationary distribution of $W$ marginalized over $\delta$, namely $W(x,t) = \int_{\delta=-\infty}^\infty W(x,\delta,t)$. Marginalizing Eq.~\ref{eq:FPMP1} over $\delta$ leads, after introducing the reduced moments
\al
m_k(x,t) = \int_{\delta=-\infty}^\infty \delta^k W(x,\delta,t),
\eal
to the following hierarchy:
\al
\p_t m_0 & = \p_x \left( U'(x) m_0 \right) + D \p^2_x m_0 + \kappa \p_x m_1, \label{eq:m_0_MP_perturbativeexp_smalltime} \\
\p_t m_1 & = \p_x \left( U'(x) m_1 \right) + D \p^2_x m_1 + \kappa \p_x m_2 - (\kappa+\alpha) m_1 - U'(x) m_0 \label{eq:m_1} \quad - 2D \p_x m_0, \\
\p_t m_2 & = \p_x \left( U'(x) m_2 \right) + D \p^2_x m_2 + \kappa \p_x m_3 - 2(\kappa+\alpha) m_2 - 2U'(x) m_1 - 4D \p_x m_1 + 2D m_0, \\
\p_t m_k & = \p_x \left( U'(x) m_k \right) + D \p^2_x m_k + \kappa \p_x m_{k+1} - k(\kappa+\alpha) m_k - kU'(x) m_{k-1} \label{eq:m_k_MP_perturbativeexp_smalltime} - 2kD \p_x m_{k-1} + k(k-1) D m_{k-2},
\eal
for any $k \geq 2$. \\

\paragraph{Hierarchy in the $\{x,z\}$-space.}

The second hierarchy is derived from the Fokker–Planck representation in terms of $\{x,z\}$:
\al
\p_t W(x,z,t) & = \CL_{\rm MP}^2 W(x,z,t), \label{eq:FPMP2}
\eal 
where
\al 
\CL_{\rm MP}^2 & = \p_x \left\{ U'(x) \bullet \right\} + D \p^2_x \bullet + \kappa \p_x \left\{ (x-z) \bullet \right\} - \alpha \p_z \left\{ (x-z) \bullet \right\}. \nonumber
\eal
Introducing the reduced moments
\al
m_k(x,t) = \int_{z=-\infty}^\infty z^k W(x,z,t),
\eal
we obtain the hierarchy
\al
\p_t m_0 & = \p_x \left( U'(x) m_0 \right) + D \p^2_x m_0 + \kappa \p_x (x m_0 - m_1), \label{eq:MP_perturbativeexp_smallamp_1} \\
\p_t m_1 & = \p_x \left( U'(x) m_1 \right) + D \p^2_x m_1 + \kappa \p_x (x m_1 - m_2) + \alpha (x m_0 - m_1), \\
\p_t m_2 & = \p_x \left( U'(x) m_2 \right) + D \p^2_x m_2 + \kappa \p_x (x m_2 - m_3) + 2\alpha (x m_1 - m_2), \\
\p_t m_k & = \p_x \left( U'(x) m_k \right) + D \p^2_x m_k + \kappa \p_x (x m_k - m_{k+1}) + k\alpha (x m_{k-1} - m_k), \label{eq:MP_perturbativeexp_smallamp_k}
\eal
for any $k \geq 2$.

\subsubsection{Persistent AOUPs}

We now derive the hierarchy of moments for thermal and persistent AOUPs. The Fokker–Planck representation of Eq.~\ref{eq:AOUP} reads
\al
\p_t W(X,\xi,t) = \CL_{\rm pAOUP} W(X,\xi,t), \label{eq:FP_pAOUP}
\eal
where
\al
\CL_{\rm pAOUP} = \p_X \left\{ U'(X) \bullet \right\} + D \p^2_X \bullet - v \xi \p_X \bullet + \frac{1}{\tau} \p_\xi (\xi \bullet) + \frac{1}{\tau^2} \p_\xi^2 \bullet. \nonumber
\eal
Defining the reduced moments as
\al
M_k(X,t) = \int_{\xi=-\infty}^\infty \xi^k W(X,\xi,t),
\eal
we obtain the hierarchy
\al
\p_t M_0 & = \p_X \left( U'(X) M_0 \right) + D \p^2_X M_0 - v \p_X M_1, \label{eq:moment0_pAOUP}\\
\p_t M_1 & = \p_X \left( U'(X) M_1 \right) + D \p^2_X M_1 - v \p_X M_2 - \frac{1}{\tau} M_1, \\
\p_t M_2 & = \p_X \left( U'(X) M_2 \right) + D \p^2_X M_2 - v \p_X M_3 - \frac{2}{\tau} M_2 + \frac{2}{\tau^2} M_0, \\
\p_t M_k & = \p_X \left( U'(X) M_k \right) + D \p^2_X M_k - v \p_X M_{k+1} - \frac{k}{\tau} M_k + \frac{k(k-1)}{\tau^2} M_{k-2}, \label{eq:momentk_pAOUP}
\eal
for any $k \geq 2$.

\subsubsection{Antipersistent AOUPs}

Last, we conclude this section, by deriving the moments hierarchy for antipersistent AOUPs (as defined in Eq.~\ref{eq:antipersistentOUP}). The Fokker–Planck equation reads
\al
\p_t W(X,\xi,t) = \CL_{\rm apAOUP} W(X,\xi,t),
\eal
where
\al
\CL_{\rm apAOUP} = \p_X \left\{ U'(X) \bullet \right\} + D \p^2_X \bullet + v \xi \p_X \bullet + \frac{1}{\tau} \p_\xi (\xi \bullet) + \frac{2\sqrt{D}}{\tau} \p_X \p_\xi \bullet + \frac{1}{\tau^2} \p_\xi^2 \bullet. \nonumber
\eal
This Fokker-Planck operator differs compared to the one for persistent AOUPs in two respects which echoes our comments in the Langevin representation of AOUPs in App.~\ref{app:apAOUP}. First, it involves the cross-derivatives $\p_X \p_\xi$ representing noise cross-correlations between $X$ and $\xi$. Second, $\CL_{\rm apAOUP}$ is no longer invariant under the symmetry $(v,\xi) \to -(v,\xi)$, contrary to $\CL_{\rm pAOUP}$. This has notable consequences in presence of interactions as we discuss in the next section. Defining again the reduced moments as
\al
M_k(X,t) = \int_{\xi=-\infty}^\infty \xi^k W(X,\xi,t) ,
\eal
we find the hierarchy
\al
\p_t M_0 & = \p_X \left( U'(X) M_0 \right) + D \p^2_X M_0 + v \p_X M_1, \label{eq:moment0_apAOUP} \\
\p_t M_1 & = \p_X \left( U'(X) M_1 \right) + D \p^2_X M_1 + v \p_X M_2 - \frac{1}{\tau} M_1 - \frac{2\sqrt{D}}{\tau} \p_X M_0, \\
\p_t M_2 & = \p_X \left( U'(X) M_2 \right) + D \p^2_X M_2 + v \p_X M_3 - \frac{2}{\tau} M_2 - \frac{4\sqrt{D}}{\tau} \p_X M_1 + \frac{2}{\tau^2} M_0, \\
\p_t M_k & = \p_X \left( U'(X) M_k \right) + D \p^2_X M_k + v \p_X M_{k+1} - \frac{k}{\tau} M_k - \frac{2k\sqrt{D}}{\tau} \p_X M_{k-1} \label{eq:momentk_apAOUP} + \frac{k(k-1)}{\tau^2} M_{k-2},
\eal
for any $k \geq 2$. The two differences with respect to persistent AOUPs we noted earlier in this section at the level of the Fokker-Planck operator are found back at the level of the moments: first the presence of the cross-correlated term $-2k\sqrt{D}/\tau \p_X M_{k-1}$ and second the lack of the symmetry $(v,\{M_k\}_{k \in \mathbb{N}}) \to (-v,\{ (-1)^k M_k \}_{k \in \mathbb{N}} )$. \\

We now aim at solving these different infinite hierarchies of moments, which we do in the next section.

\subsection{Perturbation theory for small memory time}

\label{app:smalltimeperturb}

Since there is no obvious way to exactly close the hierarchies derived for MPs and AOUPs in the previous section, we now seek for an approximate closure. To this end, we develop a perturbative approach in the limit of small memory time $\tau_m$, as already done in different contexts, see~\cite{martin2021statistical} and  references therein. Given that we work at the level of marginalized Fokker-Planck equations, we rather follow here~\cite{maitra2020enhanced}.

\subsubsection{MPs}

We first close the hierarchy of moments for MPs given in Eqs.~\ref{eq:m_0_MP_perturbativeexp_smalltime}–\ref{eq:m_k_MP_perturbativeexp_smalltime} by considering the limit of small memory time. This limit is achieved by rescaling $\tau_m \to \varepsilon \tau_m$ (or equivalently $\alpha, \kappa \to \varepsilon^{-1} \alpha, \varepsilon^{-1} \kappa$) and taking $\varepsilon \to 0$.

We assume that the hierarchy can be truncated at a finite order $N$, such that $m_n \equiv 0$ for $n > N$. The first $N$ moments are then expanded as power series in $\varepsilon$:
\al
m_n(x,t) = \sum_{l=0}^\infty \varepsilon^l m_n^l(x,t), \quad \forall\, n < N,
\eal
and we solve for each $m_n^l(x,t)$ order by order in $\varepsilon$, using the stationary version of Eqs.~\ref{eq:m_0_MP_perturbativeexp_smalltime}–\ref{eq:m_k_MP_perturbativeexp_smalltime}. We start from the highest nonvanishing moment and proceed recursively to lower moments, ultimately obtaining $m_0(x)$, the key observable of interest.  

This approximation scheme is self-consistent: to reach a given precision of order $\varepsilon^l$ for $m_0$, there exists a finite $N$ such that contributions from higher-order moments only appear at $\mathcal{O}(\varepsilon^{l+1})$ or beyond.

In practice, using a computer algebra system, we solve the hierarchy up to order $\mathcal{O}(\varepsilon)$, which requires truncating at $N=5$. We find at order $\mathcal{O}(\varepsilon^0)$
\al
m_{0}^0(x) = \frac{\cB(x)}{\cZ}, \quad \cB(x) = e^{-U(x)/D_{\rm \scriptscriptstyle MP} }, \quad \cZ = \int_{\mathbb{R}} e^{-U/D_{\rm \scriptscriptstyle MP} }, \label{eq:order0_MP}
\eal
where $D_{\rm \scriptscriptstyle MP} = D / (1+\chi)$ and  $\chi = \kappa / \alpha$ as defined in App.~\ref{app:pAOUP}. At order $\mathcal{O}(\varepsilon^1)$, we find
\al
m_{0}^1(x) = m_{0}^0(x) \left( \frac{\chi}{\alpha(1+\chi)} \left[ \frac{U'^2(x)}{2 D}(1+\chi) - U''(x) \right] - \cN \right), \label{eq:order1_MP}
\eal
where
\al
\cN = \int_{x'=-\infty}^\infty m_{0}^0(x') \frac{\chi}{\alpha(1+\chi)} \left[ \frac{U'^2(x')}{2 D}(1+\chi) - U''(x') \right] \,.
\eal
The stationary distribution for MPs up to order $\mathcal{O}(\varepsilon)$ can then be written in exponential form as
\al
m_{0}(x) \propto \exp\!\left\{ -\frac{U(x) + \varepsilon \, U_1(x) }{D_{\rm \scriptscriptstyle MP}} \right\} + \mathcal{O}(\varepsilon^2) , \label{eq:pstat_MP_smalltime}
\eal
where the first perturbative correction reads
\al
U_1(x) = - \frac{\chi}{\alpha(1+\chi)} \left( \frac{U'^2(x)}{2} - D_{\rm \scriptscriptstyle MP} U''(x) \right) \,.
\eal
It is the main result of this section. It can be checked that Eqs.~\ref{eq:order0_MP},\ref{eq:order1_MP} agree with the exact harmonic results of Eq.~\ref{eq:pstat_MP_harmonic} expanded up to order $\mathcal{O}(\tau_m^2)$. We keep further discussions about this result for App.~\ref{app:perturb_comments} so as to compare them in the light of the stationary distributions for AOUPs we obtain in the next two sections.

\subsubsection{Persistent AOUPs}

We now apply the same perturbative expansion to persistent AOUPs. However, this case is less straightforward because the perturbation is singular: the self-propulsion variable $\xi$ diverges as $\tau \to 0$, as it can be observed on the two-time correlator
\al
\langle \xi(t_1)\xi(t_2) \rangle = \frac{1}{\tau} e^{-\abs{t_1-t_2}/\tau},
\eal
recalled from Eq.~\ref{eq:amplixi}. To regularize this divergence, we rescale $\xi \to \tau^{1/2}\xi$ and $t \to \tau t $, following standard treatments~\footnote{See e.g.~\cite{martin2021statistical} and the references inside.}. This rescaling modifies the Fokker–Planck operator of Eq.~\ref{eq:FP_pAOUP} into
\al
\CL_{\rm pAOUP} = \p_x \left\{ U'(x) \bullet \right\} + D \p^2_x \bullet - \frac{v \xi}{\tau^{1/2}} \p_x \bullet + \frac{1}{\tau} \p_\xi (\xi \bullet) + \frac{1}{\tau} \p_\xi^2 \bullet.
\eal
The corresponding hierarchy of moments can be obtained from Eqs.~\ref{eq:moment0_pAOUP}–\ref{eq:momentk_pAOUP} by replacing $\xi$ with $\xi/\tau^{1/2}$. We recall that $\CL_{\rm pAOUP}$ is still invariant under the symmetry $(v,\xi) \to -(v,\xi)$.

We now perform a perturbative expansion analogous to that used for MPs. To avoid fractional powers of $\varepsilon$, we rescale $\tau \to \varepsilon^2 \tau$ and use the ansatz
\al
M_n(x,t) = \sum_{l=0}^\infty \varepsilon^l M_n^l(x,t), \quad \forall\, n < N.
\eal
Truncating the hierarchy at order $N=5$, we find at order $\cO(\varepsilon^0)$
\al
M_0^0(X) = \frac{\cB(X)}{\cZ}, \quad \cB(X) = e^{-U(X)/D_{\rm \scriptscriptstyle AOUP}^p}, \quad \cZ = \int_\mathbb{R} e^{-U/D_{\rm \scriptscriptstyle AOUP}^p}, \label{eq:order0_pAOUP}
\eal
where $D_{\rm \scriptscriptstyle AOUP}^p = D(1+\zeta_p)$ and $\zeta_p = v^2/D$ as defined in App.~\ref{app:pAOUP}. The first order correction in $\cO(\varepsilon^1)$ vanishes due to the $(v,\xi) \to -(v,\xi)$ symmetry we already mentioned. The first nonvanishing perturbative correction is thus of order $\cO(\varepsilon^2)$ and reads
\al
M_0^2(X) = M_0^0(X) \left( \frac{ \tau \zeta_p }{ 1+\zeta_p } \left[ U''(X) - \frac{U'(X)^2}{ 2 D(1+\zeta_p) } \right] - \cN \right), \label{eq:order2_pAOUP}
\eal
where
\al
\cN = \int_{X'=-\infty}^\infty M_0^0(X') \frac{ \tau \zeta_p }{ 1+\zeta_p } \left[ U''(X) - \frac{U'(X)^2}{ 2 D(1+\zeta_p) } \right] \,.
\eal
The stationary distribution up to order $\mathcal{O}(\varepsilon^2)$ is thus given by
\al
M_0(X) = \exp\!\left\{ - \frac{U(X) + \varepsilon^2 U_2(X) }{D_{\rm \scriptscriptstyle AOUP}^p} \right\} + \mathcal{O}(\varepsilon^3), \label{eq:perturbative_M0_pAOUP}
\eal
where the first nonvanishing perturbative correction reads
\al
U_2(X) = \frac{\tau \zeta_p}{1+\zeta_p} \left( \frac{U'(X)^2}{2} - D_{\rm \scriptscriptstyle AOUP}^p U''(X) \right) \,.
\eal
It is the main result of this section. These perturbative results for persistent AOUPs, notably Eqs.~\ref{eq:order0_pAOUP},\ref{eq:order2_pAOUP}, yield back the exact results obtained for an harmonic trap in App.~\ref{app:OU} and are in agreement with results found in the literature~\cite{martin2021statistical,martin2021AOUP}. It allows us to comment in a fair way the results we obtained for MPs, as we do in App.~\ref{app:perturb_comments}.

\subsubsection{Antipersistent AOUPs}

Finally, we perturbatively derive the stationary distribution for antipersistent AOUPs to enable direct comparison with antipersistent MPs. As for the case of persistent AOUPs, the expansion is singular in the $\tau \to 0$ limit and we rescale again $\xi \to \tau^{1/2}\xi$. The modified stationary Fokker–Planck operator reads
\al
\CL_{\rm apAOUP} = \p_x \left\{ U'(x) \bullet \right\} + D \p^2_x \bullet + \frac{v \xi}{\tau^{1/2}} \p_x \bullet + \frac{1}{\tau} \p_\xi (\xi \bullet) + \frac{2\sqrt{D}}{\tau^{1/2}} \p_x \p_\xi \bullet + \frac{1}{\tau} \p_\xi^2 \bullet. \nonumber
\eal
To avoid fractional powers of $\varepsilon$, the persistent time is again rescaled as $\tau \to \varepsilon^2 \tau$. Using the Ansatz
\al
M_n(x,t) = \sum_{l=0}^\infty \varepsilon^l M_n^l(x,t), \quad \forall\, n < N,
\eal
truncated beyond $N=5$, we obtain at order $\cO(\varepsilon^0)$
\al
M_0^0(X) = \frac{\cB(X)}{\cZ}, \quad \cB(X) = e^{-U(X)/D_{\rm \scriptscriptstyle AOUP}^{ap}}, \quad \cZ = \int_\mathbb{R} e^{-U/D_{\rm \scriptscriptstyle AOUP}^{ap}}, \label{eq:order0_apAOUP}
\eal
where $D_{\rm \scriptscriptstyle AOUP}^{ap} = D(1-\zeta_{ap})^2$ and $\zeta_{ap} = v/D^{1/2}$ as defined in App.~\ref{app:apAOUP}. The first order correction in $\cO(\varepsilon^1)$ vanishes, although the symmetry of persistent AOUPs $(v,\xi) \to -(v,\xi)$ is absent in $\CL_{\rm apAOUP}$. The first nonvanishing perturbative correction is thus of order $\cO(\varepsilon^2)$ and reads
\al
M_0^2(X) = M_0^0(X) \left( \frac{ \tau \zeta_{ap} (\zeta_{ap}-2)}{ (\zeta_{ap}-1)^2} \left[ U''(X) - \frac{U'(X)^2}{ 2 D(1-\zeta_{ap})^2 } \right] - \cN \right), \label{eq:order2_apAOUP}
\eal
where
\al
\cN = \int_{X'=-\infty}^\infty M_0^0(X') \frac{ \tau \zeta_{ap} (\zeta_{ap}-2)}{ (\zeta_{ap}-1)^2} \left[ U''(X') - \frac{U'(X')^2}{ 2 D(1-\zeta_{ap})^2 } \right] \,.
\eal
This time, a violation of the symmetry of persistent AOUPs $(v,\{M_k\}_{k \in \mathbb{N}}) \to (-v,\{ (-1)^k M_k \}_{k \in \mathbb{N}} )$ can be observed in $M_0^2$, as expected. The stationary distribution up to order $\mathcal{O}(\varepsilon^2)$ is thus given by
\al
M_0(X) = \exp\!\left\{ - \frac{U(X) + \varepsilon^2 U_2(X) }{D_{\rm \scriptscriptstyle AOUP}^{ap}} \right\} + \mathcal{O}(\varepsilon^3), \label{eq:perturbative_M0_apAOUP}
\eal
where the first nonvanishing perturbative correction reads
\al
U_2(X) = \frac{ \tau \zeta_{ap} (\zeta_{ap}-2)}{ (\zeta_{ap}-1)^2} \left( \frac{U'(X)^2}{2} - D_{\rm \scriptscriptstyle AOUP}^{ap} U''(X) \right) \,. 
\eal
It is the main result of this section and we note that $\tau \zeta_{ap} (\zeta_{ap}-2)/(\zeta_{ap}-1)^2$ has a negative sign for any (antipersistent) value of $\zeta_{ap}$. These perturbative results for antipersistent AOUPs, notably Eqs.~\ref{eq:order0_apAOUP},\ref{eq:order2_apAOUP}, agree with the exact results obtained for an harmonic trap in App.~\ref{app:OU}. It allows us to comment in a fair way the results we obtained for antipersistent MPs, as we do in the next section.

\subsubsection{Discussion of the perturbative results}

\label{app:perturb_comments}

We now discuss the analytical expressions we obtained for the stationary distribution of the different models of particles (MPs, AOUPs) in a generic confining potential $U$. The first nonvanishing perturbative corrections of these stationary measures can be cast under the same functional form
\al
p(x) \propto \exp\!\left\{ -\frac{1}{D_{\rm eff}} \left( U(x) + \varepsilon \, \cC_{\rm eff} \left[ \frac{U'(X)^2}{2} - D_{\rm eff} U''(X) \right] \right) \right\} + \mathcal{O}(\varepsilon^2) , \label{eq:perturb_smalltime}
\eal
where for MPs
\al
D_{\rm eff} & = D_{\rm \scriptscriptstyle MP} = \frac{D}{1+\chi}, \\
\cC_{\rm eff} & = - \frac{\chi}{\alpha(1+\chi)},
\eal
while for persistent AOUPs ($\chi<0$)
\al
D_{\rm eff} & = D_{\rm \scriptscriptstyle AOUP}^{p} = D (1+\zeta_p) = \frac{D}{(1+\chi)^2}, \\
\cC_{\rm eff} & = \frac{\tau \zeta_p}{1+\zeta_p} = - \frac{\chi (2+\chi)}{\alpha(1+\chi)},
\eal
and for antipersistent AOUPs ($\chi>0$)
\al
D_{\rm eff} & = D_{\rm \scriptscriptstyle AOUP}^{ap} = D (1-\zeta_{ap})^2 = \frac{D}{(1+\chi)^2}, \\
\cC_{\rm eff} & = \frac{ \tau \zeta_{ap} (\zeta_{ap}-2)}{ (\zeta_{ap}-1)^2} = - \frac{\chi (2+\chi)}{\alpha(1+\chi)} \,. \label{eq:eff_apAOUP}
\eal

For this comparison to be fair, two remarks are in order. First one needs to  match the scaling $\tau \to \varepsilon^2 \tau$ for  (anti)persistent AOUPs and the scaling $\tau \to \varepsilon \tau$ that we used for MPs. Given that the first perturbative correction of AOUPs is vanishing, their first nonvanishing correction turns out to be of the same order as the one for MPs. Second, we chose to express any parameter of these stationary measures in terms of the ones of MPs, $\chi$ and $\alpha$, as in App.~\ref{app:matching}, to highlight that these different types of particles are matched in the absence of interaction.
From Eqs.~\ref{eq:perturb_smalltime}-\ref{eq:eff_apAOUP}, we find that the stationary measures of MPs and AOUPs quantitatively differ (via $D_{\rm eff},\cC_{\rm eff}$) in presence of interactions at first order in perturbation  $\varepsilon$, although there are indistinguishable in free space. More precisely,  in the persistent case ($\chi<0$), the correction ($\cC_{\rm eff}$) is positive both for persistent MPs and AOUPs, while it is negative in the antipersistent case ($\chi>0$)  for both antipersistent MPs and AOUPs. The analysis beyond first order in perturbation is left for future works.

\subsection{Perturbation theory for small polarity amplitude}

\label{app:smallamplitudepertub}

We underline in the main text that subtle effects concerning MPs are arising in a confining box of size $L$, notably in the limit where $L \ll \ell_{D} = \sqrt{D \tau_m}$ or $\ell_{\star} = \sqrt{D_\star \tau_m}$. This notably means that the perturbation theory in small nonmarkovian time $\tau_m$ we discuss in the previous section might not be best suited to capture such effects. In this section, we thus develop an alternative perturbation theory in small nonmarkovian amplitude. Although the development of such a perturbation theory is a priori conceivable for generic potentials, we only deal in this section with the simpler case of a hard box potential, in accordance with what we need for the main text. Again, both MPs and AOUPs are treated.

\subsubsection{MPs}

\label{app:squarewell_MP_smallamp}

We derive in this section the stationary probability distribution of positions for MPs in an infinite depth square well defined as
\al
U(x) = 
\begin{cases}
	0, \ \text{for} \ x \in [0,L], \\
	\infty, \ \text{otherwise} \,.
\end{cases}
\eal 
To do so, we perturbatively close the hierarchy of moments of Eqs.~\ref{eq:MP_perturbativeexp_smallamp_1}-\ref{eq:MP_perturbativeexp_smallamp_k} for $x \in [0,L]$, where the potential term now translates into the following null-flux or reflecting boundary conditions at the walls
\al
\begin{cases}
J_x(x=0/L,z) = 0, \ \text{for} \ z \in \mathbb{R}, \ \\
J_x(x,z) = - \kappa\left(x-z\right) p(x,z) - D \p_x p(x,z),
\end{cases}
\eal
which gives the following conditions for the reduced moments in $x=0$ or $L$
\al
D \p_x m_n + \kappa (x m_n - m_{n+1}) = 0, \ \text{for} \ n \geq 0 \,. \label{eq:NF_BC}
\eal
We choose this time to work in the regime of small polarity $p$, that we achieve by rescaling $\kappa$ by $\varepsilon$ and taking the limit $\varepsilon \to 0$. We expand all moments in a series of $\varepsilon$ 
\al
m_n(x,t)=\sum_{l=0}^{\infty} \varepsilon^l m_n^l(x,t),
\eal
and look for the coefficients $m_n^l(x,t)$ solving the stationary state of Eqs.~\ref{eq:MP_perturbativeexp_smallamp_1}-\ref{eq:MP_perturbativeexp_smallamp_k} with the null-flux boundary conditions of Eq.~\ref{eq:NF_BC}. \\

At order $\mathcal{O}(\varepsilon^0)$, we find that $m_0^0$ solves $D \p_x^2 m_0^0 = 0$, which gives, together with the reflecting boundary conditions at $\varepsilon=0$ and the normalization condition, $m_0^0(x)=1/L$. $m_0^0$ follows the equilibrium Boltzmann measure as expected. For $m_1^0$, we obtain from the equation
\al
D \p_x^2 m_1^0 - \alpha m_1^0 = - \frac{\alpha x}{L},
\eal
completed by the null-flux condition, that it reads
\al
m_1^0(x) = \frac{x}{L} + \frac{\ell_{D}^{0}}{L} \sech\left(\frac{L}{2\ell_{D}^{0}}  \right)\sinh\left(\frac{L/2-x}{\ell_{D}^{0}}\right),
\eal
where $\ell_{D}^{0}=(D/\alpha)^{1/2}$ is the $\kappa \to 0$ limit of $\ell_D = \sqrt{D \tau_m}$. At order $\mathcal{O}(\varepsilon^1)$, the first order correction, $m_0^1$, obeys
\al
D \p_x^2 m_0^1 + \kappa \p_x (x m_0^0 - m_1^0) = 0, \label{eq:order1_MP_hardbox} \,.
\eal
Enforcing again the null-flux boundary conditions and the normalization of the probability distribution at any order in $\varepsilon$, we find
\al
m_0^1(x) = - \frac{\chi }{L} \sech \left(\frac{L/2}{\ell_{D}^{0}}\right)  \left[\cosh \left(\frac{L/2-x}{\ell_{D}^{0}}\right) - \frac{\ell_{D}^{0} }{L/2} \sinh \left(\frac{L/2}{ \ell_{D}^{0}}\right)\right],
\eal
where $\chi=\kappa/\alpha$ is a dimensionless number which captures the strength of the memory that drives the MP as defined in App.~\ref{app:matching}. The expression of $m_0^0 + \varepsilon m_0^1$ is the main result of this section, it corresponds to Eq.~12 of the main text and we comment it in App.~\ref{app:discussion_hardbox}, once similar expressions for the AOUPs are obtained.

It is moreover possible to reinterpret this result in term of the forces felt by the MPs and to make the connection with Sec.~\ref{app:conditionalproba}. More precisely, using the marginalized form of the Fokker-Planck equation for $x$ derived in Sec.~\ref{app:conditionalproba} and expanding it order by order in $\varepsilon$ with
\al
& p(x) = m_0(x) = m_0^0(x)+\varepsilon m_0^1(x) + ... \, , \\
& \mu^{\rm \scriptscriptstyle MP}(x) = 0 + \varepsilon \mu^{\rm \scriptscriptstyle MP}_1(x) + \varepsilon^2 \mu^{\rm \scriptscriptstyle MP}_2(x) + ... \, ,
\eal
we find at order $\mathcal{O}(\varepsilon)$
\al
D \p_x^2 m_0^1 - \p_x ( \mu_1^{\rm \scriptscriptstyle MP}(x) m_0^0 ) = 0 \label{eq:order1_marginalizedFP}
\eal
By identifying Eq.~\ref{eq:order1_marginalizedFP} with Eq.~\ref{eq:order1_MP_hardbox}, we find the expression for the mean force due to memory at first order in $\varepsilon$
\al
\mu_1^{\rm \scriptscriptstyle MP}(x) = - \kappa \left( x - \frac{m_1^0}{m_0^0} \right) = \kappa \ell_{D}^{0} \sech{\left(\frac{L/2}{\ell_{D}^{0}} \right)}\sinh\left(\frac{L/2-x}{\ell_{D}^{0}}\right) \,.
\eal

\subsubsection{Persistent AOUPs}

We derive in this section the stationary profile for a persistent AOUP in a hard confining box in the limit of a small driving velocity $v$, in order to compare it to the ones obtained for MP. 

To do so, we rescale the speed $v$ by $\varepsilon v$ in the limit $\varepsilon \to 0$ and we again solve perturbatively the hierarchy of moments in Eqs.~\ref{eq:moment0_pAOUP}–\ref{eq:momentk_pAOUP}, order by order in powers of small $\varepsilon$. The null-flux condition at the walls imposed on the current $J_X(X,\xi)= \left( v \xi  - D \p_X \right) p(X,\xi)$ by the confining potential translates into the following condition for the moments
\al
D \p_X M_k(X) - v M_{k+1}(X) = 0, \ \text{for} \ X = 0 \ \text{or} \ L \,.
\eal
Proceeding similarly as in Sec.~\ref{app:squarewell_MP_smallamp}, we find at order $\mathcal{O}(\varepsilon^0)$
\al
M_0^0(X) = \frac{1}{L}, \quad M_1^0(X) =0 , \quad M_2^0(X) = \frac{1}{L \tau},
\eal
while at order $\mathcal{O}(\varepsilon^1)$
\al
M_0^1(X) = 0, \quad M_1^1(X) = - \frac{v}{L \ell_D} \sech{\left(\frac{L/2}{\ell_D}\right)} \sinh{\left(\frac{L/2-X}{\ell_D} \right)} ,
\eal
where $\ell_D=\sqrt{D \tau}$ is the standard diffusion length for a persistent AOUP. The first-order correction in $v$ to $M_0$ is thus vanishing as it is known in the literature~\cite{martin2021statistical} and which is a consequence of the symmetry $(v,\{M_k\}_k) \to (-v,\{ (-1)^k M_k \}_k )$ that we already pointed out. It it worth to note that this does not happen for MPs where the first-order perturbative correction does not vanish (because there is no such symmetry). The second-order correction is however nonvanishing and we find
\al
M_0^2(X) = \frac{\zeta_p}{L} \sech \left(\frac{L/2}{\ell_D}\right)  \left[\cosh \left(\frac{L/2-X}{\ell_D}\right) - \frac{\ell_D}{L/2} \sinh \left(\frac{L/2}{\ell_D}\right) \right],
\eal
where $\zeta_p=v^2/D$ is a dimensionless number which compares the strength of the selfpropulsion with respect to the one thermal noise, as introduced in App.~\ref{app:matching}. This is the main result of this section and we comment it in App.~\ref{app:discussion_hardbox}.

Again, to make connection with App.~\ref{app:conditionalproba}, we can find the mean selfpropulsion force that acts on the position $X$ of an AOUP via the perturbative solution of the moments hierarchy. Going to order $\mathcal{O}(\varepsilon^2)$ because the first order is vanishing, we obtain 
\al
\mu_1^{\rm \scriptscriptstyle pAOUP}(X) = 0, \ \mu_2^{\rm \scriptscriptstyle pAOUP}(X) = v \frac{M_1^1}{M_0^0} = -\frac{v^2}{\ell_D} \sech{\left(\frac{L/2}{\ell_D}\right)} \sinh{\left(\frac{L/2-X}{\ell_D} \right)} \,.
\eal
This perturbative expansion matches to first order numerical simulations as shown in the inset of Fig.~$2$a.

\subsubsection{Antipersistent AOUPs}

In this section, we similarly derive the stationary distribution for antipersistent AOUPs in the hard confining box, in the limit of a small driving velocity $v$, in order to make a fair comparison with antipersistent MPs. 

Adapting \emph{mutatis mutandis} the reasoning of the previous sections, we rescale the speed $v$ by $\varepsilon v$ in the limit of $\varepsilon \to 0$ to perturbatively solve the hierarchy in Eqs.~\ref{eq:moment0_apAOUP}-\ref{eq:momentk_apAOUP}, completed by the null-flux boundary condition at the walls for the current $J_X(X,\xi) = - \left(v \xi + D \p_X + 2\sqrt{D}/\tau \p_\xi \right) p(X,\xi) $, which translates for the reduced moments into
\al
\begin{cases}
    D \p_X M_0 + v M_{1} = 0, \\
    D \p_X M_k + v M_{k+1} - \frac{ 2\sqrt{D} }{ \tau } k M_{k-1} = 0,
\end{cases}
\ \text{for} \ X = 0 \ \text{or} \ L, \ \text{and for} \ k \geq 1 \,.
\eal
This null-flux boundary condition is different than the one for persistent AOUPs and it is in fact the main difference at first order in perturbation, which is thus responsible for depletion instead of accumulation in the vicinity of the hard walls. Following the same steps as in the previous sections, we find at order $\mathcal{O}(\varepsilon^0)$
\al
M_0^0(X) = \frac{1}{L}, \quad M_1^0(X) = - \frac{2\sqrt{D}}{L \ell_D} \sech{\left(\frac{L/2}{\ell_D}\right)} \sinh{\left(\frac{L/2-X}{\ell_D}\right)},
\eal
while at order $\mathcal{O}(\varepsilon^1)$
\al
M_0^1(X) = - \frac{2 \zeta_{ap}}{L} \sech \left(\frac{L/2}{\ell_D}\right)  \left[ \cosh \left(\frac{L/2-X}{\ell_D}\right) - \frac{\ell_D}{L/2} \sinh \left(\frac{L/2}{\ell_D}\right) \right] \,.
\eal
It is the main result of this section and we comment it in the next section. Contrary to persistent AOUPs where it is necessary to go to order $2$ in perturbation due to the symmetry $(v,\xi) \to -(v,\xi)$, the mean force $M_1^0(X)$ felt by $X$ due to the motor $\xi$ is not vanishing at zeroth order, which creates an order-one perturbation of the position profile $M_0^1$. Moreover, compared to the perturbative limit of antipersistent AOUPs at small $\tau$, the first-order perturbative correction does not vanish due to the finite null-flux boundary condition.

\subsubsection{Discussion of the perturbative results}

\label{app:discussion_hardbox}

We discuss in this section the perturbative expressions we obtained for the stationary probability distribution of the positions of MPs and AOUPs (persistent and antipersistent) in a confining hard box. The first perturbative corrections of these stationary measures can be cast under the same functional form
\al
p(x) = \frac{1}{L} + \varepsilon \, \frac{\cC_{\rm eff}}{L} \sech \left(\frac{L/2}{\ell_{\rm eff}}\right)  \left[\cosh \left(\frac{L/2-x}{\ell_{\rm eff}}\right) - \frac{\ell_{\rm eff}}{L/2} \sinh \left(\frac{L/2}{ \ell_{\rm eff}}\right)\right],
\eal
where for MPs
\al
& \ell_{\rm eff} = \ell_{D}^{0}, \\
& \cC_{\rm eff} = - \chi,
\eal
while for persistent AOUPs
\al
& \ell_{\rm eff} = \ell_D = \frac{\ell_{D}^{0}}{\sqrt{1+\chi}}, \\
& \cC_{\rm eff} = \zeta_p = - \frac{\chi (2+\chi)}{(1+\chi)^2} \varepsilon,
\eal
and for antipersistent AOUPs
\al
& \ell_{\rm eff} = \ell_{D} = \frac{\ell_{D}^{0}}{\sqrt{1+\chi}}, \\
& \cC_{\rm eff} = -2\zeta_{ap} = -\frac{2\chi}{1+\chi} \ \text{or} \ -\frac{2(2+\chi)}{1+\chi} \,.
\eal
Here, given that the different types of particles are matched in the absence of interaction, we choose to work as in App.~\ref{app:matching} by expressing any parameter in terms of the ones of MPs, $\chi$ and $\ell_D^0$, so that a fair comparison is possible. 

We can now observe that, for the persistent case ($\chi<0$), the correction ($\cC_{\rm eff}$) is positive both for persistent MPs and AOUPs, while it is negative in the antipersistent case ($\chi>0$) both for antipersistent MPs and AOUPs. This means that persistent behavior yields to accumulation at the walls, while antipersistence to depletion, as already discussed in the main text. The detailed shapes are displayed against numerical simulations in the insets of Figs.~$2$a,b of the main text.

\section{Simulations details}

\label{app:simu}

We provide in this section the simulation details for all figures in the main text and in the supplementary material. \\

All Langevin dynamics are integrated with an Euler–Maruyama scheme. Unless stated otherwise, persistent MPs and AOUPs are parameter-matched in the non-interacting limit through Eq.~\ref{eq:matching_pAOUP_pMP}. Antipersistent MPs, antipersistent AOUPs and equilibrium dimers are analogously matched via Eq.~\ref{eq:matching}.

\subsection*{Fig.~$1$} 

For Figs.~$1$c–f, we simulate (anti)persistent MPs, persistent AOUPs and equilibrium dimers according to their evolution equations, Eqs.~$7,8$ and $10$ of the main text and Eq.~\ref{eq:antipersistentOUP} of the supplementary material. The dynamics are integrated with a time step $\Delta t = 10^{-4}$, and averages are taken over $N = 10^{6}$ independent trajectories. The noise strength is fixed to $D = 0.1$.\\

The parameters defining each type of particles are chosen as follows. For persistent AOUPs and MPs, we set $v^{2}=100$ and $\tau=10$. For antipersistent MPs and equilibrium dimers, we use $\alpha = 0.2$ and $\kappa = 1$. \\

Each system is initialized in its stationary state (with or without a harmonic trap) by evolving it for $100,\tau_m$, where $\tau_m = (\alpha + \kappa)^{-1}$. In the trap-and-release protocols we impose a harmonic potential of stiffness $k=10$, while $k=0$ is used in the control protocols. \\

After stationarity is reached, the trap is removed ($k=0$) and we record two observables: $(i)$ the polarity correlator $c(t,T)=\langle p(T)p(t+T) \rangle$ measured over a window of duration $5\tau_m$~; $(ii)$ the mean-squared increments $\Delta(t,T)=\langle (x(T+t)-x(T))^2 \rangle$ monitored over $500 \, \tau_m$.

\subsection*{Fig.~$2$}

All simulations are performed in a 1D square-well potential of width $L$, implemented numerically using reflecting boundary conditions at the walls at positions $x=0$ and $L$. \\

\paragraph*{Fig.~$2$a.} We set $L=1$. Persistent AOUPs and MPs are simulated with $v^2=10, \ \tau=10$ and $D=0.1$. For the inset, to compare with the small-nonmarkovian perturbative theory, we use weaker self-propulsion amplitudes $v^2=0.02$ with $\tau=1$. In both cases, stationary distribution $p(x)$ are obtained from histograms over $N=10^7$ independent particles, using respectively $100$ and $20$ bins. We take $\Delta t=10^{-2}$ and allow relaxation for $100 \, \tau_m$, where $\tau_m=(\alpha+\kappa)^{-1}$. \\

\paragraph*{Fig.~$2$b.} We again set $L=1$. Antipersistent AOUPs, MPs, and equilibrium dimers are simulated with $v^2=0.05, \ \tau=0.5$ and $D=0.1$. For the inset, we decrease self-propulsion to $v^2=0.0005$ with $\tau=0.5$. Stationary distributions are computed as in Fig.~$2$ and similarly stationarized. \\

\paragraph*{Fig.~$2$c.} Same numerical parameters as in Fig.~$2\mathbf{(a)}$ except for the box size. We report results for $L=0.2$ (dark green), $L=1$ (medium green) and $L=5$ (orange). \\

\paragraph*{Fig.~$2$d.} Same numerical parameters as in Fig.~$2\mathbf{(b)}$, with box sizes $L=0.2$ (dark green), $L=1$ (brown) and $L=5$ (magenta). \\

\paragraph*{Fig.~$2$e.} Persistent AOUPs and MPs are simulated with $v^2=1, \tau=10$ and $D=0.1$. In the main figure, we determine via simulations the stationary polarity correlator $c(t)$ for $L=0.2, 5$ and $50$. MP curves appear in fluo green, yellow-green and orange respectively. All AOUP curves are in red and coincide. The black line shows the analytical AOUP result of Eq.~\ref{eq:correlator_polarity_AOUP}.

In the inset, we include additional box sizes $L=0.2, 1., 2., 5., 10., 20. 50.$ and $100$. The red line shows $c(0)$ for AOUPs as given by Eq.~\ref{eq:correlator_polarity_AOUP}. The orange-to-green line shows the analytical prediction of $c(0)$ in Eq.~\ref{eq:correl_pp_0} for MPs confined in an harmonic trap. To relate the trap stiffness $k$ to the box size $L$, we match the variance of a Brownian particle in both geometries:
\al
\langle x^2 \rangle = \frac{L^2}{12} = \frac{D}{k} \,. \label{eq:match_harmonic_hardbox}
\eal
Although only qualitative agreement is expected, the correspondence is semi-quantitatively accurate. \\

\paragraph*{Fig.~$2$f.} Antipersistent AOUPs and MPs are simulated with $v^2=0.05, \tau=0.5$ and $D=0.1$. In the main figure, we determine via simulations $c(t)$ for $L=0.2, 0.5, 1.$ and $10.$ that are respectively drawn in fluo green, green dark green and magenta. All antipersistent AOUP curves are in blue and coincide. The black line shows the analytical AOUP result of Eq.~\ref{eq:correlator_polarity_AOUP}.

In the inset, we use the same parameters and box sizes, with additional box sizes $L=0.2, 0.5, 1., 2., 5.$ and $10$. The blue line shows the analytical AOUP value of $c(0)$ and the magenta-to-green points follow the MP analytical prediction of Eq.~\ref{eq:correl_pp_0} with the same harmonic-hard-box matching as above (Eq.~\ref{eq:match_harmonic_hardbox}). The agreement is even more quantitative in this antipersistent regime.

\subsection*{Fig.~$3$}

All simulations in this figure use periodic boundary conditions and in what follows forces originating from the potential landscape are accordingly computed with their periodic images.

\subsubsection*{First row}

We first consider one-dimensional systems. The isolated trap is modeled by the double-barrier potential
\al
U(x) = U_0 \left( \exp{-\frac{(x-\chi L)^2}{2 \sigma^2}} + \exp{-\frac{(x-(1-\chi)L)^2}{2 \sigma^2}} \right) \,.
\eal
with parameters $U_0=1.2, \chi=0.45$ and $\sigma=0.05$. The system size is set to $L=5$, yielding an effective trap width $\ell_{tr}=(1-2\chi)L = L/10 = 0.5$. \\

We use a time step $\Delta t = 10^{-3}$ to resolve the steep potential profile. Simulations are run until a stationary state is reached, i.e. for times much larger than all relevant dynamical timescales, including the barrier-crossing time. To accelerate convergence, particles are initially sampled from the Boltzmann distribution associated with $U(x)$. To ensure statistical independence of barrier-crossing events for these nonmarkovian dynamics, we check that the memory time $\tau_m=(\alpha+\kappa)^{-1}$ is shorter than the approximate Brownian barrier-crossing time $\exp(U_0/D)$ associated to the barrier heights. \\

\paragraph*{Fig.~$3$a.} For persistent AOUPs and MPs, we use $v^2=2500$, $\tau=5$ and $D=0.3$. Stationary distributions $p(x)$ are obtained from histograms over $N=10^6$ independent trajectories, comprising $250$ bins. In the inset, we compute the ratio of probabilities of being inside the trap (integrated between $x=\chi L$ and $x=(1-\chi)L$) for persistent MPs versus persistent AOUPs as a function of $\tau\in\{0.1,0.5,1,2,5\}$, keeping $v^2=2500$ and $D=0.3$. \\

\paragraph*{Fig.~$3$b.} For antipersistent MPs and equilibrium dimers, we choose $\kappa=3$, $\alpha=\kappa/2$, and $D=0.3$ to enhance antipersistent effects. Distributions $p(x)$ are obtained from histograms over $N=10^6$ independent trajectories with $200$ bins. In the inset, we compute the trap-occupation probability ratio for antipersistent MPs versus equilibrium dimers as a function of $\kappa\in\{0.01,0.02,0.05,0.1,0.2,0.35,0.7,3\}$, keeping $\alpha=\kappa/2$ and $D=0.3$.

\subsubsection*{Second row}

The second row shows two-dimensional systems in a corrugated potential landscape composed of Gaussian pillars of amplitude $U_0$ and width $\sigma$, located on a square lattice of spacing $4\sigma$:
\al
U(\mathbf{r}) = U_0 \sum_{i=1}^{2n+1} \sum_{j=1}^{2n+1} \exp{-\frac{(\mathbf{r}-\mathbf{c}_{i,j})^2}{2 \sigma^2}},
\eal
where
\al
\mathbf{c}_{i,j} = \begin{pmatrix} c_x^{(i)} \\ c_y^{(i)} \end{pmatrix} = \begin{pmatrix} L/2 + 4(i-n-1)\sigma \\ L/2 + 4(j-n-1)\sigma \end{pmatrix} \,.
\eal
We choose $L=5$ and $n=2$. The barrier height between the saddle points and the central wells is approximately
\al
\Delta E = E_s - E_w \approx 2 U_0 e^{-2} - 4 U_0 e^{-4} \approx 0.198 U_0 \,.
\eal
To maintain a reasonable barrier-crossing probability, we choose $U_0=5$ and $\sigma=0.05$. \\

\paragraph*{Fig.~$3$d.} For persistent AOUPs and MPs, we use $v^2=2500$, $\tau=3$ and $D=0.3$. The two-dimensional heat maps are obtained from histograms over $N=10^6$ independent particles, binned into $150\times150$ bins. \\

\paragraph*{Fig.~$3$e.} For antipersistent MPs and equilibrium dimers, we use $\kappa=0.8$, $\alpha=\kappa/2$ and $D=0.3$. Heat maps are obtained from $N=5\times 10^8$ trajectories binned into $500\times500$ bins.

\subsection*{Fig.~$4$}

We simulate $N$ interacting MPs in two dimensions with periodic boundary conditions. The overdamped Langevin equations for particle positions $\mathbf r_i(t)$ read
\al
\begin{cases}
\dot{\mathbf r}_i & \! \! = -\kappa (\mathbf{r}_i - \mathbf{z}_i) - \nabla_i U(\{\mathbf r\}) + \sqrt{2D }\,\boldsymbol\eta_i(t),\\
\dot{\mathbf z}_i & \! \! = \alpha (\mathbf{r}_i - \mathbf{z}_i),
\end{cases}
\qquad \forall i \in \llbracket 1, N \rrbracket \,.
\eal
The noise terms $\boldsymbol\eta_i$ are Gaussian white noises whose first and second moments read
\al
& \langle\eta_{i,\alpha}(t)\rangle=0, \\
& \langle\eta_{i,\alpha}(t)\eta_{j,\beta}(t')\rangle=\delta_{ij}\delta_{\alpha\beta}\delta(t-t') \,.
\eal
The pair-wise interaction potential is 
\al
U(\{\mathbf r\})=\sum_{i<j} u(r_{ij}),\qquad r_{ij}=|\mathbf r_i-\mathbf r_j|,
\eal
where $u$ is a soft-sphere harmonic repulsive potential of the form
\al
u(r) = 
\begin{cases}
\frac{k}{2} \left(2 r_0-r\right)^2 & \ \text{if} \ r < r_0,\\
0 & \ \text{otherwise,}
\end{cases} \label{eq:softsphere}
\eal
with particle radius $r_0=1$ and stiffness $k=20$.

For comparison, we also simulate AOUPs in two dimensions evolving via
\al
\begin{cases}
\dot{\mathbf r}_i & \! \! = v \boldsymbol\xi_i - \nabla_i U(\{\mathbf r\}) + \sqrt{2D }\,\boldsymbol\eta_i(t),\\
\tau \dot{\boldsymbol\xi}_i & \! \! = - \boldsymbol\xi_i + \sqrt{2}\,\boldsymbol \lambda_i,
\end{cases}
\qquad \forall i \in \llbracket 1, N \rrbracket,
\eal
where $\boldsymbol\eta_i$ and $\boldsymbol\xi_i$ are two independent Gaussian white noises with zero mean and unit variance:
\al
& \langle\eta_{i,\alpha}(t)\rangle=\langle\xi_{i,\alpha}(t)\rangle=0, \\
& \langle\eta_{i,\alpha}(t)\eta_{j,\beta}(t')\rangle=\delta_{ij}\delta_{\alpha\beta}\delta(t-t'), \\
& \langle\xi_{i,\alpha}(t)\xi_{j,\beta}(t')\rangle=\delta_{ij}\delta_{\alpha\beta}\delta(t-t') \,.
\eal
We set $D=0.1$ to operate in a low-noise regime. The system size is $L \times L=400 \times 400$, which determines the number of particles $N$ through the global packing fraction $\overline{\phi}$. We use a time step $\Delta t = 0.01$ and evolve the system up to $t=10^4$. \\

\paragraph*{Figs.~$4$a–d.} The local packing fraction $\phi$ is computed from disk-averaged particle counts within a radius $10\,r_0$.\\ 

\paragraph*{Fig.~$4$e.} The purple, gold, and salmon curves correspond to Gaussian fits of the distributions obtained numerically. The black curve is a smoothing spline of the numerical data, computed using the function \verb|make_smoothing_spline| in the \verb|scipy| library \verb|interpolate|, with \verb|lam|\(=10^{-5}\). These fits are used solely to locate the maxima of $P(\phi)$ and as a guide for the eye. \\

\paragraph*{Fig.~$4$f.} We report the positions of the local maxima of $P(\phi)$, extracted from the spline fits described above, as a function of the control parameters $\overline{\phi}$ and $\tau$.

\subsection*{Fig.~$5$}

In Fig.~$5$, interacting antipersistent MPs are simulated in the same manner as in Fig.~$4$. Interacting equilibrium dimers are also simulated, and their overdamped Langevin dynamics read
\al
\begin{cases}
\dot{\mathbf r}_i & \! \! = -\kappa' (\mathbf{r}_i - \mathbf{z}_i) - \nabla_i U(\{\mathbf r\}) + \sqrt{2D }\,\boldsymbol\eta_i(t),\\
\dot{\mathbf z}_i & \! \! = r \kappa' (\mathbf{r}_i - \mathbf{z}_i) + \sqrt{2 D r}\,\boldsymbol\eta_{Z,i}(t),
\end{cases}
\qquad \forall i \in \llbracket 1, N \rrbracket,
\eal
where $\boldsymbol\eta_i$ and $\boldsymbol\eta_{Z,i}$ are two independent Gaussian white noises with
\al
& \langle\eta_{i,\alpha}(t)\rangle=\langle\eta_{Z,i,\alpha}(t)\rangle=0, \\
& \langle\eta_{i,\alpha}(t)\eta_{j,\beta}(t')\rangle=\delta_{ij}\delta_{\alpha\beta}\delta(t-t'), \\
& \langle\eta_{Z,i,\alpha}(t)\eta_{Z,j,\beta}(t')\rangle=\delta_{ij}\delta_{\alpha\beta}\delta(t-t') \,.
\eal
We set $D=0.1$ to work in a regime of low thermal noise. The systems are made of $N=1000$ particles, which fixes the linear size of the simulation box through the global packing fraction.\\

\paragraph*{Fig.~$5$a–d.} The global packing fraction is set to $\overline{\phi}=0.72$. Both antipersistent MPs and equilibrium dimers use parameters $\alpha=0.5$ and $\kappa=1.0$. Trajectories are plotted using logarithmically spaced time sampling. \\

\paragraph*{Fig.~$5$c.} The MSD is computed as $\overline{\left(x(t)-x(0)\right)^2}$ where the average $\overline{\ \cdot \ }$ denotes an average over all antipersistent MPs. 
The curves correspond to global packing fractions $\overline{\phi}=0.6, 0.62, 0.64, 0.66, 0.67, 0.68, 0.69, 0.7$. \\

\paragraph*{Fig.~$5$d.} The effective diffusion $D_{\rm eff}$ is extracted from a 
linear fit of the MSD in the interval $t \in [10^3, 5\times10^5]$, corresponding to the long-time diffusive regime. Critical densities are defined as those values of $\overline{\phi}$ for which 
$D_{\rm eff}$ decreases by two orders of magnitude relative to its dilute value (i.e. $D_{\rm eff}$ for the smallest density available in the figure). 

Because thermal noise is weak, the liquid–solid transition is dominated by geometric crowding. 
Confined antipersistent MPs behave as if they had a larger effective interaction radius $r_0+\delta r$, 
whereas equilibrium dimers (following Boltzmann measure) keep their physical radius $r_0$. Consequently, if the critical packing fractions are matched
\al
\overline{\phi}_{{\rm \scriptscriptstyle Brownian},c} = \overline{\phi}_{{\rm \scriptscriptstyle eq. dim.},c} = \pi r_0^2 \rho_c = \overline{\phi}_{{\rm \scriptscriptstyle apMP, c}} = \pi (r_0+\delta r)^2 \rho_c,
\eal
where $\rho_c$ stands for the number density at the transition in each case, this defines a $\delta r$, whose values are reported in the inset. 
The colored curves in Fig.~$5$d are obtained via spline interpolation (as in Fig.~$4$) and serve as guides for the eye.

\subsection*{Supplementary Material: Fig.~\ref{fig:matching}}

\paragraph*{Fig.~$2$a.} Persistent AOUPs and MPs are simulated with parameters $\alpha=3.0$, $\kappa=-2.9$ and $D=0.1$. The systems are first stationarized for $10 \, \tau_m$, where $\tau_m=(\alpha+\kappa)^{-1}$. The mean-squared increments $\Delta(t)=\langle \left( x(t)-x(0))^2 \right) \rangle$ are then measured over a window of $2 \, \tau_m$ and $100$ uniformly spaced time points are displayed in the figure. The dynamics is integrated using a time-stepping $\Delta t =10^{-4}$ and averaged over $10^4$ independent particles. \\

\paragraph*{Fig.~$2$b.}  Antipersistent AOUPs, MPs and equilibrium dimers are simulated with $v=0.28$, $\tau=1.$ and $D=0.1$. The systems are first stationarized for $10 \, \tau_m$, where $\tau_m=(\alpha+\kappa)^{-1}$. The mean-squared increments $\Delta(t)=\langle \left( x(t)-x(0))^2 \right) \rangle$ are then measured over a window of $5 \, \tau_m$, using again $100$ uniformly spaced time points. We use a time step $\Delta t=10^{-3}$ and average over $10^5$ independent particles. 

\subsection*{Supplementary Material: Fig.~\ref{fig:SRL_protocols}}

The shaken Rouse polymer is simulated using the overdamped Langevin dynamics of Eq.~\ref{eq:SRP}. The chain contains $51$ monomers, with only the central monomer subjected to Gaussian white noise of variance $2 k_{\mathrm B}T$ (we set $k_{\mathrm B}T=1$). The friction coefficient $\zeta$ and all spring constants between the monomers are set to unity. The system is first stationarized up to $t = 10^6$ before sampling the mean-squared increments (MSI). \\

For the trap–release protocol, the central monomer is confined in the past by a harmonic trap of stiffness $k=100$. After release, the dynamics is integrated up to $t = 2\times 10^4$, and $\Delta(t,T)$ is recorded. The time-stepping is set to $\Delta t = 3\times 10^{-2}$. Ensemble averages are taken over $10^4$ independent realizations of the shaken Rouse polymer. \\

The grey curves correspond to logarithmic fits of the MSI in the intermediate regime between the two crossover times, $t=1$ and $t=828$. The fit is obtained using the \verb|polyfit| function of the \verb|numpy| library with \verb|deg=1| applied to the pair $(\log{t}, \Delta(t,T))$.

\bibliography{biblio.bib}